\documentclass[12pt]{article}
\usepackage[utf8]{inputenc}
\usepackage{amsmath}
\usepackage{amsfonts}

\usepackage{amssymb}

\usepackage{float}

\def\lb{\label}
\def\be{\begin{equation}}
\def\ee{\end{equation}}
\def\qed{\rule{5pt}{5pt}}

\newcommand{\mcP}{\mathcal{P}}
\newcommand{\hC}{\widehat{C}}
\newcommand{\hc}{\hat{c}}
\newcommand{\bc}{\bar{c}}
\newcommand{\bI}{\mathbf{I}}
\newcommand{\bP}{\mathbf{P}}
\newcommand{\bK}{\mathbf{K}}

\newcommand{\mfg}{\mathfrak{g}}

\newcommand{\oproj}{\overline{\rm P}}

\newcommand{\p}[1]{(\ref{#1})}
\newcommand{\bea}{\begin{eqnarray}}
\newcommand{\eea}{\end{eqnarray}}

\newcommand{\ba}{\begin{array}} \newcommand{\ea}{\end{array}}

 \newcommand{\nn}{\nonumber}
 
 \newcommand{\cP}{{\cal P}}
\newcommand{\aA}{{{\mathbb{A}\mathbb{C}}}}
\newcommand{\sS}{{{\mathbb{S}\mathbb{C}}}}

\DeclareMathOperator{\ad}{ad}
\DeclareMathOperator{\tr}{\bf Tr}

\DeclareMathOperator{\proj}{P}

\begin{document}




\begin{center}
{\huge\bf Split Casimir operator for simple Lie algebras,
 solutions of Yang-Baxter equations and Vogel parameters}
\end{center}
\vspace{1cm}

\begin{center}
{\Large \bf  A.P.Isaev${}^{a,b,c}$, S.O.Krivonos${}^{a}$}
\end{center}

\vspace{0.2cm}

\begin{center}
{${}^a$ \it
Bogoliubov  Laboratory of Theoretical Physics,\\
Joint Institute for Nuclear Research,
141980 Dubna, Russia}\vspace{0.1cm} \\
{${}^b$ \it
St.Petersburg Department of
 Steklov Mathematical Institute of RAS,\\
Fontanka 27, 191023 St. Petersburg,  Russia}\vspace{0.1cm} \\
{${}^c$ \it Faculty of Physics,
Lomonosov Moscow State University, Moscow, Russia}\vspace{0.3cm}

{\tt isaevap@theor.jinr.ru, krivonos@theor.jinr.ru}
\end{center}
\vspace{3cm}

\begin{abstract}\noindent
  We construct characteristic identities for the split (polarized)
  Casimir operators of the simple Lie algebras in defining
  (minimal fundamental) and adjoint representations.
  By means of these characteristic identities,
  for all simple Lie algebras
  we derive explicit formulae for invariant projectors onto
  irreducible subrepresentations in $T^{\otimes 2}$ in two cases, when
   $T$ is the defining and the  adjoint
   representation. In the case when $T$ is the defining representation,
   these projectors and the split Casimir operator are used
   to explicitly write down invariant
   solutions of the Yang-Baxter equations. In the case when $T$ is the adjoint representation, these projectors and characteristic identities are considered from the viewpoint of the universal description of
   the simple Lie algebras in terms of the Vogel parameters.
\end{abstract}

\newpage
\setcounter{page}{1}
\setcounter{equation}{0}

\tableofcontents


 \vspace{0.5cm}

\section{Introduction}

It is known that the special invariant operator, the split (or polarized)
Casimir operator $\hC$ (see definition in Section {\bf \ref{splkaz}}), plays an important role both in the description of the Lie algebras $\mathfrak{g}$ themselves and in the studies    of their representation theory.
On the other hand, the split Casimir operator $\hC$ is the building block (see e.g. \cite{ChPr}, \cite{Ma} and references therein)  for constructing $\mathfrak{g}$-invariant solutions $r$ and $R$
of semiclassical and quantum Yang-Baxter equations
 \be
 \lb{cYBE}
 [r_{12}(u), \, r_{13}(u+v)] + [r_{13}(u+v), \, r_{23}(v)] +
 [r_{12}(u), \, r_{23}(v)] = 0 \; ,
 \ee
 \be
 \lb{YBE}
 R_{12}(u) \, R_{13}(u+v) \, R_{23}(v) =
 R_{23}(v) \, R_{13}(u+v) \, R_{12}(u) \; .
 \ee
 We use here the standard matrix notation  which will be explained below. Recall \cite{Drin0} that
 $\mathfrak{g}$-invariant rational solutions of the Yang-Baxter   equations (\ref{YBE}) allow one to define the Yangians $Y(\mathfrak{g})$  within the so-called $RTT$-realization.

In this  paper, we demonstrate the usefulness of  the $\mathfrak{g}$-invariant split Casimir operator $\hC$ in the
 representation theory of Lie algebras. Namely,  for all simple Lie algebras $\mathfrak{g}$, explicit formulas are found for invariant projectors onto irreducible representations  that appear in the expansion of the tensor product $T \otimes T$ of two representations $T$. These projectors  are constructed  in terms of the   operator $\hC$ for two cases, when $T$ is the defining (minimal fundamental) and when  $T \equiv \ad$ is the adjoint representation of $\mathfrak{g}$.

It is natural to find such invariant projectors in terms of $\mathfrak{g}$-invariant operators, which in turn
 are images of special elements of the so-called  centralizer algebra. The idea of this approach is not new. For example,
invariant projectors acting in tensor representations of $s\ell(N)$ algebras are called Young symmetrizers and constructed as images of special elements (idempotents) of group algebra $\mathbb{C}[S_r]$ of the symmetric group
 $S_r$. The algebra $\mathbb{C}[S_r]$  centralizes the action of the $s\ell(N)$  in the space of tensors of rank $r$ (in the spaces of   representations $T^{\otimes r}$). In this paper, we consider a very particular problem of constructing
invariant projectors in representation spaces of $T^{\otimes 2}$, where $T$ is the defining, or adjoint representation but for all simple Lie algebras $\mathfrak{g}$. Our approach is closely related to the one outlined in \cite{Cvit}.
In \cite{Cvit}, such invariant projectors were obtained in terms of several special  invariant operators and the calculations were performed using a peculiar diagram technique. In our approach, we try to construct invariant projectors in the representation space $V^{\otimes 2}$ of $T^{\otimes 2}$ by using only one $\mathfrak{g}$-invariant operator which is the split Casimir operator $\hC$. It turns out that for all simple Lie algebras
$\mathfrak{g}$ in the defining representations  all invariant projectors in $V^{\otimes 2}$ are constructed as polynomials
in $\hC$. It is not the case for the adjoint representation,  i.e. not for all algebras $\mathfrak{g}$ the invariant projectors  in $V_{\ad}^{\otimes 2}$ are constructed as polynomials  of only one operator $\hC_{\ad} \equiv \ad^{\otimes 2}\hC$. Namely, in the case of $s\ell(N)$ and $so(8)$ algebras  there are additional $\mathfrak{g}$-invariant operators which are independent of $\hC_{\ad}$ and act, respectively, in the antisymmetrized and symmetrized parts of the space $V_{\ad}^{\otimes 2}$. We construct such additional operators explicitly in Sections {\bf \ref{slad}} and {\bf 3.2.1}.

Our study of the split Casimir operator $\hC$ was motivated by the works  \cite{MkSV}, \cite{MkrV}, \cite{MMM}
 and \cite{MMM1}, and by the idea of finding formulas for solutions of the Yang-Baxter equation expressed in terms of only the operator $\hC$. For defining (minimal fundamental) representations of the simple Lie algebras  $\mathfrak{g}$ (except for the algebra $\mathfrak{e}_8$), such formulas were derived in this paper (see equations (\ref{ybesl}), (\ref{best2}),   (\ref{Rg2}), (\ref{Rf4}), (\ref{Rce6}) and (\ref{Rce7}) below). Note that these formulas are obtained by using well-known
 \cite{OgWieg}, \cite{NJMK} spectral decompositions for   rational $R$-matrices\footnote{All these spectral decompositions
 can also be obtained from the spectral decompositions of trigonometric $R$-matrices (see e.g. \cite{Ma} (section 7.2.4),
 \cite{Kuni}, \cite{Serg})  in the special limit $q \to 1$.}.  For the adjoint representations of the simple Lie algebras $\mathfrak{g}$,  as it was argued in \cite{Drin1},  there are no such formulas  (we need to extend the adjoint representations of algebras $\mathfrak{g}$; see \cite{ChPr1}, \cite{Nicol}).   However, in the case of the adjoint representation
  of $\mathfrak{g}$,    the knowledge of the characteristic identities for $\hC_{\ad}$ turns out to be a key point for understanding the so-called universal formulation of the simple Lie algebras \cite{Fog} (see also the historical notes in \cite{Cvit}, section 21.2). Though some characteristic identities
 and formulas for certain    $\mathfrak{g}$-invariant projectors can be found in a different form in \cite{Cvit},
  we believe that the methods we used and the results obtained can be useful for future research,
e.g. from the viewpoint of technical applications  of the split Casimir operator.

In our paper, to simplify the notation, we everywhere write $s\ell(N)$, $so(N)$  and $sp(2n)$ instead of $s\ell(N,\mathbb{C})$, $so(N,\mathbb{C})$  and $sp(2n,\mathbb{C})$.

\section{Split Casimir operator for simple Lie algebras\label{splkaz}}
\setcounter{equation}0

\subsection{General definitions}

Let $\mathfrak{g}$ be a simple Lie algebra with  the basis $X_a$ and defining relations
\be\lb{lialg}
[X_a, \; X_b] = C^d_{ab} \; X_d \; ,
\ee
 where $C^d_{ab}$ are the structure constants. The Cartan-Killing  metric is defined in the standard way
\be\lb{li04}
   {\sf g}_{ab} \equiv C^{d}_{ac} \, C^{c}_{bd} =   \tr(\ad(X_a)\cdot \ad(X_b)) \; ,
\ee
where $\ad$ denotes adjoint representation: $\ad(X_a)^d_b=C^{d}_{ab}$. Recall that the
 structure constants $C_{abc} \equiv C^{d}_{ab} \, {\sf g}_{dc}$  are antisymmetric under permutation of indices
 $(a,b,c)$. We denote an enveloping algebra of the Lie algebra  $\mathfrak{g}$ as ${\cal U}(\mathfrak{g})$.
 Let ${\sf g}^{df}$ be the inverse matrix to the Cartan-Killing  metric (\ref{li04}). We use this matrix and construct the operator
 \be
 \lb{kaz-01}
\hC  = {\sf g}^{ab} X_a \, \otimes \,
  X_b \;\; \in \;\; \mathfrak{g} \, \otimes \,  \mathfrak{g}
   \;\; \subset \;\; {\cal U}(\mathfrak{g})\, \otimes \, {\cal U}(\mathfrak{g}) \; ,
 \ee
 which is called the {\it split (or polarized) Casimir operator} of the Lie algebra $\mathfrak{g}$.
This operator is related to the usual quadratic Casimir operator
 \be
 \lb{kaz-c2}
 C_{(2)} = {\sf g}^{ab} \; X_a \cdot X_b \;\; \in \;\;
 {\cal U}(\mathfrak{g}) \; ,
  \ee
 by means of the formula
 \be
 \lb{adCC1}
 \Delta(C_{(2)}) = C_{(2)} \otimes I + I \otimes C_{(2)} + 2 \, \hC  \; ,
 \ee
 where $\Delta$ is the standard comultiplication
 for enveloping algebras ${\cal U}(\mathfrak{g})$:
 \be
 \lb{mrep2}
 \Delta(X_a) = (X_a \otimes I + I \otimes X_a) \; .
 \ee

The following statement holds (see, for example,
 \cite{Book1}, \cite{Okub}, \cite{ToHa}).
  \newtheorem{pro1}{Proposition}[subsection]
\begin{pro1}\label{pro1}
The operator $\hC$, given in (\ref{kaz-01}), does not depend on
the choice of the basis in $\mathfrak{g}$
and satisfies the condition (which is called $\ad$-invariance or
$\mathfrak{g}$-invariance):
 \be
 \lb{kaz-02}
[\Delta(A), \, \hC  ]=
[(A \otimes I + I \otimes A), \, \hC  ]= 0 \; , \;\;\;\;
\forall A \in \mathfrak{g} \; ,
 \ee
 where $\Delta$ is comultiplication (\ref{mrep2}). In addition, the operator
 $\hC$ obeys the equations
 \be
 \lb{kaz-03}
 [\hC _{12}, \, \hC _{13} + \hC _{23} ] = 0 \;\;\; \Rightarrow \;\;\;
[\hC _{13}, \, \hC _{23} ]
= \frac{1}{2} \; [\hC _{12}, \, \hC _{13} - \hC _{23} ] \; ,
 \ee
which use the standard notation
  \be
 \lb{kaz-01ya}
 \hC _{12} = {\sf g}^{ab} X_a \, \otimes \,  X_b \, \otimes \, I \; , \;\;\;
 \hC _{13} = {\sf g}^{ab} X_a \, \otimes \, I \, \otimes \,  X_b  \; , \;\;\;
 \hC _{23} = {\sf g}^{ab} I \, \otimes \,  X_a \, \otimes \,  X_b  \; .
 \ee
 Here $I$ is the unit element in ${\cal U}(\mathfrak{g})$ and
 $\hC _{ij} \in {\cal U}(\mathfrak{g}) \, \otimes \,
{\cal U}(\mathfrak{g}) \, \otimes \, {\cal U}(\mathfrak{g})$.
  \end{pro1}

 \vspace{0.2cm}

Relations (\ref{kaz-03}) indicate that the split Casimir operator (\ref{kaz-01}) realizes the Kono-Drinfeld Lie algebra
  and can be used as a building block for constructing solutions to the quasi-classical (\ref{cYBE}) and quantum (\ref{YBE}) Yang-Baxter equations. In particular, the solution for the quasi-classical Yang-Baxter equation (\ref{cYBE})
 is the operator $r(u) = \hC/u$ (see e.g. \cite{Ma}).

 \noindent
 {\bf Remark.} Let the normalization of the generators $H_i, E_\alpha$   in the Cartan-Weyl basis of the algebra $\mathfrak{g}$ be chosen so that  we have for (\ref{kaz-c2}) and (\ref{kaz-01}):
 \be
 \lb{kazeh}
 C_{(2)} = g^{ij} (H_i \, H_j) +
 \sum_{\alpha} (E_\alpha \, E_{-\alpha}) \;\;\; \Rightarrow \;\;\;
 \hC = g^{ij} (H_i \otimes H_j) +
 \sum_{\alpha} (E_\alpha \otimes E_{-\alpha}) \; ,
 \ee
 where the sum goes over all roots $\alpha$  and $g^{ij}$ is the
 inverse matrix to the metric in the root space
 \be
 \lb{metrKK}
 g_{ij} = \sum\limits_{\alpha} \alpha_i \alpha_j \; .
 \ee
Then the split Casimir operator is decomposed into the sum $\hC = (r_{+} + r_{-})$   of two solutions $r = r_{+}$ and $r = r_{-}$
 of a constant (i.e. independent of the spectral parameter) semiclassical Yang-Baxter equation
 $[r_{12},r_{13}]+[r_{12},r_{23}]+[r_{13},r_{23}]=0$. These solutions
 are written in the form (see e.g. \cite{ChPr},\cite{Ma})
 $$
 r_{+} = \frac{1}{2} g^{ij} (H_i \otimes H_j) +
 \sum_{\alpha >0} (E_\alpha \otimes E_{-\alpha}) \; , \;\;\;
  r_{-} = \frac{1}{2} g^{ij} (H_i \otimes H_j) +
 \sum_{\alpha >0} (E_{-\alpha} \otimes E_{\alpha}) \; ,
 $$
 where the sum goes over all positive roots  $\alpha > 0$ of the algebra $\mathfrak{g}$.

\subsection{The split Casimir operator for simple  Lie algebras in the adjoint representation\label{hCad}}
 \setcounter{equation}0

The generators  $X_a$ of a simple Lie algebra $\mathfrak{g}$ satisfy   the defining relations (\ref{lialg}) and, in the adjoint representation,  $X_a$ are implemented as matrices  ad$(X_a)^d_{\; b} = C^d_{ab}$. In this case
 the split Casimir operator (\ref{kaz-01}) is written as
 \be
 \lb{adCC}
 (\hC_{\ad})^{a_1 a_2}_{\; b_1 b_2} \equiv
 (\ad \otimes \ad)^{a_1 a_2}_{\; b_1 b_2}  (\hC) =
 C^{a_1}_{h b_1} \, C^{a_2}_{f b_2} \, {\sf g}^{h f} \; .
\ee
By definition this operator satisfies identities (\ref{kaz-03}). Below we need one more $\ad$-invariant operator
 \be
 \lb{adK}
 ({\bf K})^{a_1 a_2}_{b_1 b_2} =
 {\sf g}^{a_1 a_2} \; {\sf g}_{b_1 b_2} \; .
\ee
The operators (\ref{adCC}) and (\ref{adK}) act in the tensor product  $V_{\ad} \otimes V_{\ad}$ of two spaces
 $V_{\ad} = \mathfrak{g}$ of the adjoint representation   and have the symmetry properties
 $(\hC_{\ad})^{a_1 a_2}_{\; b_1 b_2} = (\hC_{\ad})^{a_2 a_1}_{\; b_2 b_1}$
 and $\bK^{a_1 a_2}_{\; b_1 b_2} = \bK^{a_2 a_1}_{\; b_2 b_1}$, which are conveniently written in the form
$$
 (\hC_{\ad})_{21} = \bP(\hC_{\ad})_{12}\bP =
  (\hC_{\ad})_{12} \; , \;\;\;
 {\bf K}_{21} = \bP \, {\bf K}_{12} \, \bP = {\bf K}_{12} \; ,
 $$
where $1,2$  are numbers of spaces $V_{\ad}$  in the product $(V_{\ad} \otimes V_{\ad})$ and
 $\bP$ is permutation matrix in $(V_{\ad} \otimes V_{\ad})$:
 \be
 \lb{perm0}
 \bP (X_{a_1} \otimes X_{a_2}) = (X_{a_2} \otimes X_{a_1}) =
 (X_{b_1} \otimes X_{b_2}) \bP^{b_1 b_2}_{a_1 a_2} \; , \;\;\;
 \bP^{b_1 b_2}_{a_1 a_2} = \delta^{b_1}_{a_2} \, \delta^{b_2}_{a_1} \; .
 \ee
 Here $(X_{a} \otimes X_{b})$ is the basis in the space  $(V_{\ad} \otimes V_{\ad})$. Define the symmetrized and
 antisymmetrized parts of the operator $\hC_{\ad}$
 \be
 \lb{adCpm}
 (\hC_{\pm})^{a_1 a_2}_{b_1 b_2} =
 \frac{1}{2} ((\hC_{\ad})^{a_1 a_2}_{b_1 b_2} \pm
 (\hC_{\ad})^{a_2 a_1}_{b_1 b_2}) \; ,
 \;\;\;\;\;
  \hC_{\pm} = \bP^{(ad)}_{\pm} \; \hC_{\ad}  =
  \hC_{\ad}\; \bP^{(ad)}_{\pm} \; ,
 \ee
 where $\bP^{(ad)}_{\pm} = \frac{1}{2} (\bI \pm \bP)$ and  $\bI$ is the unit operator in $(V_{\ad})^{\otimes 2}$.
  \newtheorem{proCK}[pro1]{Proposition}
\begin{proCK}\lb{proCK}
 The operators $\hC_{\ad}$, $\hC_{\pm}$ and ${\bf K}$, given in  (\ref{adCC}), (\ref{adK})
 and (\ref{adCpm}), satisfy the identities
 \be
 \lb{idCC}
 \hC_{-}^2 = - \frac{1}{2}  \hC_{-}  \; ,
 \ee
  \be
 \lb{idK}
 \hC_{-} \, {\bf K} = 0 = {\bf K} \, \hC_{-}  \; , \;\;\;
 \hC_{\ad} \, {\bf K} = {\bf K} \, \hC_{\ad} = - {\bf K} \; ,
 \ee
 \be
 \lb{idK2}
 \hC_{+} \, {\bf K} = {\bf K} \, \hC_{+} = - {\bf K} \; .
 \ee
 \end{proCK}
 {\bf Proof.} To prove equality (\ref{idCC}), we note that $\hC_{-}$ has a
 useful expression followed from  the Jacobi identity  $C^{d}_{a b} C^{r}_{dc}  + {\rm cycle}(a,b,c)  = 0$  (see e.g. \cite{Cvit}):
 \be
 \lb{idkm}
 (\hC_{-})^{a_1 a_2}_{b_1 b_2} = - \frac{1}{2}
  \, C^{a_1 a_2}_d \; C^d_{b_1 b_2} \; , \;\;\;\;\;
 C^{a_1 a_2}_d  \equiv C^{a_1}_{d \, b_2} \; {\sf g}^{b_2 a_2} \; .
 \ee
 Using this expression and identities
 \be
 \lb{casad}
 C^d_{b_1 b_2} C^{b_1 b_2}_a = \delta^d_a \;\;\;\;\;\;\;
 \Leftrightarrow \;\;\;\;\;\;\;
  {\rm ad}(C_{(2)})^f_{\;\; r} =
 {\sf g}^{ab} \, C^{f}_{a\, d} C^{d}_{b\, r} = \delta^f_r\;,
 \ee
 which are equivalent to the definition (\ref{li04}) of the Cartan-Killing  metric, we calculate $\hC_{-}^2$ and obtain (\ref{idCC}).  The first equality in (\ref{idK}) follows from the evident relations
$(\bI - \bP){\bf K} = 0 = {\bf K} (\bI - \bP)$. The second  equality in (\ref{idK}) is proved with the
help of  identities (\ref{casad}) and complete antisymmetry  of the constants $C_{a b c} = C^{d}_{a b}\, {\sf g}_{dc}$.
 Relations (\ref{idK2}) are derived from (\ref{idK}). \hfill \qed

 \vspace{0.2cm}

Now we take into account definitions (\ref{adCC}), (\ref{adK}), (\ref{perm0}) and relations (\ref{idCC}), (\ref{idkm}),
 (\ref{casad}), and  $C_{ba}^a = 0$, which is valid for all simple Lie algebras, and obtain general formulas for the traces
 \be
 \lb{trac1}
 \begin{array}{c}
 {\bf Tr}(\hC_{\ad}) = 0\,  , \;\;\;
 {\bf Tr}(\hC_{\pm}) = \pm \frac{1}{2} \dim \mathfrak{g} \, , \;\;\;
 {\bf Tr} (\hC_{\ad}^2) = \dim \mathfrak{g} \, , \\ [0.3cm]
 {\bf Tr}(\hC_{-}^2) = - \frac{1}{2}  {\bf Tr}(\hC_{-}) =
   \frac{1}{4}  \dim \mathfrak{g}\; , \;\;\;
   {\bf Tr}(\hC_{+}^2) = {\bf Tr}(\hC_{\ad}^2 - \hC_{-}^2) =
   \frac{3}{4}  \dim \mathfrak{g}\; , \\ [0.3cm]
    {\bf Tr}(\bK) = \dim \mathfrak{g} \, , \;\;\;
    {\bf Tr}(\bI ) = (\dim \mathfrak{g})^2 \, , \;\;\;
      {\bf Tr}(\bP) = \dim \mathfrak{g} \; .
 \end{array}
 \ee
 where ${\bf Tr} \equiv {\rm Tr}_{1}{\rm Tr}_{2}$ is the trace  in the space $V_{\ad} \otimes V_{\ad}$
 (indices $1$ and $2$ are attributed to factors in the product  $V_{\ad} \otimes V_{\ad}$). These formulas will be
  used in what follows.

Using the characteristic identity (\ref{idCC}) for the operator $\hC _{-}$, one can construct two mutually orthogonal projectors
\be
 \lb{XX12}
 \proj_1 = - 2 \, \hC_{-} \; , \;\;\;\;
 \proj_2 =  2 \, \hC_{-} + \bP^{(\ad)}_{-} \;\;\;\;\; \Rightarrow \;\;\;\;\;
  \proj_i \proj_k = \proj_i \; \delta_{ik} \; ,
 \ee
which decompose the antisymmetrized  part $\bP^{(ad)}_{-} (\ad \otimes \ad)$
of the representation $(\ad \otimes \ad)$ into two subrepresentations ${\sf X}_{1,2} = \proj_{1,2}  (\ad \otimes \ad)$.
Dimensions of these subrepresentations are equal to the  traces
  of corresponding projectors (\ref{XX12})
 \be
 \lb{XX123}
 \dim {\sf X}_1 = {\bf Tr}(\proj_1) =
 \dim \mathfrak{g} \; , \;\;\;\;\;\;
 \dim {\sf X}_2 = {\bf Tr}(\proj_2) =
 \frac{1}{2} \dim \mathfrak{g}\;  (\dim \mathfrak{g} - 3) \; ,
 \ee
 where we use the general formulae (\ref{trac1}). Since the constants  $C^d_{b_1 b_2}$ play the role of the Clebsch-Gordan coefficients for the  fusion $\ad^{\otimes 2} \to \ad$, we see from the explicit form
 (\ref{idkm}) of the operator $\hC_{-}$ that the projector $\proj_1$, given in (\ref{XX12}), extracts  the adjoint representation  ${\sf X}_1 = \ad$ in $\bP^{(ad)}_{-} (\ad^{\otimes 2})$. Thus, the adjoint representation
 is always contained in  the antisymmetrized part $\bP^{(ad)}_{-} (\ad^{\otimes 2})$.
 The first formula in (\ref{XX123}) confirms  the  equivalence of ${\sf X}_1$ and $\ad$.
Note also that ${\sf X}_2$   is not necessarily irreducible representation for all simple Lie algebras.
  As we will see below (see Remark after Proposition  {\bf \em \ref{prdop}}), the representation
 ${\sf X}_2$ is reducible for algebras of the series  $A_n = s\ell(n+1)$.

 \subsection{The split Casimir operator for highest weight representations}
 \setcounter{equation}0

The invariant metric for a simple Lie algebra $\mathfrak{g}$ is uniquely determined up to a normalization constant
(see e.g. \cite{Book1}), i.e. such metric is always proportional to the Cartan-Killing metric (\ref{li04}). Therefore, for any irreducible representation $T$ of the simple Lie algebra $\mathfrak{g}$ we have
  \be
  \lb{kaz-11b}
  {\rm Tr}\bigl(T(X_a) \cdot T(X_b)\bigr)  =
  {\sf d}_2(T) \, {\sf g}_{ab} \; ,
  \ee
where the coefficient ${\sf d}_2(T)$ characterizes the representation $T$.
 Indeed, ${\sf d}_2(T)$ is expressed in terms of values $c_2^{(T)}$ of the quadratic Casimir (\ref{kaz-c2}) in the irreducible representation of $T$ using the well-known relation
 $$
 c_2^{(T)} \; {\rm dim}(T) =
 {\sf d}_2(T)\; {\rm dim}(\mathfrak{g}) \; ,
 $$
 which is obtained from (\ref{kaz-11b}) by contraction with  the inverse metric ${\sf g}^{ab}$.

Let $T_{\lambda_1}$ and $T_{\lambda_2}$ be two irreducible  representations with the highest weights $\lambda_1$ and $\lambda_2$  acting in the spaces  ${\cal V}_{\lambda_1}$ and ${\cal V}_{\lambda_2}$. Let
 the representation  $T_{\lambda_1} \otimes T_{\lambda_2}$ be decomposed
 into irreducible representations $T_\lambda$ with the highest weights $\lambda$
 as follows: $T_{\lambda_1} \otimes T_{\lambda_2}
 = \sum_\lambda n_\lambda  T_\lambda$, where $n_\lambda$ is the multiplicity of occurrence of $T_\lambda$ in the expansion
 of $T_{\lambda_1} \otimes T_{\lambda_2}$. Denote the  space of the representation $T_\lambda$ as ${\cal V}_\lambda$.
 Then, from (\ref{adCC1}) and expansion  ${\cal V}_{\lambda_1}\otimes {\cal V}_{\lambda_2} =
 \sum_\lambda n_\lambda {\cal V}_\lambda$, we obtain
 \cite{Okub}
 \be
 \lb{adCC2}
 \begin{array}{c}
 T_{(\lambda_1\times \lambda_2)} (\hC )\cdot
 ({\cal V}_{\lambda_1}\otimes {\cal V}_{\lambda_2}) =
 \frac{1}{2} \sum_\lambda n_\lambda \;
 (c_{2}^{(\lambda)} -c_{2}^{(\lambda_1)} -c_{2}^{(\lambda_2)})
  {\cal V}_\lambda \; \Leftrightarrow \; \\ [0.2cm]
 T_{(\lambda_1\times\lambda_2)}(\hC )\cdot {\cal V}_\lambda =   \frac{1}{2}
 (c_{2}^{(\lambda)} -c_{2}^{(\lambda_1)} -c_{2}^{(\lambda_2)}) {\cal V}_\lambda  \; ,
 \end{array}
 \ee
 where we use the concise notation  $T_{(\lambda_1\times\lambda_2)}:=  (T_{\lambda_1} \otimes T_{\lambda_2})$. Here
 $c_{2}^{(\lambda)}$ is the value of the quadratic Casimir operator
 $C_{(2)}$, defined in (\ref{kazeh}), in the representation  with the highest weight $\lambda$
 \be
 \lb{kvkaz}
 c_{2}^{(\lambda)} = (\lambda,\lambda + 2 \, \delta) \; , \;\;\;\;
 \delta  := \sum_{f=1}^r \lambda_{(f)} =
 \frac{1}{2} \sum_{\alpha >0} \alpha \; ,
 \ee
 $\lambda_{(f)}$ are the fundamental weights of the rank $r$ Lie algebra  $\mathfrak{g}$, $\alpha$ are the roots of $\mathfrak{g}$ and   summation is over positive roots $(\alpha >0)$.  Note that the operator $T(\hC )$ is diagonalizable
 for simple Lie algebras  and in general its spectrum is degenerate. Therefore, formula (\ref{adCC2}) implies the characteristic identity
 \be
 \lb{char01}
 {\prod_{\lambda}}' \Bigl(T_{(\lambda_1 \times \lambda_2)}(\hC ) -
 \hat{c}^{\; \lambda}_{\lambda_1,\lambda_2} \Bigr) = 0 \; ,
 \;\;\;\; \hat{c}^{\;\lambda}_{\lambda_1,\lambda_2} := \frac{1}{2}
 (c_{2}^{(\lambda)} -c_{2}^{(\lambda_1)}
 -c_{2}^{(\lambda_2)}) \; ,
 \ee
where the prime in $\prod'_\lambda$ means that the product  does not run over all weights $\lambda$ that participate
  in the expansion: $T_{\lambda_1} \otimes T_{\lambda_2}
 = \sum_\lambda n_\lambda  T_\lambda$ but only  those  $\lambda$ that correspond to unequal eigenvalues
 $\hat{c}^{\;\lambda}_{\lambda_1,\lambda_2}$.

\vspace{0.2cm}

In the next sections, we obtain explicit   expressions for the split Casimir operator
 $T_{(\lambda_1\times \lambda_2)}(\hC ) \equiv  (T_{\lambda_1}\otimes T_{\lambda_2})(\hC )$ for all simple
 Lie algebras in the case when both representations  $T_{\lambda_1}$ and $T_{\lambda_2}$ are either defining or adjoint.
 In the case when $T_{\lambda_1}$ and $T_{\lambda_2}$ are adjoint   representations
 $\lambda_1 = \lambda_2 = \lambda_{\ad}$,  the characteristic identity \eqref{char01} takes the form
 \begin{equation}
 \lb{char02}
 {\prod_{\lambda}}' \Bigl(\ad^{\otimes 2}(\hC)-
 \frac{1}{2}(c_{2}^{(\lambda)} - 2 c_{2}^{(\lambda_{\ad})}
 )\Bigr) \equiv  {\prod_{\lambda}}' \Bigl(\ad^{\otimes 2}(\hC)-
 \hat{c}_{2}^{(\lambda)} \Bigr)= 0 \; ,
 \end{equation}
 \begin{equation}
 \lb{char03}
 \hat{c}_{2}^{(\lambda)} :=
 \frac{1}{2}(c_{2}^{(\lambda)} - 2 c_{2}^{(\lambda_{\ad})})
 = \frac{1}{2} c_{2}^{(\lambda)} - 1  \; ,
 \end{equation}
 where $\lambda_{\ad}$ is the highest weight of the adjoint  representation of the algebra $\mathfrak{g}$, which is equal to   the highest root $\theta$ of $\mathfrak{g}$.  In the definition (\ref{char03}) of
 values $\hat{c}_{2}^{(\lambda)}$ of the operator  $\ad^{\otimes 2}(\hC)$  we use the condition (see (\ref{casad}))
  \be
  \lb{adjc2}
  c_{2}^{(\lambda_{\ad})}\equiv c_{2}^{(\theta)} = 1 \; .
  \ee
Note that this formula is consistent with (\ref{kvkaz}) only if the metric in the root space of the algebra
$\mathfrak{g}$ is given by   (\ref{metrKK}), which corresponds to the condition
 \be
 \lb{ducox}
 (\theta,\theta)= t^{-1} \; ,
 \ee
   where $t$ is the dual Coxeter number  of the algebra $\mathfrak{g}$.
In the next sections, we demonstrate this fact  explicitly and also find an explicit form for the
  characteristic identities (\ref{char02}) for all finite-dimensional simple Lie algebras.

 \section{Split Casimir operator for
  Lie algebras of classical series}

 \subsection{Split Casimir operator $\hC$ for Lie algebra $s\ell(N)$}
 \setcounter{equation}0

 \subsubsection{Operator $\hC$ for $s\ell(N)$ in the defining representation.\label{defsl}}

In this Subsection, to fix the notation, we give the standard definitions of the Lie
 algebra  $s\ell(N):= s\ell(N,\mathbb{C})$ and the corresponding operator $\hC$ in the defining representation.
 One can find these definitions  in many monographs and textbooks (see e.g.  \cite{Cvit}, \cite{Book1}).

We denote the space $\mathbb{C}^N$ of the defining representation  $T$ of the algebra $s\ell(N)$ as $V_N$.
Choose the basis  in $s\ell(N)$ consisting of traceless matrices
 \be
 \lb{defrep}
 T_{ij} = e_{ij} - \delta_{ij} \, I_N/N
 \;\;\;\;\;\;\; \Rightarrow \;\;\;\;\;\;\;
 T_{ij} = e_{ij} \;\;\;\; (i \neq j) \; , \;\;\;
 T_{ii} = e_{ii} - I_N/N \; ,
 \ee
 where $e_{ij}$ are $N \times N$ matrix units, and $I_N$ is the unit  $N \times N$ matrix. The elements (\ref{defrep})
 satisfy the defining relations
  \be
 \lb{defSL}
[T_{ij} , \, T_{km}] = \delta_{jk} \, T_{im} - \delta_{im} \, T_{kj} \equiv
 C^{rs}_{ij,km}\, T_{rs} \; ,
 \ee
 where for structure constants we have the explicit expression
 \be
 \lb{strSL}
C^{rs}_{ij,km} = \delta_{jk}\delta^{r}_{i}\delta^{s}_{m} -
 \delta_{im}\delta^{r}_{k}\delta^{s}_{j} \; .
 \ee
 Using this expression and definition (\ref{li04}), we  find the Cartan-Killing metric for $s\ell(N)$:
 \be
 \lb{metSL}
 {\sf g}_{ij,k\ell} =
 C^{rs}_{ij,mn} \, C^{mn}_{k\ell,rs} =
 2 (N \delta_{jk} \delta_{i\ell} -  \delta_{ij} \delta_{k\ell})
 \; = \; {\sf g}_{k\ell,ij} \; ,
 \ee
and, for the basis (\ref{defrep}) in the defining representation, equality (\ref{kaz-11b}) gives
 \be
 \lb{kaz-d}
 {\rm Tr} (T_{ij} \, T_{k\ell}) =
 \frac{1}{2N} \, {\sf g}_{ij,k\ell} \;\;\;\; \Rightarrow \;\;\;\;
 {\sf d}_2(T) = \frac{1}{2N} \; .
 \ee
 The inverse to (\ref{metSL}) metric ${\sf g}^{ij,k\ell}$  is defined by the traceless conditions
 ${\sf g}^{ii,k\ell} = 0 = {\sf g}^{ij,kk}$ and relations
 \be
 \lb{metSL0}
 {\sf g}_{mn,ij} \; {\sf g}^{ij,k\ell} = \bar{I}^{k\ell}_{mn} \; , \;\;\;\;\;\;\;\;
 \bar{I}^{k\ell}_{mn} \equiv
 \delta^{k}_m \delta^{\ell}_{n} -
 \frac{1}{N}\delta_{mn} \delta^{k\ell} \; ,
 \ee
 where the projector $\bar{I}^{k\ell}_{mn}$ plays the role  of the identity operator in the adjoint
 representation space  $V_{\ad}$; we identify $V_{\ad}$ with the  subspace of traceless tensors in $V_N \otimes V_N$, i.e.
 $V_{\ad} = \bar{I} \cdot V_N^{\otimes 2} $.
 As a result, we obtain
 \be
 \lb{metSL1}
 {\sf g}^{ij,k\ell} =
 \frac{1}{2N} \left( \delta^{jk} \delta^{i\ell} -
 \frac{1}{N} \delta^{ij} \delta^{k\ell} \right)
 \; =  \; {\sf g}^{k\ell,ij}   \; .
 \ee
 {\bf Remark.}
 Strictly speaking, $V_{\ad}$ is the subspace of second
 rank traceless tensors in $V_N \otimes \bar{V}_N$, where $\bar{V}_N$ is the space of the contragradient representation of $s\ell(N)$. In other words,  $V_{\ad}$ is the space of tensors with the components $\psi^{i}_{\; k}$ that satisfy the traceless property  $\psi^{i}_{\; i}=0$. Further, for technical reasons, we treat $\bar{V}_N$ as
  $V_N$ and consider $V_{\ad}$ as the space of traceless tensors in $V_N \otimes V_N$ with the components $\psi^{ik}$ such that   $\psi^{ii}=0$. The cases where the difference between $V_N$  and $\bar{V}_N$ is important are specially negotiated.

  \vspace{0.2cm}

 The matrix $T^{\otimes 2}(\hC)$ for the split Casimir operator   (\ref{kaz-01}) of the algebra $s\ell(N)$
  with basis (\ref{defrep}) in the defining representation   $T$ is written as
   \be
 \lb{kaz-gg2}
T^{\otimes 2}(\hC)^{i_1i_2}_{j_1j_2} = {\sf g}^{ij,k\ell} \,
(T_{ij} \otimes T_{k\ell})^{i_1i_2}_{j_1j_2} =
{\sf g}^{ij,k\ell} \,
(T_{ij})^{i_1}_{j_1} \; (T_{k\ell})^{i_2}_{j_2} =
\frac{1}{2N} \Bigl( \delta^{i_1}_{j_2}\delta^{i_2}_{j_1} -
    \frac{1}{N} \, \delta^{i_1}_{j_1}\delta^{i_2}_{j_2} \Bigr) \; ,
 \ee
 and in the index-free notation we have
 \be
 \lb{kaz-gg3}
 T^{\otimes 2}(\hC) = \frac{1}{2N} \, (P - \frac{1}{N} I) \equiv
  \hC_{_T} \; ,
 \ee
 where $I = I_N^{\otimes 2}$ is the unit operator in  $V_N^{\otimes 2}$ and $P$ is the permutation operator
 acting in the space  $V_N^{\otimes 2}$. Let $e_i$ $(i=1,...,N)$ be the basis vector in $V_N$, then the operator $P$ is defined as follows
 \be
 \lb{perm}
 P (v\otimes u) =  u\otimes v \;\;\;\;\;\; (\forall v, u \in V_N)
 \;\; \Rightarrow \;\; P (e_k\otimes e_m) =  e_m\otimes e_k =
 (e_i\otimes e_j) P^{ij}{}_{km} \; ,
 \ee
 i.e. the operator $P$ has the components  $P^{ij}{}_{km}=\delta^i_m\delta^j_k$ in the basis
 $(e_m\otimes e_k) \in V_N \otimes V_N$. We note that in view of (\ref{metSL0}) the equality holds
 \be
 \lb{kaz-05}
 {\rm dim}_{\mathbb{C}}\bigl( s\ell(N)\bigr) \equiv
 {\sf g}^{ij,k\ell}{\sf g}_{ij,k\ell}=(N^2-1) \; .
 \ee
 In addition, setting $ j_1 = i_2 $ in (\ref{kaz-gg2})  and summing over $ i_2 $, we obtain the value of the quadratic Casimir operator in the defining representation.  Besides, setting $j_1=i_2$  in (\ref{kaz-gg2}) and summing over $i_2$,
 we obtain the value of the quadratic Casimir operator in the defining representation
 \be
 \lb{sl33}
 T(C_{(2)}) =
 {\sf g}^{ij,k\ell} \,
(T_{ij} T_{k\ell})^{i_1}_{j_2} =
\frac{N^2-1}{2N^2}\; \delta^{i_1}_{j_2} \;\;\;\;\;
\Rightarrow \;\;\;\;\; c_2^{(T)} =  \frac{N^2-1}{2N^2}  \; .
 \ee
 This corresponds to (\ref{kvkaz}) if we fix metric (\ref{metrKK}) in the root space of the algebra
   $s\ell(N)$ so that the square of the lengths of all roots
   of  $s\ell(N)$ is equal to $1/N$; in particular
 for the highest root $\theta := \lambda_{\ad}$ we also have
 (cf. (\ref{ducox}))
 \be
 \lb{cst01}
 (\theta, \theta) = 1/N \; .
 \ee
 Finally, for the split Casimir operator (\ref{kaz-gg3}) of the algebra $s\ell(N)$ in the defining representation
  we obtain the characteristic identity
  \be
 \lb{slN}
 \hC_{_T}^2 + \frac{1}{N^2}\, \hC_{_T} + \frac{1-N^2}{4N^4}  = 0
 \;\;\;\;  \Leftrightarrow   \;\;\;\;
 \Bigl(\hC_{_T} +\frac{1+N}{2N^2}\Bigr)
 \Bigl(\hC_{_T} +\frac{1-N}{2N^2}\Bigr) = 0 \; ,
 \ee
 here and below the identity operator $I$ in $V_N \otimes V_N$ is replaced with $1$ for simplicity.
Identity (\ref{slN}) is consistent with formula  (\ref{char01}) when the root space metric
 is normalized in accordance with (\ref{cst01}).  In view of (\ref{slN}) the projectors onto eigen-spaces
 of the operator $\hC_{_T}$ in $V_N \otimes V_N$ have the form
 \be
 \lb{slN01}
P_{\pm} = \pm  \Bigl(N \, \hC_{_T} + \frac{1 \pm N}{2 N}  \Bigr) =
\frac{1}{2}(1 \pm P) \; ,
\ee
where $P_{+}$ and $P_{-}$ denote a symmetrizer and an antisymmetrizer, respectively.
Finally, the $s\ell(N)$-symmetric solution of the Yang-Baxter equation
  (\ref{YBE}) in the defining representation (which is called the Yang solution) is written in several equivalent ways (including the   form of $R(u)$ written in terms of the operator $\hC_{_T}$):
   \be
 \lb{ybesl}
 \begin{array}{c}
 \displaystyle
 R(u) = \frac{u + P}{1-u} =
 \frac{(u+1)}{(1-u)}P_{+} - P_{-}
 \;\; \Leftrightarrow  \;\;
 R(u) =
 \frac{P_{+} + u}{P_{+} - u} =
 \frac{N  \hC_{_T} + \frac{1 + N}{2 N} + u}{N  \hC_{_T} +
 \frac{1 + N}{2 N} - u} \, ,
 \end{array}
 \ee
 where $u$ is the spectral parameter. Solution (\ref{ybesl}) is unitary
 $P \, R(u) \, P \, R(-u) = R(u) \, R(-u) = 1$ and is defined up to
 multiplication by an arbitrary function $f(u)$ that satisfies  $f(u)f(-u)=1$.

 \subsubsection{Operator $\hC$ for $s\ell(N)$ in the adjoint representation.\label{slad}}

In this paper we use the standard index-free matrix notation.
Namely, let $A$ be an operator in
$V_N\otimes V_N$, where $V_N=\mathbb{C}^N$ is the space
of the defining representation of the algebra $s\ell(N)$.
The operator $A$ is defined by the relations
 $$
 A \; (e_k\otimes e_l)=e_i\otimes e_j A^{ij}{}_{kl} \; ,
 $$
 where $A^{ij}{}_{kl}$ are the components of $A$ in the basis
$\{e_i\otimes e_j\}_{i,j=1}^N \in V_N^{\otimes 2}$.
Then, $A_{ab}$ denotes the action of the operator $A$ in
the space $V_N^{\otimes 4}$ so that it is nontrivial
 only in the $a$-th and $b$-th factors in the product
 $V_N\otimes V_N\otimes V_N\otimes V_N$ (cf. (\ref{kaz-01ya})).
 For example, the operator $A_{13}$ has the components
 $$
 (A_{13})^{i_1i_2i_3i_4}_{\ \ j_1j_2j_3j_4}=A^{i_1i_3}{}_{j_1j_3}
 \delta^{i_2}_{j_2}\delta^{i_4}_{j_4}
 $$
 in the basis $e_{i_1} \otimes e_{i_2} \otimes e_{i_3} \otimes e_{i_4}
  \in V_N^{\otimes 4}$ and so on.
  In addition to the permutation operator $P$,
  which is defined in (\ref{perm}), we need one more operator $K$
   acting in $V_N^{\otimes 2}$:
\be
\lb{kkk}
K \cdot (e_m\otimes e_n) = (e_i\otimes e_j) \delta^{ij} \delta_{mn}
\;\;\; \Rightarrow \;\;\; K^{ij}_{\;\; mn} =
\delta^{ij} \delta_{mn} \; .
\ee
The operators $P_{ab}$ and $K_{ab}$ acting
in the space $V_N^{\otimes 4}$ satisfy the following
useful relations:
\be
\lb{kkpp}
\begin{array}{c}
K_{ab} = K_{ba} \; , \;\;\; K_{ab}^2 = N \, K_{ab}
\; , \;\;\; P_{ab} K_{ab} = K_{ab}
\; , \;\;\; K_{ab} K_{bc}  = K_{ab} P_{ac} =
P_{ac} K_{bc} \; , \\ [0.3cm]
K_{ab} K_{bc} K_{ab} = K_{ab} = K_{ab} P_{bc} K_{ab}\; , \;\;\;
 P_{ab} K_{\ad} K_{bc} = P_{cd} K_{\ad} K_{bc}  \; ,
\end{array}
\ee
which are specific to matrix representations of the Brauer algebra generators (see e.g. \cite{Book2}, \cite{Molev}).

The split Casimir operator $\hC_{\ad}$ for the algebra $s\ell(N)$ in the adjoint representation is given by formula (\ref{adCC}) and acts in
 $V_{\ad} \otimes V_{\ad}\subset V_N^{\otimes 4}$.
Taking into account the definitions (\ref{strSL})
and (\ref{metSL1}), we have
\be
\lb{kazSL}
\begin{array}{c}
 (\hC_{\ad})^{i_1 i_2 i_3 i_4}_{j_1 j_2 j_3 j_4} =
 {\sf g}^{k_1 k_2, k_3 k_4} \; {\rm ad} (T_{k_1 k_2})^{i_1 i_2}_{j_1 j_2}
 \; {\rm ad} (T_{k_3 k_4})^{i_3 i_4}_{j_3 j_4} = \\ [0.3cm]
 = {\sf g}^{k_1 k_2, k_3 k_4} \; C^{i_1 i_2}_{k_1 k_2,j_1 j_2}
 \; C^{i_3 i_4}_{k_3 k_4 ,j_3 j_4}  =
 \frac{1}{2N} (P_{13} + P_{24}
 - K_{14} - K_{23})^{i_1 i_2 i_3 i_4}_{j_1 j_2 j_3 j_4} \; .
 \end{array}
\ee
Note that here the operators ${\rm ad} (T_{i k})$
  act in the adjoint representation space $V_{\ad}$,
  which we identify with the space of the second rank traceless tensors
 $V_{\ad} \equiv \bar{I}\cdot V_N^{\otimes 2} =
 (I - \frac{1}{N}K) \cdot V_N^{\otimes 2}$, where the projector
 $\bar{I}$ was introduced in (\ref{metSL0}).
 It means that the indices of the adjoint representation
 are associated in formula (\ref{kazSL}) with
 the pairs of indices $(i_1 i_2)$, $(j_3j_4)$ etc.,
 possessing the traceless property, i.e. the contraction
  of the indices in each such pair gives zero.
  Setting in (\ref{kazSL}) $j_1 = i_3$ and $j_2 = i_4$ and summing over
  $i_3$ and $i_4$, we obtain the value of the quadratic Casimir operator
  (\ref{kaz-c2}) of the algebra $s\ell(N)$ in the adjoint representation
 $$
 \ad(C_{(2)})_{12} = {\rm Tr}_{34}(\hC_{\ad}P_{13} P_{24}) =
 \frac{1}{2N} {\rm Tr}_{34}\bigl( (P_{13} + P_{24}
 - K_{14} - K_{23})P_{13} P_{24}\bigr) = \bar{I}_{12} \; ,
 $$
 i.e. $c_{(2)}^{\ad}=1$, which is consistent with the general formula (\ref{casad}).

 Next, we need three more operators ${\bf K}$,
 ${\bf P}^{(ad)}$ and ${\bf P}$, where
 the first two act in $V_{\ad}^{\otimes 2}$, and
 the last one acts in $V_N^{\otimes 4}$.
 The operator ${\bf K}$ is defined as follows (cf. (\ref{kkk})):
\be
\lb{ksl}
\begin{array}{c}
{\bf K}^{i_1 i_2 i_3 i_4}_{j_1 j_2 j_3 j_4} =
{\sf g}^{i_1 i_2 i_3 i_4}\, {\sf g}_{j_1 j_2 j_3 j_4} =
\left( \delta^{i_2 i_3} \delta^{i_1 i_4} -
 \frac{1}{N} \delta^{i_1 i_2} \delta^{i_3 i_4} \right)
(\delta_{j_2 j_3} \delta_{j_1 j_4} -  \frac{1}{N} \delta_{j_1 j_2} \delta_{j_3 j_4}) = \\ [0.3cm]
= \left( K_{23} K_{14} - \frac{1}{N} P_{24} K_{12} K_{34}
- \frac{1}{N} P_{13} K_{23} K_{14}
+ \frac{1}{N^2} K_{12} K_{34}
\right)^{i_1 i_2 i_3 i_4}_{j_1 j_2 j_3 j_4} \; .
 \end{array}
\ee
The operator ${\bf P}$ permutes in the tensor product
$V_N^{\otimes 4}$
the first factor with the third one
and the second factor with the fourth one  and has an explicit form
\be
\lb{Psl}
{\bf P}^{i_1 i_2 i_3 i_4}_{j_1 j_2 j_3 j_4} =
\delta^{i_1}_{j_3}  \delta^{i_3}_{j_1}
\delta^{i_2}_{j_4}  \delta^{i_4}_{j_2} \;\;\;
\Rightarrow \;\;\; {\bf P} = P_{13} P_{24} \;\;\;
\Rightarrow \;\;\; {\bf P}^2 = I \; ,
\ee
where $I$ is the unit operator in $V_N^{\otimes 4}$.
Finally, the operator
 \be
 \lb{Pad}
 {\bf P}^{(ad)} \equiv \bar{I}_{12} \bar{I}_{34}{\bf P} \; ,
 \ee
 plays the role of the permutation operator in the space
$V_{\ad} \otimes V_{\ad} \subset V_N^{\otimes 4}$.
We stress that ${\bf P}$ commutes with both $\hC_{\ad}$
and ${\bf K}$
 \be
 \lb{cpk}
{\bf P} \; \hC_{\ad} = \hC_{\ad} \; {\bf P} \; , \;\;\;\;
{\bf P} \, {\bf K} = {\bf K} ={\bf K} \, {\bf P} \; ,
 \ee
 and therefore ${\bf P}$ can be diagonalized simultaneously
 with $\hC_{\ad}$ and ${\bf K}$.

 We also note that one cannot choose in the definition
  of the permutation (\ref{Pad}) in $V_{\ad}^{\otimes 2}$
  instead of $\bP= P_{13}P_{24}$
  another operator ${\bf P}' = P_{14} P_{23}$.
 This is because actually
$V_{\ad}  = \bar{I}_{12} \cdot
 (V_N \otimes \bar{V}_N)$, where $\bar{V}_N$ is the space of
 the contragradient representation of $s\ell(N)$
  (see Remark after (\ref{metSL1})), and the
element $A \in SL(N)$ acts in the space
$V_{\ad} \otimes V_{\ad}$ as follows:
 \be
 \lb{acti}
 V_{\ad} \otimes V_{\ad} \;\; \to \;\;
 (A \otimes A^{-1 \sf T}\otimes A \otimes A^{-1 \sf T})
 \; V_{\ad} \otimes V_{\ad} \; ,
 \ee
 and this action commutes with ${\bf P}$
 but does not commute with ${\bf P}'$.

Using the permutations ${\bf P}$ and ${\bf P}^{(ad)}$, we
define the symmetrizer ${\bf P}_{+}^{(ad)}$
 and the antisymmetrizer  ${\bf P}_{-}^{(ad)}$ in the space
 $(V_{\ad})^{\otimes 2}$:
 \be
 \lb{syma}
 \begin{array}{c}
 {\bf P}_{\pm}^{(ad)} \equiv \frac{1}{2}
 \,  ({\bf I} \pm {\bf P}^{(ad)}) =  \frac{1}{2} (I \pm {\bf P})
 \bar{I}_{12} \bar{I}_{34} = \frac{1}{2} \bar{I}_{12} \bar{I}_{34}
 \,  (I \pm {\bf P}) \; , \\ [0.3cm]
 {\bf P}_{-}^{(ad)} =  \frac{1}{2}(1 - P_{13}P_{24})(1 - \frac{1}{N}(K_{12}+K_{34}))  \; , \\ [0.3cm]
 {\bf P}_{+}^{(ad)} =  \frac{1}{2}(1 + P_{13}P_{24})(1 - \frac{1}{N}(K_{12}+K_{34})+\frac{1}{N^2} K_{12}K_{34} )  \; ,
 \end{array}
 \ee
where ${\bf I} = \bar{I}_{12} \bar{I}_{34}$ is the unit operator
in $(V_{\ad})^{\otimes 2}$
 (we often write unit $1$ instead of the unit operator
 $I$ in $V_N^{\otimes 4}$). We define, respectively,
the  symmetrized and antisymmetrized parts of the Casimir
  operator (\ref{kazSL})
\be
\lb{symc}
\begin{array}{c}
\hC_{+} = {\bf P}_{+}^{(ad)}\hC_{\ad} =
\frac{1}{2} (1 + {\bf P}) \hC_{\ad} =
 \frac{1}{4N} (1 + P_{13} P_{24})(2 P_{13}  - K_{14} - K_{23})=
 \\ [0.3cm]
= \frac{1}{4N}\Bigl(2 P_{13} + 2 P_{24} - (1 + P_{13} P_{24}) K_{14}
- (1 + P_{13} P_{24}) K_{23}\Bigr) \; ,
\end{array}
\ee
\be
\lb{asymc}
\hC_{-} = {\bf P}_{-}^{(ad)}\hC_{\ad} =
\frac{1}{2} (I - {\bf P}) \hC_{\ad} =
\frac{1}{4N}(P_{13}P_{24} - 1 )( K_{14}+ K_{23}) \; ,
\ee
where we used the equations
 $\bar{I}_{12} \hC_{\ad} = \hC_{\ad} =
\bar{I}_{34} \hC_{\ad}$, which are easily checked with the help
of explicit formula (\ref{kazSL}).
For $\hC_{+}$ and $\hC_{-}$ in view of
(\ref{cpk}) we have the following relations:
\be
\lb{plmiP}
\hC_{+}+\hC_{-} = \hC_{\ad} \; , \;\;\;
{\bf P} \, \hC_{\pm} = \hC_{\pm} \, {\bf P}
 \; , \;\;\; \hC_{+} \, \hC_{-} = 0 = \hC_{-} \, \hC_{+} \; .
\ee
 In addition, we have
\be
\lb{plmiK}
 \begin{array}{c}
 {\bf K} \, \hC_{-} = 0 = \hC_{-} \, {\bf K}
 \; , \;\;\; {\bf K} \, \hC_{+}  = - {\bf K}
  =\hC_{+} \, {\bf K} \; , \\ [0.3cm]
 {\bf K} \, \hC_{\ad} =
 - {\bf K} =\hC_{\ad} \, {\bf K} \; ,
 \end{array}
\ee
which are nothing but  formulas (\ref{idK})
 valid for all simple Lie algebras.
The second chain of equalities in (\ref{plmiK}) is derived
by means of relations (\ref{kkpp}).
 \newtheorem{prosl}[pro1]{Proposition}
\begin{prosl}\lb{prosl}
The antisymmetrized $\hC_{-}$ and symmetrized
 $\hC_{+}$ parts of the split Casimir operator of the Lie algebra
  $s\ell(N)$ (defined in (\ref{asymc})
 and (\ref{symc})) satisfy the identities
 \be
\lb{cha}
\hC_{-}^2 + \frac{1}{2} \hC_{-} = 0 \;\; \Rightarrow \;\;
\hC_{-}(\hC_{-}+ \frac{1}{2}) = 0 \; .
\ee
 \be
\lb{chcp2}
\hC_{+}^3 +\frac{1}{2} \hC_{+}^2  - \frac{1}{N^2} \hC_{+}
 - \frac{1}{4 N^2} (\bI^{(ad)} + \bP^{(ad)} -2 \bK) = 0\; ,
\ee
 \be
\lb{chs1}
  \hC_{+}(\hC_{+} +\frac{1}{2}) (\hC_{+} -\frac{1}{N})
   (\hC_{+} +\frac{1}{N}) = \frac{1}{2N^2} {\bf K} \; ,
\ee
 \be
\lb{chs2}
  (\hC_{+}+1) (\hC_{+} +\frac{1}{2}) (\hC_{+} -\frac{1}{N})
   (\hC_{+} +\frac{1}{N}) \bP^{(ad)}_{+} = 0 \; ,
\ee
\be
\lb{chs}
\hC_{+}(\hC_{+} +1) (\hC_{+} +\frac{1}{2}) (\hC_{+} -\frac{1}{N})
   (\hC_{+} +\frac{1}{N}) =  0 \; ,
\ee
The split Casimir operator $\hC_{\ad} =\hC_{+} +\hC_{-}$
satisfies the characteristic identity (cf. (\ref{char02}))
 \be
\lb{chad1}
\hC_{\ad}(\hC_{\ad}+\frac{1}{2})(\hC_{\ad}+1)(\hC_{\ad}-\frac{1}{N})
(\hC_{\ad}+\frac{1}{N}) = 0 \; .
\ee
\end{prosl}
{\bf Proof.} Identity (\ref{cha}) follows from the general
statement (\ref{idCC}), which is valid for all simple Lie algebras.
Further, the identities (\ref{chcp2}) -- (\ref{chs})
 can be proved  by using the diagram technique developed in \cite{Cvit}.
Nevertheless, we give here a direct algebraic
proof of these identities that uses relations (\ref{kkpp}) arising in the matrix representations of the Brauer algebra (see e.g. \cite{Book2},
\cite{Molev}). First, we calculate
 \be
 \lb{idslN}
\begin{array}{c}
\hC_{+}^2 = \frac{1}{4} ({\bf I} + {\bf P}) \hC_{\ad}
 ({\bf I} + {\bf P}) \hC_{\ad} =
 \frac{1}{2} ({\bf I} + {\bf P}) \hC_{\ad}^2 = \\ [0.3cm]
 = \frac{1}{8 N^2}(1 + P_{13}P_{24})\Bigl(4
 - 2 P_{13}\, (K_{12,34} + K_{14,23})
 +N K_{14,23}
 + 2 K_{14}K_{23}\Bigr) ,
\end{array}
 \ee
 where we introduce the notation
  $K_{ij,k\ell} \equiv K_{ij} + K_{k\ell}$.
Multiply the left and right sides of (\ref{idslN})
by $\hC_{+}$. As a result, we get
$$
\begin{array}{c}
\hC_{+}^3  =
  \frac{1}{16 N^3}(1 + \bP) \Bigl(8 P_{13}
 - 4 (K_{12,34}+ K_{14,23})
 + \\ [0.3cm]
 +2 N P_{13} (K_{12,34}+ K_{14,23})
 + 4 P_{13} (K_{12}K_{34}
 + K_{14} K_{23})
  - N^2 K_{14,23}  - 6 N K_{14} K_{23} \Bigr) = \\ [0.3cm]
 = -\frac{1}{2} \hC_{+}^2  + \frac{1}{N^2} \hC_{+}
 + \frac{1}{4 N^3} (1 + \bP) \Bigl(
 N - K_{12,34}
 + P_{13} (K_{14} K_{23}+ K_{12}K_{34})
  -  N K_{14} K_{23} \Bigr) = \\ [0.3cm]
 = -\frac{1}{2} \hC_{+}^2  + \frac{1}{N^2} \hC_{+}
 + \frac{1}{4 N^2} (\bI^{(ad)} + \bP^{(ad)})
 - \frac{1}{2 N^2} \bK ,
\end{array}
$$
where in the last equality we used formulas
 (\ref{ksl}), (\ref{syma}) and
 $$
  (N - K_{12,34}) = (N \bar{I}_{12}\bar{I}_{34}  -
 \frac{1}{N} K_{12} K_{34}) \, , \;\;\;\;
 \bar{I} \equiv I - \frac{1}{N} K \, , \;\;\;\;
 \bP \equiv P_{13}P_{24} \; .
 $$
 Thus, identity (\ref{chcp2}) is proved.
 Again, we multiply both sides of equality (\ref{chcp2})
  by $\hC_{+}$ and obtain
 \be
 \lb{chs11}
\hC_{+}^4 +\frac{1}{2} \hC_{+}^3  - \frac{1}{N^2} \hC_{+}^2
   - \frac{1}{2N^2}\hC_{+} = \frac{1}{2N^2} {\bf K}
 \ee
that is equivalent to (\ref{chs1}). Substitution of the
expression (\ref{chs11}) of the
operator $\bK$ into relation (\ref{chcp2}) gives (\ref{chs2}).
Identity (\ref{chs}) is obtained either by multiplying
(\ref{chs2}) by $\hC_{+}$, or by multiplying both sides of  equality (\ref{chs1}) by $(\hC_{+}+1)$ and taking into account relation ${\bf K} (\hC_{+}+1) =0$, which follows
from (\ref{plmiK}).

Note that the antisymmetric part $\hC_{-}$ of the Casimir operator, in view of the relation (\ref{cha}), satisfies the same
relation (\ref{chs}) as the symmetric part of $\hC_{+}$. Hence, taking into account the last relation in (\ref{plmiP}),
it follows that the complete Casimir operator
 $\hC_{\ad} = \hC_{+}+\hC_{-}$  for the algebra $s\ell(N)$
  will obey a characteristic identity (\ref{chad1})
   similar to (\ref{chs}). \hfill \qed

 \vspace{0.2cm}

Now we show that in the $s\ell(N)$ case, in addition to the invariant operators
in $V_{\ad}^{\otimes 2}$ represented as polynomials
in $\hC_{\ad}, \hC_{\pm}$, there is one more invariant
operator ${\sf Q}_-$ in $V_{\ad}^{\otimes 2}$, which commutes
with $\hC_{\ad}, \hC_{\pm}$ but is not expressed as a polynomial in $\hC_{\ad}, \hC_{\pm}$.
To construct such an operator, note that the generators
$T_a=T(X_a)$ of $s\ell(N)$ in the defining representation $T$ together with the unit matrix $I_N $
form a basis in the space of all $N \times N$ matrices; therefore, along with the defining relation (\ref{lialg}), there is one more relation
 \be
 \lb{ddd}
 [ T_a , \; T_b]_+ = D_{ab}^c \; T_c + \alpha \, I_N \, {\sf g}_{ab} \; , \;\;\;\;\; \alpha \equiv \frac{2 c_2(T)}{N} \; .
 \ee
 Here the parameter $\alpha$ is fixed by the condition
  (\ref{kaz-11b}), and $D_{ab}^c$ are new structure
  constants of the algebra $s\ell(N)$, symmetric with respect to permutation of subscript indices $a$ and $b$.
 Now we define the operators
 ${\sf Q}$ and ${\sf Q}_{-}$ in the space
 $V_{\ad}^{\otimes 2}$ (cf. (\ref{adCC}) and (\ref{adCpm}))
 \be
\lb{doper1}
{\sf Q}^{a_1 a_2}_{\;\; b_1 b_2} \equiv
{\sf g}^{d f} \, C^{a_1}_{d  b_1} \, D^{a_2}_{f b_2}
\,  , \;\;\;\;
 {\sf Q}_{-} \equiv \frac{N}{4} \,
 (\bI - \bP) \, {\sf Q} \, (\bI - \bP)  \; ,
\ee
 where $\bI$ and $\bP$ are the  unit matrix and the permutation matrix in $V_{\ad}^{\otimes 2}$, respectively. Note that operator
 ${\sf Q}_{-}$ acts nontrivially in the antisymmetric part
 $\bP_{-}^{(ad)} \; V_{\ad}^{\otimes 2}$ of
 the space $V_{\ad}^{\otimes 2}$.

\newtheorem{prdop}[pro1]{Proposition}
\begin{prdop}\lb{prdop}
The operator ${\sf Q}_{-}$ given in (\ref{doper1}) is written as an operator in the space $V_{\ad}^{\otimes 2} =
\bar{I}_{12} \bar{I}_{34}(V_N^{\otimes 4})$ as follows:
\be
\lb{doper}
({\sf Q}_{-})_{1234} = \frac{1}{2} (P_{13} - P_{24})
\Bigl(1 - \frac{1}{N}(K_{12} + K_{14} +K_{23} + K_{34} )\Bigr) \; .
\ee
The operator ${\sf Q}_{-}$ satisfies the relations
 \be
\lb{doper2}
\bP_{+}^{(ad)} \, {\sf Q}_{-} = 0 =
{\sf Q}_{-} \, \bP_{+}^{(ad)} \; , \;\;\;\;\;\;
{\sf Q}_{-} \; \hC_{-} =
\hC_{-} \; {\sf Q}_{-}   = 0 \; , \;\;\;\;\;\;
{\sf Q}_{-}^2 = 2 \hC_{-} + \bP_{-}^{(ad)}  \; ,
\ee
 \be
\lb{doper3}
{\sf Q}_{-} ({\sf Q}_{-} +1)({\sf Q}_{-} -1)=0 \; ,
\ee
 where $\hC_{-}$ and $\bP_{\pm}^{(ad)}$ are given in
 (\ref{asymc}) and (\ref{syma}).
\end{prdop}
{\bf Proof.}
Relations (\ref{ddd}) in the basis (\ref{defrep}) are represented as
 \be
 \lb{ddd1}
 [T_{ij} , \, T_{km}]_+ = D^{rs}_{ij,km}\, T_{rs} +
 \frac{1}{N^2} \, {\sf g}_{ij,k m} \, I_N \; ,
 \ee
where for the structure constants we have the explicit expressions
 $$
  D^{rs}_{ij,km} \equiv
  (\delta^{r}_{\ell}\delta^{s}_{n} -
 \frac{1}{N}K^{rs}_{\ell n}) \bar{D}^{\ell n}_{ij,km}
 \, , \;\;\;
 \bar{D}^{rs}_{ij,km} \equiv
  \delta_{jk}\delta^{r}_{i}\delta^{s}_{m} +
 \delta_{im}\delta^{r}_{k}\delta^{s}_{j} - \frac{2}{N}
 \bigl( \delta_{ij}\delta^{r}_{k}\delta^{s}_{m} +
 \delta_{km}\delta^{r}_{i}\delta^{s}_{j} \bigr)\; .
 $$
 Then the operators ${\sf Q}$
 and ${\sf Q}_{-}$, given in (\ref{doper1}),
 are equal to
 $$
 \begin{array}{c}
 {\sf Q}^{i_1 i_2 i_3 i_4}_{j_1 j_2 j_3 j_4} =
{\sf g}^{k_1 k_2, k_3 k_4} \; C^{i_1 i_2}_{k_1 k_2,j_1 j_2}
 \, D^{i_3 i_4}_{k_3 k_4 ,j_3 j_4}  = \\ [0.2cm]
 = \frac{1}{2N} \bigl( \bar{I}_{34}
 \bigl(P_{13} - P_{24} + K_{14} - K_{23}
 + \frac{2}{N}(P_{24} - P_{13})K_{34} \bigr)
 \bigr)^{i_1 i_2 i_3 i_4}_{j_1 j_2 j_3 j_4} \, ,
  \end{array}
  $$
  \be
 \lb{ddd2}
 \begin{array}{c}
 {\sf Q}_{-} = \frac{N}{4} \, (1-\bP) \bar{I}_{12}\bar{I}_{34}
  \; {\sf Q} \; \bar{I}_{12} \bar{I}_{34} (1-\bP)  =
  \\ [0.2cm]
 =  \frac{1}{8} (1-\bP) \bigl(P_{13} - P_{24} -
 \frac{2}{N} K_{34}(P_{13} - P_{24}) -
  \frac{2}{N}  (P_{13} - P_{24})K_{34} \bigr) (1-\bP) \, ,
 \end{array}
 \ee
 and the right-hand side of (\ref{ddd2}) after the substitution
 $\bP = P_{13} P_{24}$ matches the right-hand side
 of (\ref{doper}).
Relations (\ref{doper2}) and (\ref{doper3}) are
verified by direct calculations. \hfill \qed

\vspace{0.2cm}

{\bf Remark.} Characteristic identity (\ref{doper3})
for the operator ${\sf Q}_{-}$ allows us to build
three mutually orthogonal projectors:
 \be
 \lb{tildc}
 \begin{array}{c}
 \widetilde{\proj}_{0}^{(-)}= -
 ({\sf Q}_{-} +1)({\sf Q}_{-} -1)\bP^{(ad)}_{-}
 = - 2 \hC_{-} \; , \\ [0.3cm]
 \widetilde{\proj}_{\pm 1}^{(-)}= \frac{1}{2}
 {\sf Q}_{-}({\sf Q}_{-} \pm 1) =
 \hC_{-} + \frac{1}{2} \bP^{(ad)}_{-}
 \pm \frac{1}{2} {\sf Q}_{-} \; .
 \end{array}
 \ee
 Due to the relation
 $\bP^{(ad)}_{-} = \widetilde{\proj}_{+1}^{(-)} +
 \widetilde{\proj}_{-1}^{(-)}+ \widetilde{\proj}_{0}^{(-)}$,
 the projectors $\widetilde{\proj}_{0}^{(-)},
 \widetilde{\proj}_{+1}^{(-)},\widetilde{\proj}_{-1}^{(-)}$
decompose the space $\bP_{-}^{(ad)} (V_{\ad} \otimes V_{\ad})$
 of the antisymmetric part $\mathbb{A} (\ad \otimes \ad)$ of the
 representation $(\ad)^{\otimes 2}$ into eigenspaces
of the operator ${\sf Q}_{-}$ with eigenvalues
$0$, $+1$, $-1$
(which are the roots of the equation (\ref{doper3})).

 The characteristic identity (\ref{chad1}) also allows one to construct projectors (invariant with respect to the action
 (\ref{acti})) onto eigen-subspaces of the operator
  $\hC_{\ad}$ in $(V_{\ad} \otimes V_{\ad})$
  by means of the standard methods
  (see Section 3.5 in \cite{Cvit} and Section 4.6.4 in \cite{Book2}):
 \begin{equation}\lb{slProj}
\proj_{(a_j)}=\prod_{\substack{i=1\\i\neq j}}^5
\frac{\hC_{\ad} -a_i\bI}{a_j-a_i} \; ,
\end{equation}
where $a_i$ are the roots of the characteristic
equation (\ref{chad1})
$$
a_1 = 0 \, , \;\;\; a_2 = -1/2 \, , \;\;\; a_3 = -1 \, , \;\;\;
a_4 = 1/N \, , \;\;\; a_5 = - 1/N \, .
$$
Note that the case $N=2$ is special since in this case we
have $(a_2-a_5) = 0$, and projectors
$\proj_{(a_2)}=\proj_{(-1/2)}$ and
$\proj_{(a_5)}=\proj_{(-1/N)}$ are not defined
(see below (\ref{slProj1})). In addition,
we note that in general the projectors (\ref{slProj})
are not primitive and extract invariant subspaces in
$(V_{\ad})^{\otimes 2} \subset V_N^{\otimes 4}$, which are
not the spaces of the irreducible representations of $s\ell(N)$.
First of all, this is due to the presence of the
invariant permutation operator ${\bf P}$ that
 commutes with $\hC_{\ad}$ (see (\ref{cpk})) and allows us to split the projectors into two parts
$\proj_{(a_j)}^{(\pm)} =  {\bf P}_\pm^{(ad)} \cdot \proj_{(a_j)}$,
 where ${\bf P}_\pm^{(ad)} \equiv
 \frac{1}{2}({\bf I} \pm {\bf P})\bar{I}_{12} \bar{I}_{34}$.
 From the condition (\ref{cha}), which can be written as
 ${\bf P}_{-}^{(ad)}\hC_{\ad}(\hC_{\ad} + \frac{1}{2})=0$, it immediately follows that
$$
\proj_{(-1)}^{(-)} = {\bf P}_-^{(ad)} \, \proj_{(-1)} = 0 \; , \;\;\;\;\;
\proj_{(\pm 1/N)}^{(-)} =
{\bf P}_-^{(ad)} \, \proj_{(\pm 1/N)} =0 \; ,
$$
Moreover, due to  relation (\ref{chs2})
  for the symmetrized part of $\proj_{(0)}$ we obtain
 $$
    \proj_{(0)}^{(+)} = \frac{(\hC_{\ad} + 1)(\hC_{\ad} + \frac{1}{2})
 (\hC_{\ad} + \frac{1}{N})(\hC_{\ad} - \frac{1}{N})}{(+1)(+\frac{1}{2})(\frac{1}{N})(-\frac{1}{N})}
 {\bf P}_+^{(ad)}  = 0 , \\ [0.3cm]
 $$
 while the antisymmetrized part
 $$
   \proj_{(0)}^{(-)} = \frac{(\hC_{\ad} + 1)(\hC_{\ad} + \frac{1}{2})
 (\hC_{\ad} + \frac{1}{N})(\hC_{\ad} - \frac{1}{N})}{(+1)(+\frac{1}{2})(\frac{1}{N})(-\frac{1}{N})}
 {\bf P}_-^{(ad)} =
 2\hC_{-} + {\bf P}_-^{(ad)} \; \equiv \; \widetilde{\proj}_{(+1)}^{(-)}
 + \widetilde{\proj}_{(-1)}^{(-)},
 $$
 is not primitive since it is equal to the sum of the
 projectors
 $\widetilde{\proj}_{(\pm 1)}^{(-)}$ from (\ref{tildc}).

As a result, for $N> 3$ we have 7 nontrivial projectors
$$
\widetilde{\proj}_{(+1)}^{(-)} \, , \;\;\;
\widetilde{\proj}_{(-1)}^{(-)} \, , \;\;\;
\proj_{(-\frac{1}{2})}^{(-)} \, , \;\;\;
\proj_{(-\frac{1}{2})}^{(+)} \, , \;\;\;
\proj_{(-1)}^{(+)} = \proj_{(-1)} \, , \;\;\;
 \proj_{(-\frac{1}{N})}=\proj_{(-\frac{1}{N})}^{(+)} \, , \;\;\; \proj_{(+\frac{1}{N})}= \proj_{(+\frac{1}{N})}^{(+)} \, ,
$$
which extract invariant subspaces in
$(V_{\ad})^{\otimes 2}$ and by construction form a complete
  and mutually orthogonal system.
Due to (\ref{tildc}) and (\ref{slProj}), these projectors
have the form  (cf. projectors in \cite{Cvit}, Section 9.12)
 \be
 \lb{slProj1}
 \begin{array}{lr}
 \widetilde{\proj}_{(+1)}^{(-)} =
 \hC_{-} + \frac{1}{2} {\bf P}_-^{(ad)} + \frac{1}{2} {\sf Q}_- ,
&
\dim = \frac{(N^2 - 1)(N^2 - 4)}{4} ,\\ [0.3cm]
\widetilde{\proj}_{(-1)}^{(-)}=
\hC_{-} + \frac{1}{2} {\bf P}_-^{(ad)} -
\frac{1}{2} {\sf Q}_-  , &
\dim = \frac{(N^2 - 1)(N^2 - 4)}{4} ,\\ [0.3cm]
  \proj_{(-\frac{1}{2})}^{(-)} = \frac{\hC_{\ad}(\hC_{\ad} + 1)
 (\hC_{\ad} + \frac{1}{N})(\hC_{\ad} - \frac{1}{N})}{(-\frac{1}{2})(\frac{1}{2})(-\frac{1}{2}+\frac{1}{N})
 (-\frac{1}{2}-\frac{1}{N})}{\bf P}_-  =
 - 2\hC_{-}  \; \equiv \; \widetilde{\proj}_{(0)}^{(-)} ,
 & \dim = N^2 -1 ,\\ [0.3cm]
   \proj_{(-\frac{1}{2})}^{(+)} = \frac{\hC_{\ad}(\hC_{\ad} + 1)
 (\hC_{\ad} + \frac{1}{N})(\hC_{\ad} - \frac{1}{N})}{(-\frac{1}{2})(\frac{1}{2})(-\frac{1}{2}+\frac{1}{N})
 (-\frac{1}{2}-\frac{1}{N})}{\bf P}_+
 = \frac{4}{N^2-4} (N^2 \hC_{+}^2 - \bP_{+}^{(ad)} - {\bf K}) ,
 & \dim = N^2 -1 , \\ [0.3cm]
    \proj_{(-1)}^{(+)} = \frac{\hC_{\ad}(\hC_{\ad} + \frac{1}{2})
 (\hC_{\ad} + \frac{1}{N})(\hC_{\ad} - \frac{1}{N})}{(-1)(-\frac{1}{2})(-1+\frac{1}{N})
 (-1-\frac{1}{N})}{\bf P}_+  =
 \frac{1}{(N^2-1)}{\bf K} , & \dim = 1 \; , \\ [0.3cm]
 \proj_{(\frac{1}{N})}^{(+)} = \frac{\hC_{\ad}(\hC_{\ad} + 1)(\hC_{\ad} + \frac{1}{2}) (\hC_{\ad} + \frac{1}{N})
 }{(\frac{1}{N})(\frac{1}{N}+1)(\frac{1}{N}+\frac{1}{2})
 (\frac{2}{N})}{\bf P}_+ = &
   \\ [0.3cm]
 \quad\quad  = -\frac{N}{2(N+1)(N+2)}{\bf K} + \frac{N^2}{(N+2)}\hC_{+}^2
 + \frac{N}{2}\hC_{+} +  \frac{N}{2(N+2)}\bP_{+}^{(ad)} ,
 &  \dim = \frac{N^2(N-1)(N+3)}{4}, \\ [0.3cm]
  \proj_{(-\frac{1}{N})}^{(+)} = \frac{\hC_{\ad}(\hC_{\ad} + 1)(\hC_{\ad} + \frac{1}{2}) (\hC_{\ad} - \frac{1}{N})
 }{(-\frac{1}{N})(-\frac{1}{N}+1)(-\frac{1}{N}+\frac{1}{2})
 (-\frac{2}{N})}{\bf P}_+  =
 &  \\ [0.3cm]
 \quad\quad\quad  = \frac{N}{2(N-1)(N-2)}{\bf K} - \frac{N^2}{(N-2)}\hC_{+}^2
 - \frac{N}{2}\hC_{+} +  \frac{N}{2(N-2)}\bP_{+}^{(ad)} ,
 & \dim = \frac{N^2(N+1)(N-3)}{4} ,
 \end{array}
 \ee
  where we used the property
 $\hC_{\ad} {\bf P}_{\pm}^{(ad)}
 =  \hC_{\ad} {\bf P}_{\pm}= \hC_{\pm}$ and identities (\ref{cha}), (\ref{chcp2}), (\ref{chs1}), (\ref{chs11}).
 Note that the projector $\proj_{(-\frac{1}{2})}^{(-)}$ is the same as the projector
 $\widetilde{\proj}_{(0)}^{(-)}$ given in (\ref{tildc}).
 The right column in formula (\ref{slProj1}) shows the dimensions of the invariant
  subspaces in $V_{\ad}^{\otimes 2}$, which are extracted by the corresponding projectors.
The way to calculate these dimensions is shown below.

It is well known that the tensor product of two adjoint representations of the algebra $s\ell(N)$ for $N>3$ decomposes into the sum of seven irreducible representations, which can be illustrated in terms of the Young diagrams
 $$
 [2,1^{N-2}] \otimes [2,1^{N-2}] =
 [\emptyset] + [2^2,1^{N-4}] + [3,1^{N-3}]
 + [3^2,2^{N-3}] + [4,1^{N-2}] + 2 \cdot  [2,1^{N-2}] \; ,
 $$
 where the diagram $[2,1^{N-2}]$ corresponds to the adjoint
 representation and
 \be
 \lb{dimyo}
 \begin{array}{c}
 \dim \, [2,1^{N-2}] = N^2-1, \;\;\; \dim \, [\emptyset] = 1, \;\;\;
 \dim \, [2^2,1^{N-4}] = \frac{N^2(N+1)(N-3)}{4} , \\ [0.3cm]
 \dim \, [3,1^{N-3}] =
 \dim \, [3^2,2^{N-3}] = \frac{(N^2-1)(N^2-4)}{4} , \;\;\;
 \dim \, [4,1^{N-2}] = \frac{N^2(N-1)(N+3)}{4} .
 \end{array}
 \ee
 Comparing the dimensions in (\ref{slProj1}) and (\ref{dimyo}), we conclude that seven mutually orthogonal
  and nontrivial projectors (\ref{slProj1}),
 which form a complete system in the space
  $(V_{\ad})^{\otimes 2}$, select in $(V_{\ad})^{\otimes 2}$ the subspaces of all irreducible representations of $s\ell(N)$.
 To verify this fact, we need to calculate the dimensions
 of the invariant subspaces
 $\widetilde{V}_{(b_i)}^{(-)} =
 \widetilde{\proj}_{(b_i)}^{(-)} (V_{\ad} \otimes V_{\ad})$ and
$V_{(a_i)}^{(+)} = \proj_{(a_i)}^{(+)} (V_{\ad} \otimes V_{\ad})$,
which are given in (\ref{slProj1}),
and compare these dimensions with (\ref{dimyo}).
A way to calculate these dimensions is to find
 traces of the projector (\ref{slProj1}):
 \be
 \lb{dimsl}
  {\rm dim}(\widetilde{V}_{(b_i)}^{(-)}) =
  {\bf Tr}(\widetilde{\proj}_{(b_i)}^{(-)}) \; , \;\;\;\;\;\;
 {\rm dim}(V_{(a_i)}^{(+)}) =
 {\bf Tr} (\proj_{(a_i)}^{(+)}) \; ,
 \ee
 where we introduce the notation ${\bf Tr}
 \equiv  {\rm Tr}_{1}{\rm Tr}_{2}{\rm Tr}_{3}{\rm Tr}_{4}$
 for the trace in $(V_{\ad})^{\otimes 2} \subset (V_{N})^{\otimes 4}$.
 For this calculation, we use the traces (\ref{trac1}) of the basic operators that make up the projectors (\ref{slProj1}):
 $$
 {\bf Tr}  ({\bf K}) = N^2 -1 \, , \;\;\;
  {\bf Tr}  ({\bf P}_-^{(ad)}) = \frac{1}{2}(N^2 - 1)(N^2 -2)
  \, , \;\;\;
  {\bf Tr} ({\bf P}_+^{(ad)}) = \frac{1}{2}N^2(N^2 - 1)   \, ,
   $$
   $$
   {\bf Tr}  (\hC_\pm) = \pm \frac{(N^2-1)}{2} \, , \;\;
 {\bf Tr} (\hC_{\ad}^2) =(N^2-1) \, , \;\;
{\bf Tr} (\hC_+^2) = \frac{3}{4}(N^2-1) \, , \;\;\;
  {\bf Tr}  (\widetilde{C}_{-}) =0  \, .
 $$
 Substituting these traces into expressions (\ref{dimsl}), where the projectors $\widetilde{\proj}_{(b_i)}^{(-)}$,
 $\proj_{(a_i)}^{(+)}$ are defined in (\ref{slProj1}),
 we obtain the dimensions indicated in (\ref{slProj1})
 that coincide with the dimensions (\ref{dimyo}).

So using the projectors ${\bf P}_+^{(ad)}$ and ${\bf P}_-^{(ad)}$, the representation $(\ad)^{\otimes 2}$ of the algebra
$s\ell(N)$ is decomposed into the
symmetric $\mathbb{S}(\ad^{\otimes 2})$ and
 antisymmetric $\mathbb{A}(\ad^{\otimes 2})$ parts.
In turn, for all simple Lie algebras (see Section {\bf \ref{hCad}})
the antisymmetric part $\mathbb{A}(\ad^{\otimes 2})$
splits into the sum of two subrepresentations
${\sf X}_1$ and ${\sf X}_2$, which in the case of the algebra
$s\ell(N)$ correspond to the projectors
 $\proj^{(-)}_{(-\frac{1}{2})}$ and $\proj^{(-)}_{(0)}$
 and have dimensions (\ref{XX123}):
 $$
 {\rm dim}({\sf X}_1) = N^2-1 = {\rm dim}(s\ell(N)) \; , \;\;\;\;
 {\rm dim}({\sf X}_2) = \frac{1}{2}(N^2-1)(N^2-4) \; .
 $$
 Moreover, in the case of the Lie algebra $s\ell(N)$,
 the representation $ {\sf X}_2$ associated with the projector
 $\proj^{(-)}_{(0)}$ turns out to be reducible and expands into the sum of two inequivalent irreducible representations
 associated with the projectors
 $\widetilde{\proj}^{(-)}_{(+1)},
 \widetilde{\proj}^{(-)}_{(-1)}$
 and having the same dimensions:
 ${\rm dim} \, \widetilde{V}^{(-)}_{(\pm 1)} = \frac{1}{4}(N^2-1)(N^2-4)$.

For the Lie algebras $s\ell(N)$, when $ N> 3 $ (the cases $ N = 2,3 $ are special), the symmetric part $\mathbb{S}(\ad^{\otimes 2})$  decomposes into the sum of four irreducible representations,
one of which ${\sf X}_0$ associated with the projector $P^{(+)}_{(-1)}$ is trivial and has dimension $1$,
  and three other representations $Y_2$, $Y_2'$  and $Y_2^{\prime\prime}$ associated with the projectors
 $P^{(+)}_{(-1/2)}$, $P^{(+)}_{(1/N)}$ and $P^{(+)}_{(-1/N)}$
 have the corresponding dimensions
 $$
 {\rm dim} \, Y_2 = N^2-1 \; , \;\;\;
 {\rm dim} \, Y_2' = \frac{1}{4} N^2(N-1)(N+3)\; , \;\;\;
 {\rm dim} \, Y_2^{\prime\prime} = \frac{1}{4} N^2(N+1)(N-3) \; .
 $$
Note that for $N = 2$ the projectors $P^{(+)}_{(-1/2)}$
 and $P^{(+)}_{(-1/N)}$ in (\ref{slProj1}) are not defined, the dimension of the representation $Y_2^{\prime\prime}$ becomes negative and the above expansion does not work.
 For $N = 3$ we have ${\rm dim} Y_2^{\prime\prime}=0$ and, therefore, in the case of the algebra $s\ell(3)$, the representation $Y_2^{\prime\prime}$ dose not appear in
 $\mathbb{S}(\ad^{\otimes 2})$, and the corresponding
 projector $\proj^{(+)}_{-\frac{1}{N}}$ in
 (\ref{slProj1}) for $N = 3$ must vanish
 $\left. \proj^{(+)}_{-\frac{1}{N}} \right|_{N=3} = 0$, which gives
 \be
 \lb{idsl3}
 \hC_{+}^2 = -\frac{1}{6} \hC_{+}
+ \frac{1}{12} (\bI^{(ad)} + \bP^{(ad)}+ \bK) \; ,
\ee
i.e. in this case, the symmetric part $\hC_{+}$ of the split
Casimir operator satisfies the second order identity
(\ref{idsl3}), and the third order identity (\ref{chcp2})
 for $N = 3$ is a consequence of (\ref{idsl3}).
 To derive  characteristic identities in the case
 $N = 3$, we multiply
  both parts of (\ref{idsl3}) by $\hC_{+}$ and then
  we multiply the resulting relation by $(\hC_{+}+1)$ and use the equality $(\hC_{+}+1)\bK=0$. As a result, we obtain
  \be
 \lb{idsl31}
\hC_{+}^3 + \frac{1}{6} \hC_{+}^2 -\frac{1}{6} \hC_{+} =
- \frac{1}{12} \bK
 \;\;\;\; \Rightarrow \;\;\;\;
 \hC_{+}(\hC_{+}+ 1) (\hC_{+} + \frac{1}{2})
 (\hC_{+}  -\frac{1}{3}) =  0 \;\;\; \Rightarrow
\ee
 \be
\lb{chasl3}
\hC_{\ad}(\hC_{\ad}+\frac{1}{2})(\hC_{\ad}+1)(\hC_{\ad}-\frac{1}{3}) = 0 \; ,
\ee
which for the case of $s\ell(3)$ replace (\ref{chs1}), (\ref{chs}) and (\ref{chad1}).

  Finally, the decomposition of the product of two adjoint representations of the algebra  $s\ell(N)$ for $N> 3$ is written in the form
 $$
 \begin{array}{c}
 [N^2-1] \otimes [N^2-1] =  \mathbb{A}([N^2-1] \otimes [N^2-1])
 + \mathbb{S}([N^2-1] \otimes [N^2-1]) \; , \\ [0.3cm]
 \mathbb{A}([N^2-1] \otimes [N^2-1]) =
   [N^2-1] \oplus  [\frac{(N^2-1)(N^2-4)}{4}] \oplus
 [\overline{\frac{(N^2-1)(N^2-4)}{4}}]
 \; , \\ [0.3cm]
 \mathbb{S}([N^2-1] \otimes [N^2-1]) = [1] \oplus
  [N^2-1] \oplus [\frac{N^2(N-1)(N+3)}{4} ] \oplus
 [\frac{N^2(N+1)(N-3)}{4}] \; .
 \end{array}
 $$
 This decomposition is well known and
 is in accordance with the general theory of the universal description of all simple Lie algebras
 using the Vogel parameters \cite{Fog}
 (see also \cite{Lan}, \cite{MkSV}). We will discuss this universal
  description below in Section {\bf \ref{Vogel}}.

\subsection{Split Casimir operator $\hC$ for Lie algebras
$so(N)$  and $sp(2n)$}
\setcounter{equation}0

\subsubsection{Operator $\hC$ for $so(N)$  and $sp(2n)$
 in the defining representation}

In this Subsection, to fix the notation,
 we give the well-known definition of the Lie
 algebras $so(N)$ and $sp(2n)$ which
 one can find in many monographs and textbooks
 (see e.g. \cite{Cvit}, \cite{Weyl} and \cite{Book1}). We prefer to give here a natural unified definition
 \cite{Book1,IsPr} of these algebras
 since it will be useful for us below.

We introduce the metric $||c_{ij}||_{i,j=1,...,N}$
which is equal to the unit matrix
$||\delta_{ij}||$ in the case of the $so(N)$ algebras
and equal to the matrix
\begin{equation}
||c_{ij}|| =
\begin{pmatrix}
0 & I_{n}\\
-I_{n} & 0
\end{pmatrix}
\end{equation}
in the case of the algebras $sp(2 n)$.  Thus, we have
$c_{ij}=\epsilon c_{ji}$
with $\epsilon =\pm 1$ for the $so(n)/sp(2n)$ cases, respectively. The inverse metric
$\bc^{ij}$ is defined in a standard way  as $\bc^{ik}c_{kj}=\delta^i_j$.
We denote the space $\mathbb{C}^N$ of
the defining representation of
$so(N)$ and  $sp(N)$ as $V_N$.

Using the matrix units
$(e_s{}^t)^i{}_k=\delta^t_k\delta^i_s$, or $(e_{st})^i{}_k=c_{tk}\delta^i_s$ with
lowered indices, one may define the
generators of $so(N)$ and  $sp(N)$ as
\be\label{basis}
M_{ij} =e_{ij}-\epsilon e_{ji}\; , \;\;\;\;\;\;\;\;
(M_{ij})^k{}_l =c_{jl}\delta^k_i-\epsilon c_{il}\delta^k_j=2\delta_{[i}^kc_{j)l} \; ,
\ee
where the notation $[ij)$ means (anti-)symmetrization  for the ($so(N)$-)$sp(N)$ algebras.
The commutation relations for both algebras acquire the generic form
\begin{equation}\label{stru_rel}
[M_{ij},M_{kl}]=c_{jk}M_{il}-\epsilon c_{ik}M_{jl}-\epsilon c_{jl}M_{ik}+c_{il}M_{jk}=X_{ij,kl}{}^{mn}M_{mn} \; ,
\end{equation}
with the structure constants given by
\begin{equation}\label{stru_cons}
X_{ij,kl}^{\ \ mn}=c_{jk}\delta_{i}^{[m}\delta_{l}^{n)}-\epsilon c_{ik}\delta_{j}^{[m}\delta_{l}^{n)}-\epsilon c_{jl}\delta_{i}^{[m}\delta_{k}^{n)}+ c_{il}\delta_{j}^{[m}\delta_{k}^{n)} =4 \, \delta_{[i}^{[m}\, c_{_{j) [k}} \, \delta_{l)}^{n)}\; .
\end{equation}
In this basis the Cartan-Killing metric reads
\begin{equation}\label{Killing}
{\sf g}_{i_1i_2,j_1j_2}=
2(N-2\epsilon)(c_{i_2j_1}c_{j_2i_1}-\epsilon c_{i_1j_1}c_{j_2i_2})
 \; \equiv \; (N-2\epsilon) \, {\rm Tr}(M_{i_1i_2} \, M_{j_1j_2}) \; ,
\end{equation}
while the inverse metric has the form
\begin{equation}\label{invK}
{\sf g}^{i_1i_2,j_1j_2}=
\frac{1}{8(N-2\epsilon)}\; (\epsilon \, \bc^{i_1j_2}\bc^{i_2j_1}-\bc^{i_1j_1}\bc^{i_2j_2})  \; .
\end{equation}
This inverse metric is defined
by the equation ${\sf g}_{ij,k\ell}{\sf g}^{k\ell,mn}= (\mcP^{(\epsilon)})^{mn}_{ij}$,
where $(\mcP^{(\epsilon)})^{mn}_{ij}
\equiv \frac{1}{2}(\delta^m_i \delta^n_j -
 \epsilon \delta^n_i \delta^m_j)$ is the projector
 on the (anti)symmetric part of $V_N^{\otimes 2}$.

Now it is easy to calculate the split Casimir operator $\hC$ for the algebras $so(N)$ and $sp(N)$ in the defining representation  \cite{Book1}
 \be
 \lb{casdef}
 \begin{array}{c}
(\hC_T)^{k_1k_2}_{\;\; \ell_1\ell_2}  \equiv
 T^{\otimes 2} (\hC)^{k_1k_2}_{\;\; \ell_1\ell_2} = g^{ij,nm}
 (M_{ij})^{k_1}_{\;\; \ell_1}   (M_{nm})^{k_2}_{\;\; \ell_2} = \\ [0.3cm]
 = \frac{1}{2(N-2\epsilon)}\left(\delta^{k_1}_{\ell_2} \delta^{k_2}_{\ell_1} -\epsilon \; \bar{c}^{k_1k_2} c_{\ell_1 \ell_2}\right) \; ,
 \end{array}
\ee
or in the index-free matrix notation we have
\begin{equation}
\lb{casdef2}
\hC_T =\frac{1}{2(N-2\epsilon)}\left(P-\epsilon K\right) \; .
\end{equation}
Here, $P$ is the permutation operator acting in the space $V_N^{\otimes 2}$
 (see (\ref{perm})), while the operator $K$ acting in the same space $V_N^{\otimes 2}$
 has the following components:
 $$
 K^{i_1i_2}{}_{j_1j_2}=\bc^{i_1i_2}c_{j_1j_2} \; .
 $$

 \newtheorem{prso}[pro1]{Proposition}
\begin{prso}\lb{prso}
 The characteristic identity for the split Casimir operator  (\ref{casdef2}) for the algebras  $so(N)$ and $sp(N)$ in the defining representation reads
 \be
 \lb{sosp30}
 \Bigl(\hC_T +\frac{{\sf d}_2}{2}\Bigr)
 \Bigl(\hC_T -\frac{{\sf d}_2}{2}\Bigr)
 \Bigl(\hC_T +\frac{{\sf d}_2}{2}(N-\epsilon)\Bigr) = 0 \; ,
 \ee
where ${\sf d}_2 = 1/(N-2 \epsilon)$.
 \end{prso}
 The proof of this Proposition is straightforward.

In accordance with equation (\ref{sosp30}), the projectors on the eigenvalues of the operator  $\hC_{T}$ in $V_N^{\otimes 2}$ have the form
 \cite{Cvit}, \cite{Book2}
 \be
 \lb{sospN}
 \begin{array}{c}
P_{a_1} = \frac{\bigl(\hC_T - a_2\bigr)
 \bigl(\hC_T - a_3\bigr)}{(a_1-a_2)(a_1-a_3)}
 = \frac{1}{2}(I + P) - \frac{(1+\epsilon)}{2(1+N-\epsilon)} K
 \equiv P_{+}^{(\epsilon)} \; , \\ [0.3cm]
P_{a_2} = \frac{\bigl(\hC_T -a_1\bigr)
 \bigl(\hC_T -a_3\bigr)}{(a_2 -a_1)(a_2-a_3)}  =
\frac{1}{2}(I - P) - \frac{(1-\epsilon)}{2(1-N +\epsilon )} K
\equiv P_{-}^{(\epsilon)} \; , \\ [0.3cm]
P_{a_3} = \frac{\bigl(\hC_T -a_1\bigr)
 \bigl(\hC_T -a_2\bigr)}{(a_3 -a_1)(a_3-a_2)}  =
 \frac{\epsilon}{N} K \equiv P_{0}^{(\epsilon)} \; .
\end{array}
\ee
Here, $a_1 = \frac{{\sf d}_2}{2}$, $a_2 = - \frac{{\sf d}_2}{2}$,
 $a_3 = - \frac{{\sf d}_2}{2}(N-\epsilon)$ are the roots of the characteristic equation (\ref{sosp30}),
 while $P_{a_1}$ and $P_{a_2}$ are, respectively, the  symmetrization
 and antisymmetrization operator for the algebras $so(N)$ $(\epsilon=+1)$ and   $sp(N)$ $(\epsilon=-1, N=2r)$.

Finally note that   the $so(N)$ (or $sp(N)$)-symmetric solution of the Yang-Baxter
equation in the defining representation (the so-called Zamolodchikov solution) can be written as follows  (see e.g.  \cite{Book1}):
   \be
 \lb{ybesop}
 R(u) = \frac{1}{\epsilon -u} \Bigl( u +  P -
 \frac{\epsilon u}{(u+N/2-\epsilon)} K \Bigr) =
 \ee
 $$
 = \frac{(u +1)}{(\epsilon -u)} P_{+}^{(\epsilon)} +
 \frac{(u-1)}{(\epsilon -u)}P_{-}^{(\epsilon)}  +
  \frac{(N/2 -\epsilon -u)}{(N/2 -\epsilon +u)} P_{0}^{(\epsilon)}
  \; , \;\;\;\;\; P \, R(u) \, P \, R(-u) = 1 \; .
 $$

 It is quite intriguing that the solution (\ref{ybesop}) can be elegantly represented
 as a rational function of the split Casimir operator
  \be
 \lb{best2}
 R(u) =
 \frac{\hC_T +{\sf d}_2 \, (\epsilon/2 +u)}{\hC_T +
 {\sf d}_2 \, (\epsilon/2 -u)} \; ,
 \ee
 where the constant ${\sf d}_2$ was introduced in (\ref{sosp30})
 (see also (\ref{kaz-11b}) and (\ref{Killing})).

\subsubsection{Operator $\hC$ for $so(N)$  and $sp(2n)$
in the adjoint representation\label{sospad}}

The split Casimir operator  $\hC_{\ad}$ for the algebras $so(N)$ and $sp(2n)$ in the adjoint representation together with  the construction of the projectors on the irreducible representations in $\ad \otimes \ad$ were considered in  detail in  \cite{IsPr}.
Thus, for  completeness,  we will present here only a short review of the results discussed in \cite{IsPr}.

The split Casimir operator  $\hC_{\ad}$ for the algebras  $\mfg= so(N),sp(N)$ in the adjoint representation  can be expressed through the split Casimir operator in the defining representation $\hC_T$ (\ref{casdef}) as
\cite{IsPr}:
\begin{equation}
\label{rel01}
(\hC_{\ad})^{k_1k_2k_3k_4}_{j_1j_2j_3j_4}=
4\; \delta^{[k_2}_{[j_2} \; (\hC_T)^{k_1)[k_3}_{j_1)[j_3} \;
\delta^{k_4)}_{j_4)} = 4 \, \bigl(
 \mcP_{12,34}^{(\epsilon)} \; (\hC_T)_{13} \; \mcP_{12,34}^{(\epsilon)}
\bigr)^{k_1k_2k_3k_4}_{j_1j_2j_3j_4}  \; ,
\end{equation}
where we introduce the projectors
 $$
  \mcP_{12,34}^{(\epsilon)}=
  \mcP_{12}^{(\epsilon)} \,  \mcP_{34}^{(\epsilon)}
  \; , \;\;\;
 \mcP_{ab}^{(\epsilon)} \equiv \frac{1}{2}(I-\epsilon P_{ab}) \; ,
 $$
 and use the index-free matrix notation explained at the
 beginning of Section {\bf \ref{slad}}.
Using the known expression (\ref{casdef2})
 for $\hC_T$,  we obtain
\begin{equation}
 \label{Csosp}
\hC_{\ad}=\frac{2}{(N-2\epsilon)} \;
  \mcP_{12,34}^{(\epsilon)} \;  (P_{13} -\epsilon K_{13})
\; \mcP_{12,34}^{(\epsilon)} \; .
\end{equation}

Due to the existence of accidental isomorphisms   $so(3) \simeq s\ell(2)\simeq sp(2)$,
 $so(4) \simeq s\ell(2) + s\ell(2)$,
 $so(5) \simeq sp(4)$ and $so(6) \simeq s\ell(4)$, in what follows we limit ourselves
to considering the algebras  $so(N)$ with $N \geq 7$ and $sp(N)$ with $N=2n \geq 4$, only.

Let us define the following operators acting in the space
 $V_{\ad}^\epsilon \otimes V_{\ad}^\epsilon \subset V_N^{\otimes 4}$:
 \be
\begin{array}{c}
\bI \equiv \mcP_{12,34}^{(\epsilon)}=
\mcP_{12}^{(\epsilon)}\mcP_{34}^{(\epsilon)} \, , \;\;\;
\bP \equiv \mcP_{12,34}^{(\epsilon)} P_{13}P_{24}
\mcP_{12,34}^{(\epsilon)} \, , \;\;\;
\bK \equiv \mcP_{12,34}^{(\epsilon)}K_{13}K_{24}
\mcP_{12,34}^{(\epsilon)} \,.
\end{array}
\ee
These operators obey  useful relations:
\be
\lb{PKC}
\begin{array}{c}
\bI=\bI \, P_{12} P_{34} =P_{12} P_{34} \,  \bI  \, , \;\;\;\;\;\;
\bP = P_{13} P_{24} \, \bI=\bI \, P_{13} P_{24} \, , \\ [0.3cm]
\bP^2=\bI\, , \;\;\;\;  \bK\, \bP=\bP\, \bK=\bK\, , \;\;\;\; \bK^2=\frac{M(M-1)}{2}\bK \, , \quad M \equiv \epsilon N\; ,
\end{array}
\ee
 \be
 \lb{PKC2}
 \hC_{\ad} \, \bP = \bP \,  \hC_{\ad} \; , \;\;\;\;\;
 \hC_{\ad} \, \bK = \bK \,  \hC_{\ad} = - \bK  \; ,
 \ee

It proved useful to define the symmetric $\hC_+$ and antisymmetric $\hC_-$ parts of the split Casimir operator $\hC_{\ad}$ as
 \begin{equation}
\label{Cpmsosp}
\hC_+=\frac{1}{2}\left( \bI +\bP\right) \hC_{\ad}, \qquad \hC_-=\frac{1}{2}\left( \bI -\bP\right) \hC_{\ad}.
\end{equation}

In the paper \cite{IsPr}, the following  proposition was proven.
 \newtheorem{prsosp}[pro1]{Proposition}
\begin{prsosp}\lb{prsosp}
The characteristic identities for the operators  $\hC_-,
\hC_+$ and $\hC_{\ad}$ for the algebras  $so(N)$ and $sp(N)$
read
\be
\label{charCm}
 \hC_-^2+\frac{1}{2}\hC_-=0 \;\;\;\; \Leftrightarrow \;\;\;\; \hC_-(\hC_-+\frac{1}{2})=0 \; ,
\ee
\begin{equation}\label{hcp3}
\hC_+^3=-\frac{1}{2}\hC_+^2-\frac{M-8}{2(M-2)^2}\hC_+
+\frac{M-4}{2(M-2)^3}(\bI+\bP-2\bK) \; ,
\end{equation}
\begin{equation}
 \lb{cp4K2}
\hC_+ (\hC_+ +1) \Bigl(\hC_+ -\frac{1}{(M-2)} \Bigr)
\Bigl(\hC_+ +\frac{2}{(M-2)}\Bigr)
\Bigl(\hC_+ + \frac{(M-4)}{2(M-2)} \Bigr)=0 \; ,
\end{equation}
\begin{equation}\label{char-poly}
\hC_{\ad}(\hC_{\ad}+\frac{1}{2})(\hC_{\ad}+1)(\hC_{\ad}-
\frac{1}{M-2})(\hC_{\ad}+\frac{2}{M-2})
(\hC_{\ad}+\frac{M-4}{2(M-2)})=0,
\end{equation}
where the parameter $M=\epsilon N$ is supposed to obey $M \geq 7$, $M \neq 8$ for
the algebras $so(N)$, and  $M \leq -4$ for the algebras $sp(N)$.
\end{prsosp}
{\bf Remark 1.} Identity (\ref{hcp3})
 is derived from the intermediate formula \cite{IsPr}:
 $$
  \hC_+^2 = \frac{1}{(M-2)^2} \, ({\bf I} + {\bf P} +{\bf K})
   - \frac{1}{(M-2)} \hC_+ +
   \frac{(M-8)}{2(M-2)^2} \, P_{12,34}^{(\epsilon)}
   K_{13}(1 +\epsilon \, P_{24}) P_{12,34}^{(\epsilon)} \; ,
 $$
 which is simplified for $M=8$ and instead of
 (\ref{hcp3}) we obtain the identity on $\hC_+$ of the second order
 \begin{equation}
 \lb{idso8}
\hC_+^2=-\frac{1}{6}\hC_++\frac{1}{36}(\bI+\bP+\bK) \; .
\end{equation}
That is why the case $M=8$ was excluded from Proposition
{\bf \em \ref{prsosp}}.

\vspace{0.2cm}

The characteristic identity (\ref{char-poly}) can be used to construct a complete system of orthogonal projectors on the invariant subspaces in  $V_{\ad}\otimes V_{\ad}$, which are simultaneously are eigenspaces of the operator  $\hC_{\ad}$. They are defined in a standard way (see e.g. \cite{Cvit,Book2}):
\begin{equation}\label{bcProj}
\proj_j:= \proj_{a_j} =
\prod_{\substack{i=1\\i\neq j}}^6\frac{\hC_{\ad}
-a_i\bI}{a_j-a_i} \; ,
\end{equation}
where $a_i$ are the roots of the characteristic equation  (\ref{char-poly}):
\be
\label{rsosp}
a_1=0 , \,\; a_2 =-\frac{1}{2}  , \,\; a_3=-1  , \,\;
a_4 =\frac{1}{M-2}\, , \,\; a_5 =-\frac{2}{M-2} , \,\;
a_6=-\frac{M-4}{2(M-2)} ,
\ee
and the characteristic identity \p{char-poly} is
written in the form
\be\label{adSK1}
\left(\hC_{\ad}-a_1\right)\left(\hC_{\ad}-a_2\right)
\left(\hC_{\ad}-a_3\right)\left(\hC_{\ad}-a_4\right)
\left(\hC_{\ad}-a_5\right)\left(\hC_{\ad}-a_6\right)=0.
\ee

There are some special cases in which some of the roots
 (\ref{rsosp}) coincide:
\begin{itemize}
\item $M=4$ - algebra $so(4)$. In this case, $a_5=a_3=-1$ and $a_6=a_1=0$. The correct characteristic identity reads
$$
\hC_{\ad}\left(\hC_{\ad}+\frac{1}{2}\right)\left(\hC_{\ad}+1\right)
\left(\hC_{\ad}-\frac{1}{2}\right)=0.
$$
In other words, the characteristic identity contains each factor $\hC_{\ad}$ and
$\left(\hC_{\ad}+1\right)$   only once.
\item $M=6$ - algebra $so(6)$. Now, $a_5=a_2 =-\frac{1}{2}$ and, therefore, the characteristic identity is of the fifth order:
$$
\hC_{\ad}\left(\hC_{\ad}+\frac{1}{2}\right)
\left(\hC_{\ad}+1\right)
\left(\hC_{\ad}-\frac{1}{4}\right)
\left(\hC_{\ad}+\frac{1}{4}\right)=0.
$$
\item $M=8$ - algebra $so(8)$. Now $a_5=a_6 =-\frac{1}{3}$ and again the characteristic identity is of the fifth order (this case will be considered in detail below):
 \be
 \lb{idso82}
\hC_{\ad}\left(\hC_{\ad}+\frac{1}{2}\right)\left(\hC_{\ad}+1\right)
\left(\hC_{\ad}-\frac{1}{6}\right)\left(\hC_{\ad}+\frac{1}{3}\right)=0.
 \ee
\item $M=5$. In virtue of the accidental automorphism  $so(5) = sp(4)$, this case is identical
to the case with $M=-4$.
\end{itemize}

Explicitly, the six projectors $\proj_1,...,\proj_6$
 in \p{bcProj}  were calculated in \cite{IsPr,Cvit}
\begin{align}
\proj_1&\equiv \proj_1^{-} =\frac{1}{2}(\bI-\bP)+2\hC_{-} \, , \nn \\
\proj_2&\equiv \proj_2^{-}=-2\hC_{-}\, , \nn \\
\proj_3&\equiv \proj_3^{+}=\frac{2\bK}{(M-1)M} \equiv
 \frac{\bK}{\rm dim \; \mathfrak{g}}\, ,\nn \\
\proj_4&\equiv \proj_4^{+} =\frac{2}{3}(M-2)\hC_{+}^2+\frac{M}{3}\hC_{+}+
\frac{(M-4)(\bI+\bP)}{3(M-2)}-\frac{2(M-4)\bK}{3(M-2)(M-1)}\, ,
\label{star} \\
\proj_5&\equiv \proj_5^{+}=-\frac{2(M-2)^2}{3(M-8)}\hC_{+}^2
-\frac{(M-2)(M-6)}{3(M-8)}\hC_{+}+
\frac{(M-4)(\bI+\bP)}{6(M-8)}+\frac{2\bK}{3(M-8)}\, , \nn \\
\proj_6&\equiv \proj_6^{+}=\frac{4(M-2)}{M-8}\hC_{+}^2+
\frac{4}{M-8}\hC_{+}
-\frac{4(\bI+\bP)}{(M-2)(M-8)}-\frac{8(M-4)\bK}{M(M-2)(M-8)}\, . \nn
\end{align}
The dimensions of the corresponding representations
 can be easily found by means of the trace formulas (\ref{trac1}),
 and we have (see \cite{Cvit}, Table 10.3)
\begin{align}
{\rm dim}(V_{a_1}) = \tr \proj_1&=\frac{1}{8}M(M-1)(M+2)(M-3)
\nonumber \; , \\
{\rm dim}(V_{a_2}) =\tr \proj_2&=\frac{1}{2}M(M-1)\nonumber\; ,\\
{\rm dim}(V_{a_3}) =\tr \proj_3&=1\label{traces}\; ,\\
{\rm dim}(V_{a_4}) =\tr \proj_4&=\frac{1}{12}M(M+1)(M+2)(M-3)
\nonumber\; ,\\
{\rm dim}(V_{a_5}) =\tr \proj_5&=\frac{1}{24}M(M-1)(M-2)(M-3)
\nonumber\; ,\\
{\rm dim}(V_{a_6}) = \tr \proj_6&=\frac{1}{2}(M-1)(M+2)
\nonumber \; .
\end{align}
\noindent
{\bf Remark 2.}
The characteristic identities (\ref{char-poly})
and dimensions (\ref{traces}) for $so(N)$ and $sp(N)$
 are related by replacement $N \to -N$.  This fact
manifests the duality between certain formulas
in the representation theories
of the algebras $sp(N)$ and $so(N)$ (see \cite{Mkr2}, \cite{Cvit},
\cite{Mkr3}
and references therein).

\subsubsection{The algebra $so(8)$}

For  the algebra $so(8)$  we have (\ref{charCm}), (\ref{idso8}):
\begin{equation}
\label{idso81}
\hC_-^2=-\frac{1}{2}\hC_- \, ,\quad \hC_+^2=-\frac{1}{6}\hC_+
 +\frac{1}{36}(\bI+\bP+\bK) \;
\end{equation}
and, therefore, the characteristic identity has the fourth order
(cf. (\ref{cp4K2}))
 \be
\lb{idso83}
 \hC_+(\hC_++1)(\hC_+-\frac{1}{6})(\hC_++\frac{1}{3}) =0 \; .
 \ee
All these imply the existence of the characteristic identity for the full split Casimir operator $\hC_{\ad}$ of
the fifth order (\ref{idso82}).
Thus, the operator $\hC_{\ad}$ has the following eigenvalues:
$$
a_1=0 \; , \;\;\;\; a_2 =-1/2\; , \;\;\;\;a_3=-1\; ,
\;\;\;\;a_4 =1/6  \; , \;\;\;\; a_5 = -1/3\; .
$$
All projectors $\proj'_k \equiv \proj'_{a_k}$ on the eigenspaces of the operator $\hC_{\ad}$ corresponding to the eigenvalues $a_k$
 have been constructed in \cite{IsPr}. However,
 not all $\proj'_k$ are projectors onto irreducible
representations, because of the different expansion of
the tensor product $\ad^{\otimes 2}(so(8))$
into the irreducible representations
 \cite{Cvit,LieART,Yamatsu}:
\begin{equation}\label{so8Rep}
\ad^{\otimes 2}(so(8))=[28]^2 = [1]+[28]+[35]+[35']+[35^{\prime\prime}]
+[300]+[350]  \;\;\; \Rightarrow
\end{equation}
 \be
 \lb{vvv}
 V_{\ad}^{\otimes 2} = V_1+V_{28}+V_{35}+V_{35'}
 +V_{35^{\prime\prime}} +V_{300}+V_{350} \; .
\ee
The projector $\proj_5^\prime =
\left. (\proj_5+\proj_6) \right|_{M=8}$ with the eigenvalue $(-1/3)$
corresponds to the space with dimension $105$ (see (\ref{traces})).
Therefore, $\proj'_5$ is not primitive,
and it can be further split into three projectors
$$
\proj_5^\prime=\oproj_0+\oproj_+ +\oproj_- ,
$$
 each of which corresponds to the space with dimensions $35$.
Finally, for the exceptional case of the algebra $so(8)$, the complete system of primitive projectors reads \cite{IsPr}:
\begin{equation}
\label{prso8}
\begin{array}{ll}
\proj_1^\prime= \left. \proj_1\right|_{M=8} =
\frac{1}{2}(\bI-\bP)+2\hC_-\;, & \dim=350 \; , \\ [0.3cm]
\proj_2^\prime= \left. \proj_2\right|_{M=8}
=-2\hC_-\;, & \dim=28 \; , \\ [0.3cm]
\proj_3^\prime= \left. \proj_3\right|_{M=8}
=\frac{1}{28}\bK\;, & \dim=1\; ,  \\ [0.3cm]
\proj_4^\prime= \left. \proj_4\right|_{M=8}
=\frac{1}{3}(\bI+\bP)+2\hC_+ + \frac{1}{21}\bK\;,
& \dim=300 \; , \\ [0.3cm]
\oproj_0 =\proj_5^\prime - A_4= \frac{1}{6}(\bI+\bP)-2\hC_+
-\frac{1}{12}\bK-A_4\;,\quad\quad & \dim=35\; , \\ [0.3cm]
\oproj_+ = \frac{1}{2}(A_4+E_4)\;, &
 \dim=35\; , \\ [0.3cm]
\oproj_- =  \frac{1}{2}(A_4-E_4)
\;, & \dim=35 \; .
\end{array}
\end{equation}
Here, $A_4$ is the antisymmetrizer in  $V_{8}^{\otimes 4}$ and
$E_4$ is the invariant operator in $V_{8}^{\otimes 4}$ with the components
$(E_4)^{i_1\dots i_4}{}_{j_1\dots j_4} = (4!)^{-1}
\varepsilon^{i_1\dots i_4}{}_{j_1\dots j_4}$,
where  $\varepsilon^{i_1\dots i_4}{}_{j_1\dots j_4} =
\varepsilon^{i_1\dots i_4 j_1\dots j_4}$
is a fully antisymmetric rank-eight tensor
$\varepsilon^{1\,2\,3\dots 8}=1$.
The operators $A_4$ and $E_4$ obey the conditions
$$
A_4^2=A_4 \; , \;\;\;\; A_4 E_4=E_4 A_4 = E_4
\; , \;\;\;\; E_4^2=A_4 \; ,
$$
 and $\oproj_+$ and $\oproj_-$  are the projectors
onto the self-dual and anti-self-dual parts of
$V_{8}^{\wedge 4} \equiv A_4 \; (V_{8}^{\otimes 4})$.
We note that the operator $E_4$ is independent of
the operators $\hC_\pm\,$, i.e., it cannot be expressed
as a polynomial function of $\hC_\pm$.

 \subsection{Universal characteristic identities for operator $\hC_+$
 in the case of Lie algebras of classical series\label{uncha}}
 \setcounter{equation}0

For the  algebras of the classical series  $A_n,B_n,C_n,D_n$
the characteristic identities (\ref{chcp2}) and (\ref{hcp3})
 for the operator $\hC_+$ in the adjoint representation can be written in a generic form
   \be
\lb{chcp4}
\hC_+^3 +\frac{1}{2} \hC_+^2 = \mu_1 \hC_+
 + \mu_2 (\bI^{(ad)} + \bP^{(ad)} -2 \bK) \; ,
\ee
where $\mu_1$ and $\mu_2$ are the parameters we define  at the moment.
Multiplying both sides of equation \p{chcp4} by $\bK$
and using  the relations
$$
\bK\, (\bI^{(ad)} + \bP^{(ad)})= 2 \, \bK \; , \;\;\;\;
\bK \, \hC_{+} = - \bK \; , \;\;\;\;
\bK \cdot \bK = \dim \mathfrak{g} \cdot \bK \; ,
$$
one may express the dimension of the Lie algebra  $\mathfrak{g}$ through the parameters $\mu_1$ and $\mu_2$
 \be
\lb{abcd01}
\dim \mathfrak{g} = \frac{2 \mu_2 - \mu_1 + 1/2}{2 \mu_2}  \; .
\ee
Then, we multiply both sides of \p{chcp4} by $\hC_+ (\hC_+ + 1)$
and deduce
the characteristic identity for $\hC_+$:
  \be
\lb{abcd02}
\hC_+ (\hC_+ + 1) (\hC_+^3 + \frac{1}{2}\hC_+^2
- \mu_1 \hC_+ - 2 \mu_2) = 0  \; .
\ee
which can be written in a factorized form
 \be
\lb{abcd03}
\hC_+ (\hC_+ + 1) (\hC_+ + \frac{\alpha}{2t})
 (\hC_+ + \frac{\beta}{2t})(\hC_+ + \frac{\gamma}{2t}) = 0
 \;\; \Leftrightarrow \;\;
 \prod_{i=1}^5 (\hC_+ - a_i) = 0 \; .
\ee
 Here we introduce the notation for the roots
 of the identity (\ref{abcd02})
 \be
 \lb{root01}
 a_1 = 0 \; , \;\;\; a_2 = -1 \; , \;\;\;
 a_3 = -\frac{\alpha}{2t} \; , \;\;\;
 a_4 = -\frac{\beta}{2t} \; , \;\;\;  a_5 = -\frac{\gamma}{2t} \; , \quad
  t=\alpha+\beta+\gamma \, ,
 \ee
 and the last equation follows from the condition
 $(a_3+a_4+a_5) = -1/2$.
 The parameter $t$ normalizes
 the eigenvalues of the operator $\hC_+$.
 For each simple Lie algebra $\mathfrak{g}$ we
 choose $t^{-1}$ such that
 \be\label{t}
 \left( \theta, \theta\right)=\frac{1}{t} \; ,
 \ee
 where  $\theta$ is the highest   root  of $\mathfrak{g}$.
Thus, $t$ coincides with the dual Coxeter number $h^{\vee}$  of the algebra $\mathfrak{g}$.
The parameters $\alpha, \beta,\gamma$ were introduced by Vogel \cite{Fog}.
The values of these parameters for the algebras $A_n,B_n,C_n,D_n$
 are extracted from identities (\ref{chs}), (\ref{cp4K2}),
 and we summarize them in  Table 3.
  \begin{center}
Table 3. \\
\begin{tabular}{|c|c|c|c|c|}
\hline
$\;\;$ & $s\ell(n+1)$ & $so(2n+1)$ & $sp(2n)$ & $so(2n)$ \\
\hline
$t$  &\footnotesize  $n+1$ &
\footnotesize  $2n-1$ &
\footnotesize $n+1$ &\footnotesize  $2n-2$   \\
\hline
$\frac{\alpha}{2 t}$  &\footnotesize  $-1/(n+1)$ &
\footnotesize  $-1/(2n-1)$ &
\footnotesize $-1/(n+1)$ &\footnotesize  $-1/(2n-2)$   \\
\hline
$\frac{\beta}{2 t}$  &\footnotesize $1/(n+1)$ &
\footnotesize $2/(2n-1)$ &
\footnotesize $1/(2n+2)$ &\footnotesize $1/(n-1)$ \\
\hline
$\frac{\gamma}{2 t}$  &\footnotesize $1/2$ &
\footnotesize $(2n-3)/(4n-2)$ &
\footnotesize $(n+2)/(2n+2)$ &\footnotesize $(n-2)/(2 n-2)$ \\
\hline
\end{tabular}
\end{center}

\noindent
Comparison of  equations (\ref{abcd02})
and (\ref{abcd03}) implies that
the parameters $\mu_1$ and $\mu_2$ are
expressed via the Vogel parameters as
 \be
\lb{abcd04}
\mu_1 = - \frac{\alpha\beta  + \alpha\gamma  + \beta\gamma}{4t^2}
 \; , \quad
\mu_2 = - \frac{\alpha\beta \gamma}{16 t^3} \; ,
\ee
and the dimensions \p{abcd01} of the simple Lie algebras
 acquire a remarkable universal form obtained
by Deligne and Vogel \cite{Delig},\cite{Fog}:
  \be
\lb{abcd06}
 \dim \mathfrak{g} =
 \frac{(\alpha-2t)(\beta-2t) (\gamma-2t)}{\alpha\beta \gamma} \; .
\ee
 Now by using the characteristic identity (\ref{abcd03}), one
 can obtain
the universal form of the projectors $\proj_{(a_i)}^{(+)}$ on the invariant
subspaces $V_{(a_i)}$ in the symmetrized space  $\frac{1}{2}(\bI^{(\ad)} + \bP^{(\ad)})\, (V_{\ad}^{\otimes 2})$:
$$
 \begin{array}{l}
\proj^{(+)}_{(-\frac{\alpha}{2t})} = 
 \frac{4t^2}{(\beta-\alpha)(\gamma-\alpha)}
 \Bigl(  \hC_+^2 + \bigl(\frac{1}{2} -\frac{\alpha}{2t}\bigr) \hC_+
 + \frac{\beta \gamma}{8 t^2}
 \bigl( \bI^{(\ad)} +  \bP^{(\ad)}
 - \frac{2 \alpha}{(\alpha - 2t)}\bK \bigr) \, \Bigr)
 \equiv \proj^{(+)}(\alpha|\beta,\gamma) \; ,
 \end{array}
$$
$$
\proj^{(+)}_{(-\frac{\beta}{2t})} = \proj^{(+)}(\beta|\alpha,\gamma)
\; , \;\;\;\;\;\;
\proj^{(+)}_{(-\frac{\gamma}{2t})} = \proj^{(+)}(\gamma|\alpha,\beta)
\; , \;\;\;\;\;\;
\proj^{(+)}_{(-1)} = \frac{1}{\dim \mathfrak{g}}\;  \bK \; .
$$
The irreducible representations that act in the subspaces
 $V_{(-1)}$, $V_{(-\frac{\alpha}{2t})}$,
 $V_{(-\frac{\beta}{2t})}$, $V_{(-\frac{\gamma}{2t})}$
 were respectively denoted in \cite{Fog} as
 ${\sf X}_0$, $Y_2(\alpha)$, $Y_2(\beta)$, $Y_2(\gamma)$;
 see Section {\bf \ref{Vogel}} below.
Finally, we calculate (by means of trace formulas (\ref{trac1}))
 the universal expressions \cite{Fog}
 for the dimensions of the invariant eigenspaces $V_{(a_i)}$:
 \bea
  &&\dim V_{(-1)} ={\bf Tr} \, \proj^{(+)}_{(-1)} = 1\, , \nonumber \\
  &&\dim V_{(-\frac{\alpha}{2t})} =
 {\bf Tr} \, \proj^{(+)}_{(-\frac{\alpha}{2t})}
 =-\frac{(3\alpha-2t)(\beta-2t)(\gamma-2t)t(\beta+t)(\gamma+t)}{
\alpha^2(\alpha-\beta)\beta(\alpha-\gamma)\gamma} \, , \lb{unidim02a}\\
 \lb{unidim02b}
  &&\dim V_{(-\frac{\beta}{2t})} =
 {\bf Tr} \, \proj^{(+)}_{(-\frac{\beta}{2t})}
 =-\frac{(3\beta-2t)(\alpha-2t)(\gamma-2t)t(\alpha+t)(\gamma+t)}{
\beta^2(\beta-\alpha)\alpha(\beta-\gamma)\gamma} \, ,\\
&& \dim V_{(-\frac{\gamma}{2t})} =
{\bf Tr} \, \proj^{(+)}_{(-\frac{\gamma}{2t})}
 = -\frac{(3\gamma-2t)(\beta-2t)(\alpha-2t)t(\beta+t)(\alpha+t)}{
\gamma^2(\gamma-\beta)\beta(\gamma-\alpha)\alpha}  \, . \lb{unidim02c}
\eea
Here we encounter an interesting nonlinear
Diophantine problem of finding all
integer $\dim \mathfrak{g}$ in (\ref{abcd06})
 for which the parameters
$\alpha,\beta$,$\gamma$ and
$\dim V_{(a_i)}$ are integers. The partial
solutions of this problem are given in Table 3.
The analogous Diophantine problems were considered in
\cite{Rub}, \cite{RuHu}.


\section{Split Casimir operators $\hC$ for  exceptional Lie algebras}

 \setcounter{equation}0
\subsection{Characteristic identities, projectors and  $R$-matrices
in the fundamental representation}

 \subsubsection{Basic definitions}

Let $T$  be the minimal fundamental representation of the exceptional Lie algebras
$\mathfrak{g} = \mathfrak{g}_2,\mathfrak{f}_4$,
 $\mathfrak{e}_6,\mathfrak{e}_7$, $\mathfrak{e}_8$,
acting in the space $V$. Let us choose the basis elements  $X_a$ of the algebras
$\mathfrak{g}$ so that the Cartan-Killing metric
 ${\sf g}_{ab}$ is proportional  to  $\delta_{ab}$, i.e.
\be\label{01}
 {\rm Tr} \left( T_a T_b \right) = d_2 {\sf g}_{ab} = - \delta_{ab} \; ,
\ee
where $T_a \equiv T(X_a)$. For this
 normalization the split Casimir operator in the representation $T$  reads $\hat{C} = {\sf g}^{ab} T_a \otimes T_b = - d_2 T_a \otimes T_a$.
Below we omit the constant parameter $d_2$ and use the following definition for the split Casimir operator
 \be
\lb{a02}
\hat{C}_{j_1j_2}^{i_1i_2} = \frac{{\sf g}^{ab}}{d_2}
\left( T_a\right)_{j_1}^{i_1} \;
\left( T_b\right)_{j_2}^{i_2} = -
\left( T_a\right)_{j_1}^{i_1} \;
\left( T_a\right)_{j_2}^{i_2} \, .
\ee

In what follows we also need the identity $I$
and permutation $P$ operators
acting in  $(V \otimes V)$ and defined as:
\be\label{05}
I_{j_1j_2}^{i_1i_2} = \delta_{j_1}^{i_1} \delta_{j_2}^{i_2}, \qquad
P_{j_1j_2}^{i_1i_2} = \delta_{j_2}^{i_1} \delta_{j_1}^{i_2}.
\ee
Using these operators one may define the
symmetric $\sS$ and anti-symmetric $\aA$
parts of the split Casimir operator $\hat{C}$
in the fundamental representation
\be\label{03f}
\begin{array}{c}
\sS_{j_1j_2}^{i_1i_2}=
\frac{1}{2}\bigl( (I + P) \, \hat{C}\bigr)_{j_1j_2}^{i_2i_1} =
\frac{1}{2} \left( \hat{C}_{j_1j_2}^{i_1i_2}+\hat{C}_{j_1j_2}^{i_2i_1}\right),
\\ [0.3cm]
\aA_{j_1j_2}^{i_1i_2}=
\frac{1}{2}\bigl( (I - P) \, \hat{C}\bigr)_{j_1j_2}^{i_2i_1} =
\frac{1}{2} \left( \hat{C}_{j_1j_2}^{i_1i_2}-\hat{C}_{j_1j_2}^{i_2i_1}\right)
\; , \;\;\;\; \hat{C} =  \sS + \aA  \; .
\end{array}
\ee
Note that by definition we have
 \be
 \lb{sex}
P \cdot \hat{C} = \hat{C} \cdot P \; , \;\;\;\;
 (I \pm P) \cdot \hat{C} \cdot (I \mp P) = 0
\; , \;\;\;\; \sS \cdot \aA = 0=\aA \cdot \sS \; ,
 \ee
and, in accordance with (\ref{01}) and (\ref{a02}), one obtains
 \be
 \lb{trc}
 {\rm Tr}_{12}(\sS) = -  {\rm Tr}_{12}(\aA) =
 \frac{1}{2} \dim \mathfrak{g} \; .
 \ee

Besides the standard Yang-Baxter equation
for $R$ matrix (\ref{YBE}), we also need
the Yang-Baxter equation for the twisted $R$-matrix:
 $\check{R}(u) \equiv P \, R(u)$. This equation follows from
(\ref{YBE}) and is written in the form
of the braid group relations
   \be
 \lb{g2-106}
 \check{R}_{12}(u) \,  \check{R}_{23}(u+v) \,
 \check{R}_{12}(v)  = \check{R}_{23}(v) \,
 \check{R}_{12}(u+v) \,  \check{R}_{23}(u) \; ,
 \ee
or in the components
\be\label{YB}
\check{R}(u)^{i_1 i_2}_{k_1 k_2}\,
\check{R}(u+v)^{k_2 i_3}_{j_2 l_3}\,
\check{R}(v)^{k_1 j_2}_{l_1 l_2} =
\check{R}(v)^{i_2i_3}_{k_2 k_3}\,
\check{R}(u+v)^{i_1 k_2}_{l_1 j_2}\,
\check{R}(u)^{j_2 k_3}_{l_2 l_3} \, .
\ee
Below we always require the unitarity condition
\be\label{unit}
\check{R}(u)^{i_1 i_2}_{k_1 k_2} \;
\check{R}(-u)^{k_1 k_2}_{j_1 j_2} =
 \delta^{i_1}_{j_1} \delta^{i_2}_{j_2} \;\;\; \Rightarrow \;\;\;
 \check{R}(u) \check{R}(-u) =
 P \, R(u) \, P \, R(-u) = 1
\ee
for the solutions of the  Yang-Baxter equations (\ref{YB}).

\subsubsection{Split Casimir operator and $R$-matrix for the algebra
 $\mathfrak{g}_2$}

The dimension of the minimal fundamental (defining) representation of the Lie algebra
$\mathfrak{g}_2$ is equal to 7. It is known that the algebra $\mathfrak{g}_2$
is embedded into $so(7)$ (see e.g.
\cite{Georgi}, \cite{Book1}). This embedding can
 be constructed by using the definition of the algebra $\mathfrak{g}_2$ as the algebra of differentiations of the octonions.
 The procedure looks as follows \cite{OgWieg}. We
 consider the  algebra $\mathbb{O}$
of  octonions  with the generators $e_0=1$, $e_i$ $(i=1,\dots,7)$ obeying the following
multiplication rules:
 \be
\lb{li5b}
 e_i \cdot e_j = - \delta_{ij} + f_{ijk} e_k \; , \;\;\;\;\;
 i,j,k = 1,2,\dots,7 \; ,
\ee
where the structure constants $f_{ijk}$ are the
components of the  completely anti-symmetric 3-rd rank tensor.
The non-zero components (including index permutations)
of this tensor are
\be\label{f}
f_{123}=f_{145}=f_{176}=f_{246}=f_{257}=f_{347}=f_{365} =1.
\ee
Let
$D$  be the differentiation of the algebra $\mathbb{O}$:
 \be
 \lb{li5d}
D(a \cdot b) = D(a) \cdot b + a \cdot D(b) \; , \;\;\;\;
\forall a,b \in \mathbb{O} \; , \;\;\;\;\;
D(1) = 0 \; , \;\;\;\;\; D(e_i) =  e_k D_{ki} \; ,
 \ee
where $D_{ik}$ is the matrix of the differentiation operator. If we differentiate the relation (\ref{li5b}), then using
 (\ref{li5d}) we obtain
 \be
 \lb{ddf}
 D_{ij}=-D_{ji} \; , \;\;\;\;\;
 D_{im} f_{mjk}  + D_{jm} f_{imk} + D_{km} f_{ijm}  = 0 \; .
\ee
In other words, the matrix  $D_{ij}$ is antisymmetric
and, therefore, it belongs to
the algebra  $so(7)$, while the second condition in (\ref{ddf})
shows that the tensor $f_{ijk}$ is invariant
under the action of the elements $D \in so(7)$.
This second condition produces 7 additional relations
 on the matrix $D_{ij}$
 $$
 \begin{array}{c}
D_{26} +  D_{15} =  D_{73}  \, , \;\;\;
 D_{63}  + D_{27} = D_{14} \, , \;\;\;
  D_{17} + D_{24} = D_{35}  \, , \;\;\;
 D_{43} + D_{16} = D_{25} \, , \\ [0.2cm]
 D_{13} + D_{64}= D_{57}  \, , \;\;\;
 D_{56} + D_{21} = D_{47}  \, , \;\;\;
  D_{23} + D_{45} = D_{67} \, ,
 \end{array}
 $$
 which reduce the number of its independent components
  to 14. For example, one can express the components $D_{1 i}$
 and $D_{23}$ through the other ones and expand the
 antisymmetric matrix $D$ over the remaining 14 free parameters
to obtain the basis in the algebra $\mathfrak{g}_2$.

The useful identities for $f_{ijk}$ follow from the definition  (\ref{f}):
 \be
 \lb{fff}
 f_{i j k} \; f_{j k \ell} = 6 \delta_{i \ell} \; ,
 \ee
 \be
 \lb{g2-001}
f_{i j k} \; f_{k \ell m}  \; f_{m r i}  = + 3 \; f_{j \ell r}
 \; .
\ee

\vspace{0.3cm}

The split Casimir operator (\ref{a02}) of the
algebra $\mathfrak{g}_2$  in the minimal
fundamental representation $[{\sf 7}]$ acts in the reducible
49-dimensional space $[7] \times [7]$ which can be
expanded in the irreducible representations as follows:
 \be
 \lb{g2jj}
 [7] \times [7] = \mathbb{S}([7] \times [7]) + \mathbb{A}([7] \times [7])=
 ([1] + [27]) + ([7] + [14]) \; .
 \ee
 Here the fundamental $[7]$ and adjoint $[14]$ representations embed into the antisymmetric
 part of $[7] \times [7]$, while
 the representations $[1]$ and  $[27]$ compose
 the symmetric part of $[7] \times [7]$.

 In the space of  the representation
 $[7] \times [7]$ we define four operators
  \be
 \lb{g2-101}
 (I)^{i_1i_2}_{j_1j_2}  = \delta^{i_1}_{j_1} \delta^{i_2}_{j_2} ,
 \;\; (P)^{i_1i_2}_{j_1j_2}
 = \delta^{i_1}_{j_2} \delta^{i_2}_{j_1} \, , \;\;
 (K)^{i_1i_2}_{j_1j_2}  =
 \delta^{i_1 i_2} \delta_{j_1 j_2} \, , \;\;
 (F)^{i_1i_2}_{j_1j_2}  = f^{i_1 i_2 m} \, f_{m j_1 j_2}  ,
 \ee
  invariant with respect to the action of the
  algebra $\mathfrak{g}_2$, i.e.,
 for any linear combination $X$ of the operators (\ref{g2-101}) we have
 \be
 \lb{adinv}
 \left( a \otimes I + I \otimes a \right) \cdot X =
  X \cdot \left( a \otimes I + I \otimes a \right) \; ,
  \;\;\;\;\; \forall a \; \in \; \mathfrak{g}_2 \; .
 \ee
Using operators (\ref{g2-101}) one may construct
four mutually orthogonal projectors \cite{EOgie}:
  \be
 \lb{g2-102}
  \begin{array}{c}
 P^{[1]} = \frac{1}{7} K \; , \;\;\; P^{[7]} = \frac{1}{6} F \; , \;\;\;
 P^{[27]} = \frac{1}{2} \, (I + P) - \frac{1}{7} \, K \; ,
 \\ [0.2cm]
  P^{[14]} = \frac{1}{2} \, (I - P) - \frac{1}{6}\, F  \; ,
 \end{array}
 \ee
 which form the complete system  $P^{[1]}+P^{[7]}+P^{[27]}+P^{[14]}=I$ and extract irreducible representations in the tensor product of two defining representations $[7] \times [7]$.

Note that the explicit form of the projector $P^{[7]}$ leads to the interpretation
of the structure constants $f^{i_1 i_2 m}$ as
Clebsch-Gordan coefficients
describing the fusion of two fundamental
representations $[7] \times [7]$ into one such
representation $[7]$.

The symmetric part $\sS$ of the split Casimir operator is
 expressed through the
invariant structures  \p{g2-101} as
\be
 \lb{simcas}
\sS = \frac{1}{6} \left( I +P \right) - \frac{1}{3} K .
\ee
while the projectors  $\proj^{(7)}$ and $\proj^{(14)}$ are related to
 the antisymmetric part $\aA$ of the split Casimir operator
\be\lb{ProjG2}
\proj^{(7)}  =  -\aA, \quad
\proj^{(14)} =  \frac{1}{2} \left( I - P \right) + \aA \; .
\ee
Therefore, the split Casimir operator (\ref{a02}) for the algebra $\mathfrak{g}_2$
reads
 \be
 \lb{g2-110}
 \hat{C} =  \frac{1}{6}(I + P - 2 \, K - F )  \; .
 \ee
In addition, we have the following useful relation
\be
 \lb{g2-010}
 \hat{C}^{i_1 j_1}_{\;\; i_2 j_2} =
 - (T_a)^{i_1 i_2} \otimes (T_a)_{j_1 j_2} =
 - (P^{[14]})^{i_1 i_2}_{\;\; j_1 j_2} \; .
 \ee

 \newtheorem{prg2}[pro1]{Proposition}
\begin{prg2}\lb{prg2}
 The characteristic identity for the operator $\hat{C}$ reads
 \be
 \lb{chig2}
 \hat{C} (\hat{C}-1/3)(\hat{C}+1)(\hat{C}+2) = 0 \; .
 \ee
 \end{prg2}
 {\bf Proof.} The spectral decomposition for
 the split Casimir operator (\ref{g2-110}) follows
 from the definitions of the projectors in  (\ref{g2-102})
 \be
 \lb{g2-11}
 \hat{C} = \frac{1}{3} P^{[27]} - P^{[7]} - 2 P^{(1)}
  \;\; \Rightarrow \;\; \hat{C}\, P^{[14]} = 0 \; .
 \ee
Thus, the operator $\hat{C}$ has four eigenvalues
 $a_1=0$, $a_2=1/3$, $a_3=-1$, $a_4=-2$
 (which correspond to the projectors  $P^{[14]}$, $P^{[27]}$, $P^{[7]}$, $P^{[1]}$, respectively), which immediately leads to the identity  (\ref{chig2}).
 \hfill \qed

 \vspace{0.1cm}

\noindent
For completeness, we give explicit formulas
for the projectors  $P_{a_i}$ in terms of the operator $\hat{C}$:
 $$
 \begin{array}{c}
 P_{0} = - \frac{3}{2} (\hat{C}-1/3)(\hat{C}+1)(\hat{C}+2)
 \equiv P^{[14]} \; , \;\;\;
 P_{1/3}  =
  \frac{27}{28} \hat{C}(\hat{C}+1)(\hat{C}+2)
 \equiv P^{[27]} \; , \\ [0.2cm]
 P_{-1}  =
  \frac{3}{4} \hat{C}(\hat{C}-1/3)(\hat{C}+2)
 \equiv P^{[7]} \; , \;\;\;
  P_{-2}  =
  -\frac{3}{14} \hat{C}(\hat{C}-1/3)(\hat{C}+1)
 \equiv P^{[1]} \; .
 \end{array}
 $$
These formulas are obtained by means of  identity (\ref{chig2})
 via the standard procedure.

The $\mathfrak{g}_2$-invariant solution
$R(u)$ of the Yang-Baxter equation (\ref{YBE}) in the defining representation was found  in \cite{EOgie},\cite{OgWieg}.
The braid form of this solution is
 \be
 \lb{g2-105b}
 \check{R}(u) = \frac{1}{1-u} \left( I - u \, P
 - \frac{2 u}{(u -6)} \, K
 +\frac{u}{(u - 4)}  \, F \right) \; .
 \ee
 Using formulas (\ref{g2-102}) we obtain
the spectral decomposition of this solution
 \be
 \lb{g2-77}
 \check{R}(u) =
 \frac{(u + 1)(u + 6)}{(u - 1)(u - 6)}  \,  P^{[1]}
 - \frac{(u + 4)}{(u-4)} \, P^{[7]} -
 \frac{(u+1)}{(u-1)}  \,  P^{[14]} + P^{[27]}  \; ,
 \ee
 after which the fulfillment of the unitarity
 condition (\ref{unit}) for this
 $R$-matrix becomes evident.

Finally, the standard  $\mathfrak{g}_2$-invariant
$R$-matrix acquires the form \cite{OgWieg}, \cite{Kuni}
\be\label{g2R}
 \begin{array}{c}
 \displaystyle
 R(u) =  P \check{R}(u)
 =  \frac{(u+6)(u+1)}{(u-6)(u-1)} P^{[1]} +\frac{u+4}{u-4} P^{[7]} +\frac{u+1}{u-1}P^{[14]} +P^{[27]} = \\ [0.2cm]
 \displaystyle
 = \frac{1}{u-1} \left( u-P
 + \frac{2 u}{(u -6)} \, K
+ \frac{u}{(u - 4)}  \, F \right) \; .
 \end{array}
\ee
Remarkably, this $\mathfrak{g}_2$-invariant
solution of the Yang-Baxter equation
can be rewritten as the rational function of
the symmetric and antisymmetric parts of
the split Casimir operator
 \be
 \lb{Rg2}
R(u) =  \frac{(3\, \sS - u)}{(3\, \sS + u)} \; \cdot \;
\frac{(3\, \aA -1 - u)}{(3\, \aA -1 + u)} \; .
 \ee

\subsubsection{Split Casimir operator and $R$-matrix for the algebra $\mathfrak{f}_4$}\label{ff4}

The minimal fundamental (defining) representation $T$
of the Lie algebra  $\mathfrak{f}_4$ has dimension 26.
In this representation the algebra  $\mathfrak{f}_4$
can be embedded into the algebra $so(26)$.

We follow the approaches of \cite{Post} and  \cite{BCCS}
 to define the basis of the algebra $\mathfrak{f}_4$ in the  representation $T$. For this, we consider  the Jordan algebra
 $\mathbb{J}_3$ that consists of
  the hermitian $3 \times 3$ matrices
 with the elements:
 $o_{\alpha\beta} = \bar{o}_{\beta\alpha}  \in \mathbb{O}$ $(\alpha,\beta=1,2,3)$:
 \be
 \lb{matJ3}
 A = \left(
 \begin{array}{ccc}
 x_1 & o_{12} & o_{13} \\
 \bar{o}_{12} & x_2 & o_{23} \\
  \bar{o}_{13} &  \bar{o}_{23} & x_3
 \end{array}
 \right) \; , \;\;\; x_{\alpha} \equiv  o_{\alpha\alpha} =
 \bar{o}_{\alpha\alpha} \in \mathbb{R} \; .
 \ee
Thus, the matrices $A$ are defined by three octonions $\bar{o}_{12},\bar{o}_{13},\bar{o}_{23}$ and three real numbers $x_{\alpha}$, which form the $(3 \cdot 8 + 3)= 27$-dimensional
real space. Multiplication in the algebra $\mathbb{J}_3$ is defined  as
 $A \circ B \equiv \frac{1}{2} [A,B]_+$
 $(\forall A,B \in \mathbb{J}_3)$. It is easy to check that if  $A$ and
 $B$ belong to  $\mathbb{J}_3$, then $[A,B]_+$
also belongs to $\mathbb{J}_3$.

Let us choose in $\mathbb{J}_3$ the basis  $e_0=I_3$, ${\sf e}_i$ $(i=1,\dots,26)$,
where $I_3$ -- is the $3\times 3$ identity matrix,
while ${\sf e}_i$
are the basis elements in the space of traceless matrices in  $\mathbb{J}_3$
 \be
 \lb{basj3}
 {\sf e}_1 = \left(\!\!
 \begin{array}{ccc}
 \! 1 \! & \! 0 \! & 0 \\
 0 & \! -1 \! & 0 \\
 0 & 0 & 0
 \end{array}
 \!\!\right) \, , \;\;\;
 {\sf e}_{18} = \frac{1}{\sqrt{3}}
 \left(\!\!
 \begin{array}{ccc}
 \! 1 \! & \! 0 \! & 0 \\
 0 & \! 1 \! & 0 \\
 0 & 0 & -2
 \end{array}
 \!\!\right) \, , \;\;\;
 {\sf e}_{1+a} = \left(\!\!
 \begin{array}{ccc}
 \! 0  \! & \! e_a \! & 0 \\
 \bar{e}_a & \! 0 \! & 0 \\
 0 & 0 & 0
 \end{array}
 \!\!\right) \, , \;\;\;
 \ee
 $$
   {\sf e}_{9+a} = \left(\!\!
 \begin{array}{ccc}
 \! 0  \! & \! 0 \! & e_a \\
 0 & \! 0 \! & 0 \\
 \bar{e}_a  & 0 & 0
 \end{array}
 \!\!\right) \, , \;\;\;
 {\sf e}_{18+a} = \left(\!\!
 \begin{array}{ccc}
 \! 0  \! & \! 0 \! & 0 \\
 0 & \! 0 \! &e_a \\
  0 & \bar{e}_a  & 0
 \end{array}
 \!\!\right) \, , \;\;\; \;\;\;
 (a=1,2,\dots 8) \; .
 $$
Here $e_a \in \mathbb{O}$ are the basis octonions  (see (\ref{li5b})). The nomalization of elements (\ref{basj3})
is chosen so that
 ${\rm Tr}({\sf e}_{i} \circ {\sf e}_{j}) = 2 \delta_{ij}$
and the structure relations read
 \bea
 \lb{algJ3}
&& {\sf e}_i \circ {\sf e}_j \equiv \frac{1}{2} \left[ {\sf e}_i , {\sf e}_j \right]=
 \frac{2}{3} \delta_{ij} \, I_3
 - \frac{1}{2} \; d_{i,j,k} \; {\sf e}_k \;, \qquad
 (i,j,k=1,\dots,26)    \; ,\\
&&  {\rm Tr}({\sf e}_{i} \circ {\sf e}_{j}) = 2 \delta_{ij} .\nonumber
 \eea
Here, the structure constants  $d_{i,j,k}$ form the
completely symmetric 3-rd rank tensor and one can extract
the explicit values of these constants from equation (\ref{algJ3}).

Let  $D$  be the differentiation of the algebra $\mathbb{J}_3$ and
$D({\sf e}_i) =  {\sf e}_k D_{ki}$.
Acting by $D$ on the basic relation (\ref{algJ3})
we obtain
\be
 \lb{ddJ3}
 D_{ij}=- D_{ji} \; , \;\;\;\;\;
 D_{im} d_{mjk}  + D_{jm} d_{imk} + D_{km} d_{ijm} = 0 \; .
\ee
Thus, the matrices $D_{ij}$ of the
 differentiations of $\mathbb{J}_3$ belong to the algebra
$so(26)$ and their action
on the tensor  $d_{ijm}$ preserves it. The second condition in (\ref{ddJ3}) reduces
the number of  independent matrices $D_{ij}$ to 52, which is the dimension of the algebra $\mathfrak{f}_4$.

\vspace{0.3cm}

The split Casimir operator of the algebra $\mathfrak{f}_4$
in the representation $T$ acts in the
reducible 676-dimensional representation ${\sf [26]} \times [{\sf 26}]$
which can be
expanded over the irreducible representations as follows:
\be
 \lb{f4jj}
 [26] \times [26] =
 \mathbb{S}([26] \times [26]) +
 \mathbb{A}([26] \times [26])
 = ([1] + [26] + [324]) + ([52] + [273]) \, .
 \ee
 Here the antisymmetric
 part of $[{\sf 26}] \times [{\sf 26}]$ is decomposed into
the representation $[{\sf 273}]$ and the adjoint
 representation $[{\sf 52}]$, while
 the symmetric part of $[{\sf 26}] \times [{\sf 26}]$ is decomposed into representations $[{\sf 1}], [{\sf 26}]$ and  $[{\sf 324}]$.

 In the space of  the $[{\sf 26}] \times [{\sf 26}]$
  representation one may defined five operators invariant
 respect to the algebra $\mathfrak{f}_4$
 \be
 \lb{f4-101}
   \begin{array}{c}
 (I)^{i_1i_2}_{j_1j_2} = \delta^{i_1}_{j_1} \delta^{i_2}_{j_2}\; ,
 \;\;\; (P)^{i_1i_2}_{j_1j_2} = \delta^{i_1}_{j_2} \delta^{i_2}_{j_1}
 \; , \;\;\;
 (K)^{i_1i_2}_{j_1j_2} =
 \delta^{i_1 i_2} \delta_{j_1 j_2} \; , \;\;\; \\ \\
 (D)^{i_1i_2}_{j_1j_2} = d^{i_1 i_2 m} \, d_{j_1 j_2 m}  \; , \;\;\;
 (F)^{i_1i_2}_{j_1j_2} = T^{i_1 i_2}_{a} \, T_{a j_1 j_2}   \; ,
   \end{array}
 \ee
 where $a=1,2,\dots,52$ and $T_{a i j} \equiv
 \delta_{ik} (T_a)^k_{\;\; j}$ -- are the generators of the algebra $ \mathfrak{f}_4$
 in the fundamental representation. With our definitions, the structure constant
  $d_{i j k}$ and generators  $T_{a i j}$ obey the conditions:
  \be
 \lb{f4dt}
d_{i j k} \; d_{k \ell m}  \; d_{m r i}  = - 8 \; d_{j \ell r}, \quad d^{i_1 i_2 , m} \, d_{i_1 i_2 ,\ell} = \frac{56}{3} \delta^m_\ell
  , \quad   {\rm Tr}(T_a T_b)  = -\delta_{ab} \; .
 \ee

Using the operators (\ref{f4-101}), one may construct five mutually orthogonal projectors:
 \be
 \lb{f4-102}
  \begin{array}{c}
 P^{[1]} = \frac{1}{26} K \; , \;\;\; P^{(26)} =  \frac{3}{56}  D =
 \frac{1}{14} \, (I + P) - \frac{6}{7} \mathbb{SC}
  - \frac{1}{14} K \; , \;\;\; \\ [0.3cm]
 P^{[324]} =
 \frac{3}{7} \Bigl( (I + P) +2 \mathbb{SC} + \frac{1}{13} K \Bigr) \; ,
 \\ [0.3cm]
 P^{[273]} = \frac{1}{2} \, (I - P) + 2 \, \mathbb{AC} \; , \;\;\;
  P^{[52]} = F = - 2 \, \mathbb{AC}  \; ,
 \end{array}
 \ee
which form the complete system  $P^{[1]}+P^{[26]}+P^{[52]}+P^{[273]}+P^{[324]}=I$ and which single out irreducible representations in the tensor product of two defining representations  $[26] \times [26]$.

Note that  symmetric and anti-symmetric parts of the split Casimir operator can be
represented as
 \be
 \lb{f4asc}
 \mathbb{SC} = \frac{1}{12}(I +P -K- \frac{3}{4}D) \; ,
 \;\;\; \mathbb{AC} = - \frac{1}{2}F
 \ee
 while the full split Casimir operator reads
 \be
 \lb{Kazimir}
 \hC = \mathbb{SC} + \mathbb{AC} =
 \frac{1}{12}(I +P -K)  - \frac{1}{16}D - \frac{1}{2}F  \; .
 \ee

  \newtheorem{prf4}[pro1]{Proposition}
\begin{prf4}\lb{prf4}
The split Casimir operator $\hat{C}$ for the
algebra $\mathfrak{f}_4$ in the defining representation
satisfies the characteristic identity
 \be
 \lb{chif4}
 \hat{C} (\hat{C}+1)(\hat{C}+2)(\hat{C}+1/2) (\hat{C}-1/6)= 0 \; .
 \ee
 \end{prf4}
 {\bf Proof.}
 The spectral decomposition of the split Casimir operator
 (\ref{Kazimir}) follows
 from the definitions of the projectors in  (\ref{f4-102})
 \be
 \lb{f41}
 \hat{C} = \frac{1}{6} P^{[324]}
 - P^{[26]} - 2 P^{[1]} - \frac{1}{2} P^{[52]}
  \;\;\;\;\;\; \Rightarrow \;\;\;\;\;\;
  \hat{C}\, P^{[273]} = 0 \; .
 \ee
Thus, the operator $\hat{C}$ has five eigenvalues
 $a_1=0$, $a_2=-1$, $a_3=-2$,
 $a_4=-1/2$, $a_5=1/6$, which immediately
 leads to the identity (\ref{chif4}).
 \hfill \qed

 \vspace{0.2cm}

The spectral decomposition
of the $\mathfrak{f}_4$-invariant solution
$\check{R}(u)$ of the Yang-Baxter equation \p{YB} in the defining representation was obtained\footnote{There is a missprint
in the form of the $\mathfrak{f}_4$-invariant solution presented
in \cite{OgWieg}.} in \cite{OgWieg} and
has  the braid form
\be\label{r1F4}
\check{R}(u) =\frac{(u+9)(u+4)}{(u-9)(u-4)} P^{[1]} + \frac{(u+6)(u+1)}{(u-6)(u-1)} P^{[26]} -
\frac{u+4}{u-4} P^{[52]} - \frac{u+1}{u-1} P^{[273]}
 + P^{[324]} .
\ee
It is clear that the solution $\check{R}(u)$ (\ref{r1F4}) automatically obeys the unitarity condition.

Finally, the standard  $\mathfrak{f}_4$-invariant $R$-matrix
can be written in terms of the invariants (\ref{f4-101})
and acquires the form
\be\label{f4R}
 \begin{array}{l}
 \displaystyle R(u) =  P \check{R}(u) = \\ [0.2cm]
 \displaystyle  =   \frac{(u+9)(u+4)}{(u-9)(u-4)} P^{[1]} +
  \frac{(u+6)(u+1)}{(u-6)(u-1)} P^{[26]}
 + \frac{u+4}{u-4} P^{[52]} +\frac{u+1}{u-1} P^{[273]}
 + P^{[324]} = \\ [0.4cm]
 \displaystyle
 = \frac{1}{u-1} \left( u-P
 + \frac{u(u-1)}{(u-9)(u-4)} \, K
+ \frac{6 u}{(u - 4)} \, F +
\frac{3u}{4(u-6)}\, D \right) \; .
 \end{array}
\ee

Interestingly, this solution of the Yang-Baxter equation can be rewritten in the concise form as a rational function of
the symmetric and antisymmetric parts of the split Casimir operator
\be
 \lb{Rf4}
R(u) =  \frac{(6\, \sS' - u)}{(6\, \sS' + u)} \; \cdot \;
\frac{(6\, \aA' -1 - u)}{(6\, \aA' -1 + u)} \;\; .
 \ee
Here we introduced the notation $\sS' \equiv \sS + \beta P^{[1]}$,
$\aA' \equiv \aA - \beta P^{[1]}$ and $\beta=1/2$, or
$\beta=4/3$ (for both values of the parameter $\beta$ the operators (\ref{Rf4})
coincide in view of the characteristic identities
for $\aA$ and $\sS$).

\subsubsection{Split Casimir operator and $R$-matrix for the algebra
 $\mathfrak{e}_6$}

The algebra $\mathfrak{e}_6$, with dimension equal to 78, has two inequivalent
minimal fundamental representations $[{\sf 27}]$ and $[\overline{\sf 27}]$.
We only consider
the split Casimir operator of the algebra $\mathfrak{e}_6$
in the minimal
fundamental representation which acts in the reducible
$[{\sf 27}] \times [{\sf 27}]$-dimensional space. This space
 can be
expanded in the following irreducible representations
\be
 \lb{e6jj}
 [{\sf 27}] \times [{\sf 27}] =
 \mathbb{S}([{\sf 27}] \times [{\sf 27}]) +
 \mathbb{A}([{\sf 27}] \times [{\sf 27}])
 = ([\overline{\sf 27}] + [{\sf 351}]_1) + ([{\sf 351}]_2) \, .
 \ee
The mutually orthogonal projectors on these irreducible representations look as follow:
\be\label{Pre6}
P^{[27]}  = \frac{1}{15}\left( I+P\right) -\frac{3}{5} \sS , \quad
P_{(1)}^{[351]}  = -9 \aA = \frac{1}{2}\left( I- P\right), \quad
P_{(2)}^{[351]} =  \frac{13}{30}\left( I+ P\right) +\frac{3}{5} \sS,
\ee
where $\sS$ and $\aA$ are respectively the symmetric and antisymmetric parts
of the split Casimir operator $\hat{C}$ for the $\mathfrak{e}_6$ algebra in
the representation $[{\sf 27}]$.
  \newtheorem{pre6}[pro1]{Proposition}
\begin{pre6}\lb{pre6}
The operator $\hat{C}$ for $\mathfrak{e}_6$ algebra in
the representation $[{\sf 27}]$ satisfies
the characteristic identity
 \be
 \lb{chie6}
(\hat{C}+\frac{13}{9})(\hat{C}+\frac{1}{9})
 (\hat{C}-\frac{2}{9})= 0 \; .
 \ee
 \end{pre6}
 {\bf Proof.}
 The spectral decomposition for the operator $\hC$ follows
 from the definitions of the projectors in (\ref{Pre6}):
\be
 \lb{e6-11}
 \hC = \frac{1}{9} \bigl(2 P_{(2)}^{[351]} - P_{(1)}^{[351]}
 - 13 P^{[27]}  \bigr) \; .
\ee
Thus, the operator $\hat{C}$ has three eigenvalues
 $a_1=-\frac{13}{9}$, $a_2=-\frac{1}{9}$, $a_3=\frac{2}{9}$, which immediately leads to  identity (\ref{chie6}).
 \hfill \qed

The $\mathfrak{e}_6$-invariant solution
$\check{R}(u)$ of the Yang-Baxter equation \p{YB} in the defining representation $[{\sf 27}]$ has the form
\be\label{r1E6}
\check{R}(u) =\frac{u-4}{u+4} P^{[27]} +\frac{u+1}{u-1} P_{(2)}^{[351]} - P_{(1)}^{[351]} .
\ee
It is clear that the solution $\check{R}(u)$ (\ref{r1E6}) automatically obeys the unitarity condition.

Finally, the standard  $\mathfrak{e}_6$-invariant $R$-matrix
 acquires the form \cite{OgWieg}
\be\label{Re6}
R(u) = P \check{R}(u)
=\frac{u-4}{u+4} P^{[27]} +\frac{u+1}{u-1} P_{(2)}^{[351]}
+ P_{(1)}^{[351]} .
\ee

The $\mathfrak{e}_6$-invariant solution $R(u)$ in (\ref{Re6})
can be elegantly written  as a rational function of $\hC$
(compare with \p{ybesl}, \p{best2})
\be
\lb{Rce6}
R(u)=
 - \frac{(3 \hC +1/3+u)}{(3 \hC + 1/3-u)} .
\ee

\subsubsection{Split Casimir operator and $R$-matrix for the algebra
  $\mathfrak{e}_7$}

The dimension of the exceptional algebra  $\mathfrak{e}_7$ is 133.
The tensor product of its minimal 56-dimensional fundamental representations
has the following expansion into irreducible ones
\be\label{44}
[{\sf 56}] \times [{\sf 56}] = \mathbb{S}([{\sf 56}] \times [{\sf 56}])+
\mathbb{A}([{\sf 56}] \times [{\sf 56}]) =
\left( [{\sf 133}] + [{\sf 1463}]\right)  +
\left( [{\sf 1}] + [{\sf 1539}]\right) \, .
\ee
The mutually orthogonal projectors on these irreducible representations read:
\be
\label{Pre7}
\begin{array}{l}
 P^{[1]}  =  -\frac{1}{112}\,\left( I- P\right) - \frac{3}{7} \aA, \quad
P^{[133]}  = \frac{1}{16}\left( I+P\right) - \sS ,  \\ [0.3cm]
 P^{[1463]} = \frac{7}{16}\left( I+P\right) + \sS, \quad
P^{[1539]} =  \frac{57}{112}\,\left( I- P\right) +\frac{3}{7} \aA \; ,
\end{array}
\ee
where $\sS$ and $\aA$ are respectively the symmetric and antisymmetric parts
of the split Casimir operator $\hat{C}$ for the $\mathfrak{e}_7$ algebra in
the representation $[{\sf 56}]$.
  \newtheorem{pre7}[pro1]{Proposition}
\begin{pre7}\lb{pre7}
The operator $\hat{C}$ for the $\mathfrak{e}_7$ algebra in
the representation $[{\sf 56}]$ satisfies
the characteristic identity
 \be
 \lb{chie7}
\left(\hat{C}-\frac{1}{8}\right)\left(\hat{C}+\frac{7}{8}\right)\left(\hat{C}+\frac{19}{8}\right)\left(\hat{C}+\frac{1}{24}\right)= 0 \; .
 \ee
 \end{pre7}
 {\bf Proof.}
 The following spectral decomposition for the split Casimir operator follows
 from the definitions of the projectors in  (\ref{Pre7})
 \be
 \lb{spe7}
 \hC = \sS+\aA =
\frac{1}{8}\left( P^{[1463]}  - 7\, P^{[133]}\right)+
\frac{1}{24}\left( -57 \, P^{[1]}  - P^{[1539]}\right) \, ,
 \ee
Thus, the operator $\hat{C}$ has four eigenvalues
 $a_1=\frac{1}{8}$, $a_2=-\frac{7}{8}$, $a_3=-\frac{19}{8}$,  $a_4=-\frac{1}{24}$, which immediately leads to  identity (\ref{chie7}).
 \hfill \qed
\vspace{0.1cm}

The decomposition (\ref{44}) of the
antisymmetric part  $\mathbb{A}([{\sf 56}] \times [{\sf 56}])$ of the
tensor product $[{\sf 56}] \times [{\sf 56}]$
 contains the singlet
representation $[{\sf 1}]$ .
This means that the corresponding projector can be rewritten
in the form
$$
(P^{[1]})^{i_1 i_2}_{\;\; j_1 j_2} = - \frac{1}{56} \;
J^{i_1 i_2} \,
J_{\;\; j_1 j_2} \; , \;\;\;\;
J_{ik} = - J_{ki} \; , \;\;\;\; J^{ik} J_{kj} = \delta^i_j \; ,
$$
where $J_{ik}$ and $J^{ik}$ are the invariant antisymmetric metrics. With the help of these metrics one may raise and lower  indices of tensors. The existence of these
metrics indicates that the $\mathfrak{e}_7$ algebra in the representation
 $[{\sf 56}]$ is embedded as a subalgebra in the
 symplectic algebra $sp(56)$.

The $\mathfrak{e}_7$-invariant solution
$\check{R}(u)$ of the Yang-Baxter equation
(\ref{g2-106}) in the defining representation has  the form
\be\label{r1E7}
\check{R}(u)  =- \frac{(u-9)(u-5)}{(u+9)(u+5)} P^{[1]}  +
\frac{u-5}{u+5} P^{[133]} +
 \frac{u+1}{u-1}P^{[1463]} -P^{[1539]} ,
\ee
The solution $\check{R}(u)$ (\ref{r1E7}) obviously obeys the unitarity condition.

Finally, the standard  $\mathfrak{e}_7$-invariant
$R$-matrix acquires the form \cite{OgWieg}
\be\label{Re7}
R(u)  = P \, \check{R}(u)
= \frac{(u-9)(u-5)}{(u+9)(u+5)} P^{[1]}  +\frac{u-5}{u+5} P^{[133]} +
 \frac{u+1}{u-1}P^{[1463]} + P^{[1539]} .
\ee

Note that this solution of the Yang-Baxter equation can also be
written as a rational function of the "shifted"
symmetric and antisymmetric parts of the split Casimir operator
(cf. (\ref{Rf4}))
 \be
 \lb{Rce7}
 R(u)  = \frac{(u + 6 \, \sS' + 1/4)}{
 (u - 6 \, \sS' - 1/4)}
 \cdot \frac{(u + 6 \, \aA')}{(u - 6 \, \aA')} \; .
 \ee
Here, we introduced the "shifted" symmetric and antisymmetric
parts $\sS' \equiv \sS - \beta P^{[1]}$, $\aA' \equiv \aA + \beta P^{[1]}$,
where $\beta= 37/4$, or
$\beta= 21/4$. For both values of the parameter $\beta$ the operators (\ref{Re7}) coincide in view of the
characteristic identities for $\aA$ and $\sS$.

\subsubsection{Algebra $\mathfrak{e}_8$}

The exceptional Lie algebra $\mathfrak{e}_8$  has dimension 248. Its minimal fundamental
representation  has also dimension 248 and it appears to be an adjoint representation of the algebra $\mathfrak{e}_8 $.
The  split Casimir operators $\hC$ and their characteristic
identities for all exceptional Lie algebras $ \mathfrak{g}$ in the adjoint representation
are discussed in the next section {\bf \ref{adjex}}. So we postpone
the consideration of the operator $\hC$
for the Lie algebra $\mathfrak{e}_8$ in the minimal
fundamental (adjoint)
representation to subsection {\bf \ref{adje8}}.

It is known that there are no solutions of the Yang-Baxter equation ($R$ - matrices) for the
simple Lie algebras in the adjoint representation besides the $s\ell$-series of the Lie algebras. This is the consequence of the fact that the adjoint representation of the simple Lie
algebras $\mathfrak{g}$ (except $s\ell$ algebras)
 can not be extended to the representation of the Yangian $Y(\mathfrak{g})$ (see \cite{Drin1}, \cite{ChPr1}). However, the reducible representation
 $\mathfrak{g} \oplus \mathbb{C}$, which is a direct sum of the adjoint
 and trivial representations, can be extended
 to the representation of the Yangian $Y(\mathfrak{g})$.
 The $\mathfrak{g}$-invariant solution of the Yang-Baxter equation
can be constructed just within such extended
adjoint representation. For the extended adjoint
(minimal fundamental) representation $[{\sf 248}]$
of the algebra $\mathfrak{e}_8$, this solution has been constructed in  \cite{ChPr1} and \cite{Nicol}.
The explicit form of this solution, written in the form of spectral decomposition over projectors,
turns out to be  rather cumbersome and we will not present it here.
We assume that writing this solution
in terms of the symmetrized $\sS$ and
antisymmetrized $\aA $ parts of the operator $\hC$
will result in a more visual and compact formula.

  \setcounter{equation}0
\subsection{Characteristic identities for operator $\hC$  and
 invariant projectors for exceptional Lie algebras in
 the adjoint representations. \label{adjex}}

In this section, we will find characteristic identities
for the split Casimir operator $\hC $ in the adjoint representations for the exceptional algebras $\mathfrak{g} =
\mathfrak{g}_2, \mathfrak{f}_4, \mathfrak{e}_6, \mathfrak{e}_7$ and  $\mathfrak{e}_8$ .
As we noted at the end of theprevious section, the solutions of the Yang-Baxter equation that are invariant with respect to    actions of exceptional Lie algebras in the adjoint representation do not exist, so this topic is not covered here.

\subsubsection{Basic definitions}
Let us define the normalization of the generators $X_a$ of the exceptional
Lie algebra $\mathfrak{g}$ so that the Cartan-Killing metric
(\ref{li04}) looks like
\be\label{norm2}
{\rm Tr}\bigl(\ad(X_{a}) \; \ad(X_{d})\bigr) =
\sum_{c,b=1}^{{\rm dim}\;\mathfrak{g}}
(C_a)^c _{\; b} \; (C_d)^b_{\; c}
= - \delta_{a d} \; .
\ee
where $(C_d)^b_{\; c} \equiv C^b_{dc}$ are the structure constants of the Lie algebra
$\mathfrak{g}$.
The split Casimir operator in the adjoint representation reads
\be\label{02}
(\hC_{\ad})_{\;\; b_1b_2}^{a_1 a_2} =
- \sum_{d} \left( C_d\right)^{a_1}_{\;\; b_1} \;
\left( C_d\right)^{a_2}_{\;\; b_2} \; .
\ee
We will also need the identity
$\bI$ and permutation $\bP$ operators defined as:
 \be
 \lb{idper}
 \bI_{b_1 b_2}^{a_1 a_2} = \delta_{b_1}^{a_1} \delta_{b_2}^{a_2}
 \; , \;\;\;\;\;\; \bP_{b_1 b_2}^{a_1 a_2} =
 \delta_{b_2}^{a_1} \delta_{b_1}^{a_2} \; ,
 \ee
together with the operator (\ref{adK}),
which in the normalization (\ref{norm2})reads
\be\label{05a}
\bK_{b_1 b_2}^{a_1 a_2} = \delta_{b_1 b_2} \; \delta^{a_1 a_2} \; .
\ee

In what follows, similarly to the previous consideration, it proved useful to define the
symmetric $\hC_+$ and antisymmetric $\hC_-$ parts of the split Casimir operators in the adjoint representation
\be\label{03}
\begin{array}{c}
 \hC_{+} =  \bP_{+} \; \hC_{\ad} \;\; \Rightarrow \;\;
(\hC_{+})_{b_1 b_2}^{a_1 a_2}= \frac{1}{2} \left( (\hC_{\ad})_{b_1 b_2}^{a_1 a_2}+(\hC_{\ad})_{b_1 b_2}^{a_2 a_1}\right) \; , \\ [0.3cm]
\hC_{-} = \bP_{-} \; \hC_{\ad} \;\; \Rightarrow \;\;
(\hC_{-})_{b_1 b_2}^{a_1 a_2}= \frac{1}{2} \left( (\hC_{\ad})_{b_1 b_2}^{a_1 a_2}-(\hC_{\ad})_{b_1 b_2}^{a_2 a_1}\right),
\end{array}
\ee
where $\bP_{\pm} = \frac{1}{2}(\bI \pm \bP)$.

\subsubsection{Algebra $\mathfrak{g}_2$}

The tensor product of two adjoint
14-dimensional representations of the algebra
$\mathfrak{g}_2$ has the following decomposition into irreducible representations
 \cite{Cvit}, \cite{Yamatsu}:
\be\label{11}
[{\sf 14}] \times [{\sf 14}] = \mathbb{S}([{\sf 14}] \times [{\sf 14}])+
\mathbb{A}([{\sf 14}] \times [{\sf 14}]) =
\left([1]+ [27] + [77]\right)  + \left( [14] + [77^\star]\right) \, .
\ee
 The dimensions of two representations appearing in the decomposition
 of $\mathbb{A}([{\sf 14}] \times [{\sf 14}])$ are
  given by (\ref{XX123}).
The antisymmetric $\hC_-$ and symmetric $\hC_+$ parts
of the split Casimir operator $\hC_{\ad}$
in the adjoint representation obey the following identities:
\be\label{idg2}
\hC_- \Bigl(\hC_-+\frac{1}{2}\Bigr) = 0  \; , \;\;\;\;\;\;
\hC_+^2 = - \frac{1}{6} \hC_+ +
\frac{5}{96} \left( \bI + \bP +\bK \right) \; .
\ee
Here the first identity is fulfilled for all simple Lie algebras,
 while the second one has been obtained
by direct explicit calculations with the help
of the $Mathematica^{TM}$ package (for details, see \cite{Math}).

Multiplying both parts of  equation   (\ref{idg2}) by  $\hC_{+}$ and using the
relations (\ref{adCpm}),(\ref{idK2}), one can obtain
\be\label{idg3}
\hC_{+} (\hC_{+} - \frac{1}{4}) (\hC_{+} + \frac{5}{12}) =
- \frac{5}{96} \bK
\;\;\;\; \Rightarrow  \;\;\;\;
\hC_{+} (\hC_{+} + 1) (\hC_{+} - \frac{1}{4})
 (\hC_{+} + \frac{5}{12}) = 0  \; ,
\ee
where the second identity follows from the first ones after multiplying it by $(\hC_{+} + 1)$ and taking into account (\ref{idK2}).

The characteristic identity for the complete
   Casimir operator $\hC_{\ad}=\hC_{+}+\hC_{-}$ can be obtained from
 identities  (\ref{idg2}) and (\ref{idg3}):
  \be
  \lb{ch-g2}
\hC_{\ad} \left(\hC_{\ad}+1 \right) \left(\hC_{\ad}+\frac{1}{2}\right)\left(\hC_{\ad}-\frac{1}{4} \right)
\left( \hC_{\ad} + \frac{5}{12} \right) =0 \; .
\ee
Using this identity together with the relation (\ref{idg2}), one can find the projectors $\cP_{\dim(V_i)}\equiv \proj_{(a_i)}$
on the eigenspaces $V_i$ of the operator  $\hC_{\ad}$ with the eigenvalues
$a_i= (-1,-\frac{1}{2},-\frac{5}{12},\frac{1}{4},0)$
and, at the same time, on the representations \p{11}. Finally,
 we derive
\bea
&&
\cP_{14}  = \proj_{(-\frac{1}{2})} = - 2 \hC_-
\; ,\quad
\cP_{77}^\star =  \proj_{(0)} =
\frac{1}{2} \left( \bI - \bP \right) + 2 \hC_- \; , \quad \nn \\
&&
\cP_1  = \proj_{(-1)} =  \frac{1}{14}\, \bK \; , \quad
\cP_{27} =  \proj_{(-\frac{5}{12})} = \frac{3}{16} \left( \bI + \bP \right) -\frac{3}{2} \hC_+ - \frac{15}{112} \bK \; ,
\label{ProjG2a} \\
&& \cP_{77} =  \proj_{(\frac{1}{4})} =
\frac{5}{16} \left( \bI + \bP \right) + \frac{3}{2} \hC_+ +\frac{1}{16} \bK \; ,
 \nn
\eea
where the first two and the last three projectors act
on  the antisymmetrized $ \bP_{-}(14^{\otimes 2})$,
and the symmetrized $\bP_{+}(14^{\otimes 2})$ parts
of the representation $14^{\otimes 2}$, respectively.
The dimensions $\dim(V_i)$ of the representations
corresponding to  the projectors
(\ref{ProjG2a}) are calculated using formulas (\ref{trac1}).

\subsubsection{Algebra $\mathfrak{f}_4$}

The exceptional Lie algebra $\mathfrak{f}_4$ has dimension 52.
The tensor product of its two adjoint 52-dimensional representations has the following decomposition into irreducible representations
 \cite{Cvit, Yamatsu}
\be\label{11a}
[{\sf 52}] \times [{\sf 52}] = \mathbb{S}([{\sf 52}] \times [{\sf 52}])+
\mathbb{A}([{\sf 52}] \times [{\sf 52}]) =
\left([1]+ [324] + [1053]\right)  + \left( [52] + [1274]\right) \, .
\ee
 The dimensions of two representations in the decomposition of
 $\mathbb{A}([{\sf 52}] \times [{\sf 52}])$ are calculated by means
 of formulas (\ref{XX123}).
The antisymmetric $\hC_-$ and symmetric $\hC_+$ parts
of the split Casimir operator $\hC$ for the algebra $\mathfrak{f}_4$
in the adjoint representation obey the following identities\footnote{The second identity for the symmetric part
of the Casimir operator here
 and for all other exceptional algebras below
were obtained by direct explicit calculations with the help
of the $Mathematica^{TM}$ package (for details, see \cite{Math}).}:
\be\label{idf4}
\hC_- \Bigl(\hC_-+\frac{1}{2}\Bigr) = 0  \; , \;\;\;\;\;\;
\hC_+^2 = - \frac{1}{6} \hC_+ +
\frac{5}{324} \left( \bI + \bP +\bK\right) \; .
\ee
Multiplying both parts of  equation   (\ref{idf4}) by  $\hC_{+}$ and using the
relations (\ref{adCpm}),(\ref{idK2}), one obtains
\be\lb{idf5}
\hC_{+} (\hC_{+} - \frac{1}{9}) (\hC_{+} + \frac{5}{18}) =
- \frac{5}{324} \bK
\;\;\;\; \Rightarrow  \;\;\;\;
\hC_{+} (\hC_{+} + 1) (\hC_{+} - \frac{1}{9})
 (\hC_{+} + \frac{5}{18}) = 0  \; ,
\ee
Thus, the characteristic identity for the full split Casimir operator  $\hC_{\ad}=(\hC_+ + \hC_-)$ reads
  \be
  \lb{ch-f4}
\hC_{\ad} \left(\hC_{\ad}+1 \right) \left(\hC_{\ad}+\frac{1}{2}\right)\left(\hC_{\ad}-\frac{1}{9} \right)
\left( \hC_{\ad} +\frac{5}{18} \right) =0 .
\ee
Finally, the projectors  $\cP_{\dim(V_i)}\equiv \proj_{(a_i)}$ onto the representations counted in the decomposition
 \p{11a}, or in other words,
onto the representations
acting in the eigenspaces $V_i$ of $\hC_{\ad}$ with the eigenvalues
$a_i = (-1,-\frac{1}{2},-\frac{5}{18},\frac{1}{9},0)$ can be found to be
\bea
&&
\cP_{52}  = \proj_{(-\frac{1}{2})} =  - 2 \hC_-, \quad
\cP_{1274} =  \proj_{(0)} =
\frac{1}{2} \left( \bI - \bP \right) + 2 \hC_-, \quad \nn \\
&&
\cP_1  =  \proj_{(-1)} =  \frac{1}{52}\, \bK, \quad
\cP_{324} =  \proj_{(-\frac{5}{18})} =
\frac{1}{7} \left( \bI + \bP \right) -\frac{18}{7} \hC_+ - \frac{5}{91} \bK, \label{Projf4} \\
&& \cP_{1053} =  \proj_{(\frac{1}{9})} = \frac{5}{14} \left( \bI + \bP \right) + \frac{18}{7} \hC_+ +\frac{1}{28} \bK \; .
  \nn
\eea
The dimensions $\dim(V_i)$ of the representations
corresponding to the projectors
(\ref{Projf4}) are calculated using  formulas (\ref{trac1}).

\subsubsection{Algebra $\mathfrak{e}_6$}
The exceptional Lie algebra $\mathfrak{e}_6$ has dimension 78.
The tensor product of its two adjoint 78-dimensional representations has the following decomposition into irreducible representations \cite{Cvit,Yamatsu}:
\be\label{11b}
[{\sf 78}] \times [{\sf 78}] = \mathbb{S}([{\sf 78}] \times [{\sf 78}])+
\mathbb{A}([{\sf 52}] \times [{\sf 52}]) =
\left([1]+ [650] + [2430]\right)  + \left( [78] + [2925]\right) \, .
\ee
 The dimensions of two representations in the decomposition of
 $\mathbb{A}([{\sf 78}] \times [{\sf 78}])$ are
 calculated by means of (\ref{XX123}).
The antisymmetric $\hC_-$ and symmetric $\hC_+$ parts of the split Casimir operator
in the adjoint representation obey the following identities:
\be\label{ide6}
\hC_- \Bigl(\hC_-+\frac{1}{2}\Bigr) = 0  \; , \;\;\;\;\;\;
\hC_+^2 = - \frac{1}{6} \hC_+ +
\frac{1}{96} \left( \bI + \bP +\bK\right) \; .
\ee
From these identities, similarly to the previously considered cases of
the $\mathfrak{g}_2$ and $\mathfrak{f}_4$ algebras, one obtains
\be\label{ide66}
\hC_{+} (\hC_{+} + \frac{1}{4}) (\hC_{+} - \frac{1}{12}) =
- \frac{1}{96} \bK
\;\;\;\; \Rightarrow  \;\;\;\;
\hC_{+} (\hC_{+} + 1) (\hC_{+} + \frac{1}{4})
 (\hC_{+} - \frac{1}{12}) = 0  \; ,
\ee
and, therefore,
  \be
  \lb{ch-e6}
\hC_{\ad} \left(\hC_{\ad}+1 \right) \left(\hC_{\ad}+\frac{1}{2}\right)\left(\hC_{\ad}+\frac{1}{4} \right)
\left( \hC_{\ad} -\frac{1}{12} \right) =0 .
\ee
Thus, the projectors  $\cP_{\dim(V_i)}\equiv \proj_{(a_i)}$ on the representations listed in
 \p{11b} and corresponding to the eigenvalues
 $a_i = (-1,-\frac{1}{2},-\frac{1}{4},-\frac{1}{12},0)$  read
 (cf. projectors in \cite{Cvit}, Table 18.5)
\bea
&&
\cP_{78}  = \proj_{(-\frac{1}{2})} =- 2 \hC_-, \quad
\cP_{2925} =  \proj_{(0)} =\frac{1}{2}
\left( \bI - \bP \right) + 2 \hC_- , \quad \nn \\
&&
\cP_1  =  \proj_{(-1)} =\frac{1}{78}\, \bK , \quad
\cP_{650} =  \proj_{(-\frac{1}{4})} =
\frac{1}{8} \left( \bI + \bP \right) - 3\, \hC_+ - \frac{1}{24} \bK, \label{ProjE6} \\
&& \cP_{2430} =  \proj_{(\frac{1}{12})} =\frac{3}{8} \left( \bI + \bP \right) + 3\, \hC_+  +\frac{3}{104} \bK\; . \nn
\eea
These projectors are built using the standard method with the help of  identity
(\ref{ch-e6}) and relations (\ref{ide6}), (\ref{ide66}).
The dimensions $\dim V_i$ of the representations
corresponding to the projectors
(\ref{ProjE6}) are calculated using  formulas (\ref{trac1}).

\subsubsection{Algebra $\mathfrak{e}_7$}

The exceptional Lie algebra $\mathfrak{e}_7$ has dimension 133.
The tensor product of its two  adjoint 133-dimensional representations has the following decomposition into irreducible representations \cite{Cvit,Yamatsu}:
\be\label{11c}
[{\sf 133}] \times [{\sf 133}] = \mathbb{S}([{\sf 133}] \times [{\sf 133}])+
\mathbb{A}([{\sf 133}] \times [{\sf 133}]) =
\left([1]+ [1539] + [7371]\right)
+ \left( [133] + [8645]\right) \, .
\ee
 The dimensions of two representations in the decomposition of
 $\mathbb{A}([{\sf 133}] \times [{\sf 133}])$ are
  given by formula (\ref{XX123}).
The antisymmetric $\hC_-$ and symmetric $\hC_+$ parts of the split Casimir operator
in the adjoint representation obey the following identities:
\be\label{ide7}
\hC_- \Bigl(\hC_-+\frac{1}{2}\Bigr) = 0  \; , \;\;\;\;\;\;
\hC_+^2 = - \frac{1}{6} \hC_+ +
\frac{1}{162} \left( \bI + \bP +\bK\right) \; .
\ee
From these identities we obtain
 \be\label{ide77}
\hC_{+} (\hC_{+} + \frac{2}{9}) (\hC_{+} - \frac{1}{18}) =
- \frac{1}{162} \bK
\;\;\;\; \Rightarrow  \;\;\;\;
\hC_{+} (\hC_{+} + 1) (\hC_{+} + \frac{2}{9})
 (\hC_{+} - \frac{1}{18}) = 0  \; ,
\ee
and, therefore,
\be
\lb{ch-e7}
\hC_{\ad} \left(\hC_{\ad}+1 \right) \left(\hC_{\ad}+ \frac{1}{2}\right)
\left(\hC_{\ad}+\frac{2}{9} \right)
 \left( \hC_{\ad}- \frac{1}{18} \right) =0 .
\ee
Finally, the projectors  $\cP_{\dim(V_i)}\equiv \proj_{(a_i)}$ on the representations appearing in the decomposition
\p{11c} and corresponding to the eigenvalues
 $a_i = (-1,-\frac{1}{2},-\frac{2}{9},\frac{1}{18},0)$
 read
 \bea
&&
\cP_{133} = \proj_{(-\frac{1}{2})} = - 2 \hC_-, \quad
\cP_{8645} = \proj_{(0)} =
\frac{1}{2} \left( \bI - \bP \right) + 2 \hC_- \nn \\
&&
\cP_1 = \proj_{(-1)} =  \frac{1}{133}\, \bK , \quad
\cP_{1539} = \proj_{(-\frac{2}{9})} =  \frac{1}{10}
\left( \bI + \bP \right) -\frac{18}{5} \hC_+ - \frac{1}{35} \bK, \label{ProjE7} \\
&& \cP_{7371} = \proj_{(\frac{1}{18})} =  \frac{2}{5}
\left( \bI + \bP \right) + \frac{18}{5} \hC_+
+\frac{2}{95} \bK \; . \nn
\eea
The dimensions $\dim V_i$ of the representations
related to the projectors
(\ref{ProjE7}) are calculated by  formulas (\ref{trac1}).

\subsubsection{Algebra $\mathfrak{e}_8$\label{adje8}}

The exceptional Lie algebra $\mathfrak{e}_8$ has dimension 248.
The tensor product of its two adjoint 248-dimensional representations has the following decomposition into irreducible representations \cite{Cvit,Yamatsu}:
\be\label{11d}
\begin{array}{c}
[{\sf 248}] \times [{\sf 248}] = \mathbb{S}([{\sf 248}] \times [{\sf 248}])+
\mathbb{A}([{\sf 248}] \times [{\sf 248}]) = \\ [0.2cm]
= \left([{\sf 1}]+ [{\sf 3875}] + [{\sf 27000}]\right)  +
\left( [{\sf 248}] + [{\sf 30380}]\right) \, .
\end{array}
\ee
 The dimensions of two representations in the decomposition of
 $\mathbb{A}([{\sf 248}] \times [{\sf 248}])$ are
  calculated by  formula (\ref{XX123}).
The antisymmetric $\hC_-$ and symmetric $\hC_+$
parts of the split Casimir operator $\hC_{\ad}$
in the adjoint representation obey the following identities:
\be\label{ide8}
\hC_- \Bigl(\hC_-+\frac{1}{2}\Bigr) = 0  \; , \;\;\;\;\;\;
\hC_+^2 = - \frac{1}{6}\hC_+ +
\frac{1}{300} \left( \bI + \bP +\bK\right) \; .
\ee
From these identities one can obtain
 \be\label{ide88}
\hC_{+} (\hC_{+} + \frac{1}{5}) (\hC_{+} - \frac{1}{30}) =
- \frac{1}{300} \bK
\;\;\;\; \Rightarrow  \;\;\;\;
\hC_{+} (\hC_{+} + 1) (\hC_{+} + \frac{1}{5})
 (\hC_{+} - \frac{1}{30}) = 0  \; .
\ee
The characteristic identity for the full split Casimir operator
$\hC_{\ad}=(\hC_{+}+\hC_{-})$ reads
\be
 \lb{ch-e8}
\hC_{\ad} \left(\hC_{\ad}+ 1 \right) \left(\hC_{\ad}+\frac{1}{2}\right)\left(\hC_{\ad}+\frac{1}{5} \right)
\left( \hC_{\ad}- \frac{1}{30} \right) =0 .
\ee
The projectors  $\cP_{\dim(V_i)}\equiv \proj_{(a_i)}$ on the representations in the decomposition
\p{11d}, which correspond to the eigenvalues
 $a_i = (-1,-\frac{1}{2},-\frac{1}{5},\frac{1}{30},0)$,  read
\be
\label{ProjE8}
 \begin{array}{l}
\cP_{248}  = \proj_{(-\frac{1}{2})} = - 2 \hC_-, \quad
\cP_{30380} =  \proj_{(0)} = \frac{1}{2}
\left( \bI - \bP \right) + 2 \hC_- \; ,
\\ [0.3cm]
\cP_1  =  \proj_{(-1)} = \frac{1}{248}\, \bK , \quad
\cP_{3875} =  \proj_{(-\frac{1}{5})}
=\frac{1}{14} \left( \bI + \bP \right)
-\frac{30}{7} \hC_+ - \frac{1}{56} \bK, \\ [0.3cm]
 \cP_{27000} =  \proj_{(\frac{1}{30})} =
 \frac{3}{7} \left( \bI + \bP \right) + \frac{30}{7} \hC_+ +\frac{3}{217} \bK \; .
\end{array}
\ee
The dimensions $\dim V_i$ of the
representations related to the projectors
(\ref{ProjE8}) are calculated by  formulas (\ref{trac1}).

\subsection{Universal characteristic identities and general
comments\label{uniex}}
\setcounter{equation}0

In the adjoint representations the antisymmetric parts of the split Casimir operators
$hC_-$ for all simple Lie algebras obey the same identity
\be\label{idCCa}
\hC_- \left( \hC_-+\frac{1}{2}\right) =0 .
\ee
The symmetric parts of the split Casimir operators $\hC_+$ in the adjoint representation
for the exceptional Lie algebras  obey  identities  (\ref{idg2}), (\ref{idf4}),
(\ref{ide6}), (\ref{ide7}) and (\ref{ide8}), which have a
similar structure\footnote{The universal
 formulae (\ref{genid}) was obtained  in \cite{Cvit}, eq. (17.10),
under the assumption that $\hC_+^{\, 2}$ is expressed as
 a linear combination of $\mathfrak{g}$-invariant
 operators $(\bI+\bP)$, $\bK$ and $\hC_+$.
 We explicitly checked this assumption
 for all exeptional Lie algebras.}
\be
\lb{genid}
\hC_+^2 = - \frac{1}{6} \hC_+ + \mu \; \left( \bI + \bP +\bK\right) \; ,
\ee
where the universal parameter  $\mu$ is fixed as follows:
 \be
\lb{unimu}
\mu = \frac{5}{6(2 + \dim(\mathfrak{g}))} \; .
\ee
Note that  identities  (\ref{idsl3}) and
  (\ref{idso8}) for the algebras  $s\ell(3)$ and
  $so(8)$ have the same structure.

From \p{genid} one can obtain the universal characteristic identity on the symmetric part of the split Casimir operator $\hC_+$
\be
 \lb{chplu}
 \hC_+(\hC_+ +1)(\hC_+^2 + \frac{1}{6}\hC_+ - 2\mu ) \equiv
 \hC_+(\hC_+ +1)(\hC_+ + \frac{\alpha}{2 t})
 (\hC_+ + \frac{\beta}{2 t} ) = 0 \; ,
 \ee
where we introduced the notation for two eigenvalues of the $\hC_+$ :
 \be
 \lb{albe}
 \frac{\alpha}{2 t} = \frac{1 - \mu'}{12} \; , \;\;\;\;
 \frac{\beta}{2 t} = \frac{1 + \mu'}{12} \; , \;\;\;\;\;\;
  \mu' := \sqrt{1+288\mu} =
  \sqrt{\frac{\dim \mathfrak{g}+ 242}{\dim \mathfrak{g} + 2}} \; .
 \ee
These parameters are related as $$3(\alpha + \beta) = t .$$
With the fixed value of the parameter $\alpha$, this relation defines the line of the exceptional Lie algebras on the $\beta,t$ plane (see eq.(\ref{exline}) below).
Following  \cite{Cvit}, note that $\mu'$ is a rational number only
for a certain sequence of dimensions $\dim \mathfrak{g}$.
It turns out that this sequence is finite\footnote{We thank D.O.Orlov
who proved the finiteness of this sequence.}:
 \be
 \lb{diof1}
\begin{array}{c}
\dim \mathfrak{g} = 3,8,14,28,
47, 52, 78, 96, 119, 133, 190, 248, 287, 336, \\
 484, 603, 782, 1081, 1680, 3479 \; ,
\end{array}
 \ee
which includes the dimensions $14,
52, 78, 133, 248$  of the exceptional Lie algebras
$\mathfrak{g}_2,\mathfrak{f}_4,\mathfrak{e}_6,
 \mathfrak{e}_7, \mathfrak{e}_8$, and the dimensions
 $8$ and  $28$ of the algebras $s\ell(3)$ and $so(8)$, which are sometimes also
 referred to as exceptional. Thus, for these algebras,
 using (\ref{albe}), we calculate the  values of the parameters
 $\frac{\alpha}{2 t},\frac{\beta}{2 t}$ given in Table 4.
 \begin{center}
Table 4. \\
\begin{tabular}{|c|c|c|c|c|c|c|c|}
\hline
$\;\;$ & $s\ell(3)$ & $so(8)$ & $\mathfrak{g}_2$ & $\mathfrak{f}_4$
& $\mathfrak{e}_6$ &
$\mathfrak{e}_7$ & $\mathfrak{e}_8$  \\
\hline
$\frac{\alpha}{2 t}$  &\footnotesize  $-1/3$ &\footnotesize  $-1/6$ &
\footnotesize $-1/4$ &\footnotesize  $-1/9$ &\footnotesize  $-1/12$
&\footnotesize  $-1/18$ &\footnotesize  $-1/30$  \\
\hline
$\frac{\beta}{2 t}$  &\footnotesize $1/2$ &\footnotesize $1/3$ &
\footnotesize $5/12$ &\footnotesize $5/18$ &\footnotesize $1/4$ &
\footnotesize $2/9$ &\footnotesize $1/5$ \\
\hline
\end{tabular}
\end{center}
These values are in agreement with the formulas  (\ref{idsl31}), (\ref{idso83}), (\ref{idg3}), (\ref{idf5}), (\ref{ide66}),
 (\ref{ide77}), (\ref{ide88}). Taking into account
 that $\hC_-$  satisfies (\ref{idCC}) and
 $\hC_+$ satisfies (\ref{chplu}),
 we obtain the following identities for the total
 split Casimir operator
 $\hC_{\ad}=(\hC_{+}+ \hC_{-})$ in the case of the exceptional
 Lie algebras:
 \be
 \lb{chexE}
\hC_{\ad} \left(\hC_{\ad}+\frac{1}{2}\right)
\left(\hC_{\ad}+ 1 \right) \left(\hC_{\ad}^2
+ \frac{1}{6} \hC_{\ad}-2  \mu \right) =0 \;\;\; \Rightarrow
\ee
\be
 \lb{chexe}
\hC_{\ad} \left(\hC_{\ad}+\frac{1}{2}\right)
\left(\hC_{\ad}+ 1 \right) \left(\hC_{\ad}+\frac{\alpha}{2t} \right)
\left( \hC_{\ad}+ \frac{\beta}{2t} \right) =0 \; .
\ee
Here $\mu$ is defined in (\ref{unimu}) and
$\frac{\alpha}{2t},\frac{\beta}{2t}$ are given in Table 4.

\noindent
{\bf Remark.} The sequence (\ref{diof1}) contains dimensions
  $\dim \mathfrak{g}^* =(10 m -122 +360/m)$,
 $(m \in \mathbb{N})$
 referring to the adjoint representations of the so-called
 $E_8$ family of algebras $\mathfrak{g}^*$;
see \cite{Cvit}, eq. (21.1). For these dimensions we have the relation
$\mu'=|(m+6)/(m-6)|$. Two numbers $47$ and $119$ from the
sequence (\ref{diof1}) do not belong to the sequence
$\dim \mathfrak{g}^*$. Thus, the
interpretation of these two numbers as  dimensions of some algebras is missing. Moreover, for values $\dim \mathfrak{g}$
 given in (\ref{diof1}), using (\ref{albe}),
 one can calculate dimensions (\ref{unidim02a})
 of the corresponding  representations $Y(\alpha)$:
 $$
 \begin{array}{c}
 \dim V_{(-\frac{\alpha}{2t})} =
 \left\{5,27,77,300,\frac{14553}{17},1053,2430,\frac{48608}{13},
 \frac{111078}{19},7371,15504,27000, \right .
\\ [0.2cm]
\left. \frac{841279}{23},\frac{862407}{17},
107892,\frac{2205225}{13},
\frac{578151}{2},559911,\frac{42507504}{31},
\frac{363823677}{61}\right\}
    \end{array}
 $$
 Since
 $\dim  V_{(-\frac{\alpha}{2t})}$ should be integer,
 we conclude that no Lie algebras exist with
 dimensions $47,96,119,287,336,
 603,782,1680,3479$, for which we assume characteristic
 identity (\ref{chplu}) and the trace formulas (\ref{trac1}).

\section{Universal characteristic identities for
operator $\hC$ for simple Lie algebras
in the adjoint representation and Vogel parameters\label{Vogel}}
\setcounter{subsection}1
In the previous sections, the projectors were constructed onto the spaces of irreducible subrepresentations in the
representation $\ad^{\otimes 2}$ for all
simple complex Lie algebras (Lie
algebras of classical series $A_n , B_n ,  C_n , D_n$ and
the exceptional Lie algebras). In all cases  the construction was
carried out by finding the characteristic identities for the split Casimir operators. In this regard, it should be noted that certain
 results of this work, namely the construction of projectors in terms of the split Casimir operator and finding  their dimensions can be obtained
 by using the Vogel parameters
$\alpha, \beta$ and $ \gamma $,
 which were introduced in
\cite{Fog} (see also \cite{Lan, MkrV}). The specific values of
these parameters correspond to each simple complex Lie
algebra.
These values and the value of
$t =(\alpha +\beta +\gamma)$ are given in Table 5 (see below).
Since all universal formulas for the simple Lie algebras
are written as homogeneous functions of the parameters $\alpha, \beta$ and $\gamma $, and these formulas
are independent of all permutations of $\alpha, \beta, \gamma$
one can consider simple Lie algebras as points in the space
$\mathbb{RP}^3/\mathbb{S}_3$.
 It is convenient to choose  normalization in which one of the parameters
is fixed, for example $\alpha = -2 $, which is already done in Table 5.
Note that the data in the first six lines of Table 5 coincide
with the data given in Table 3 of Section {\bf \ref{uncha}}.
We indicate the Vogel parameters for the algebras
$s\ell(3)$ and $so(8)$ in the separate lines of Table 5,
 since the characteristic identities (\ref{idsl3}),
(\ref{idsl31}) and  (\ref{idso81}), (\ref{idso82}) for
the symmetric part $\hC_{+}$ of the split
Casimir operator in the adjoint representations have the same order
and the same structure as for the exceptional Lie algebras
(cf. (\ref{genid}), (\ref{chplu})).

\begin{center}
Table 5. \\ [0.1cm]
\begin{tabular}{|c|c|c|c|c|c|c|c|c|}
\hline
Type & Lie algebra & $\alpha$ & $\beta$ & $\gamma$ & $t$ &
$-\frac{\alpha}{2t}=\frac{1}{t}$ & $-\frac{\beta}{2t}$
& $-\frac{\gamma}{2t}$ \\
\hline
$A_n$ & $s\ell(n+1)$ & $-2$ & $2$ & $n+1$ & $n+1$ & $\frac{1}{n+1}$
& $-\frac{1}{n+1}$ &\footnotesize  $-1/2$ \\
\hline
$B_n$ & $so(2n+1)$ & $-2$ & $4$ & $2n-3$ & $2n-1$ & $\frac{1}{2n-1}$
& $-\frac{2}{2n-1}$ &\footnotesize  $-\frac{2n-3}{2(2n-1)}$ \\
\hline
$C_n$ & $sp(2n)$ & $-2$ & $1$ & $n+2$ & $n+1$ & $\frac{1}{n+1}$
& $-\frac{1}{2(n+1)}$ &\footnotesize  $-\frac{n+2}{2(n+1)}$ \\
\hline
$D_n$ & $so(2n)$ & $-2$ & $4$ & $2n-4$ & $2n-2$ & $\frac{1}{2n-2}$
& $-\frac{1}{n-1}$ &\footnotesize  $-\frac{n-2}{2(n-1)}$ \\
\hline
$A_2$ & $s\ell(3)$ & $-2$ & $2$ & $3$ & $3$ &\footnotesize  $1/3$
&\footnotesize  $-1/3$ &\footnotesize  $-1/2$ \\
\hline
$D_4$ & $so(8)$ & $-2$ & $4$ & $4$ & $6$ & \footnotesize $1/6$
&\footnotesize $-1/3$ &\footnotesize  $-1/3$ \\
\hline
$G_2$ & $\mathfrak{g}_2$ & $-2$ & $10/3$ & $8/3$ & $4$ &\footnotesize $1/4$
&\footnotesize $-5/12$ &\footnotesize  $-1/3$ \\
\hline
$F_4$ & $\mathfrak{f}_4$ & $-2$ & $5$ & $6$ & $9$ &\footnotesize $1/9$
&\footnotesize $ -5/18$ &\footnotesize  $-1/3$  \\
\hline
$E_6$ & $\mathfrak{e}_6$ & $-2$ & $6$ & $8$ & $12$ &\footnotesize $1/12$
&\footnotesize  $-1/4$&\footnotesize  $-1/3$ \\
\hline
$E_7$ & $\mathfrak{e}_7$ & $-2$ & $8$ & $12$ & $18$ &\footnotesize $1/18$
&\footnotesize  $-2/9$ &\footnotesize  $-1/3$ \\
\hline
$E_8$ & $\mathfrak{e}_8$ & $-2$ & $12$ & $20$ & $30$ &\footnotesize $1/30$
&\footnotesize  $-1/5$ &\footnotesize  $-1/3$ \\
\hline
\end{tabular}
\end{center}

It proved useful to split the tensor product of two adjoint representations into the symmetric and antisymmetric parts
\be\label{new1}
\ad \otimes \ad = \mathbb{S}(\ad \otimes \ad)+
\mathbb{A}(\ad \otimes \ad).
\ee
In the general case of the Lie algebras of the classical
series\footnote{The algebras $s\ell(3)$ and $so(8)$ are exeptional
cases.},
the symmetric part $\mathbb{S}(\ad^{\otimes 2})$ decomposes
into 4 irreducible representations:
a singlet, denoted as ${\sf X}_0$,
with zero eigenvalue of the quadratic
  Casimir operator $C_{(2)}$ (which corresponds to the
  eigenvalue $(-1)$ for the split operator $\hC$),
  and 3 representations
  which we denote as $Y_2(\alpha),
  Y_2(\beta), Y_2(\gamma)$.
  Their dimensions, as well as the
  corresponding values $c_{(2)}^{(\lambda)}$ and $\hat{c}_{(2)}^{(\lambda)}$
    (here $\lambda = \mu, \mu', \mu^{\prime \prime} $ are the
  highest weights of the representations
  $Y_2(\alpha), Y_2(\beta), Y_2(\gamma)$)
  of the quadratic  Casimir operator $C_{(2)}$ (defined in (\ref{kaz-c2}))
  and  split Casimir operator $\hC$ are equal to:
\begin{align}
\dim Y_2(\alpha) & = \dim V_{(-\frac{\alpha}{2t})} \, , \;\;\;
 c_{(2)}^{(\mu)} =2-\frac{\alpha}{t} \, , \;\;\;
 \hat{c}_{(2)}^{(\mu)} =-\frac{\alpha}{2t} \, , \\
\dim Y_2(\beta)& =\dim V_{(-\frac{\beta}{2t})}  \, , \;\;\;
 c_{(2)}^{(\mu')}=2-\frac{\beta}{t}\, , \;\;\;
\hat{c}_{(2)}^{(\mu')}=-\frac{\beta}{2t} \, ,
\label{dimy2b} \\
\dim Y_2(\gamma)&=\dim V_{(-\frac{\gamma}{2t})}   \, , \;\;\;
c_{(2)}^{(\mu^{\prime\prime})}=2-\frac{\gamma}{t} \, , \;\;\;
\hat{c}_{(2)}^{(\mu^{\prime\prime})}=-\frac{\gamma}{2t} \, .
\label{dimy2g}
\end{align}
where the explicit expressions for
   $\dim V_{(-\frac{\alpha}{2t})}$,
$\dim V_{(-\frac{\beta}{2t})}$, $\dim V_{(-\frac{\gamma}{2t})}$
 are given in (\ref{unidim02a})--(\ref{unidim02c})
 and the eigenvalues $c_{(2)}^{(\lambda)}$ and
$\hat{c}_{(2)}^{(\lambda)}$ of the operators $C_{(2)}$ and $\hC$ are related by the condition (\ref{char03}):
\begin{equation}
\hc^{(\lambda)}_{(2)} =\frac{1}{2}(c_{(2)}^{(\lambda)}-2c_{(2)}^{\ad})
=\frac{1}{2}c_{(2)}^{(\lambda)}-1 \; .
\end{equation}
The eigenvalues $\hat{c}_{(2)}^{(\lambda)}$ of the operator
 $\hC$ on the representations $Y_2(\alpha),Y_2(\beta),Y_2(\gamma)$ in $\mathbb{S}(\ad\times \ad)$ are presented in  three last columns of  Table 5. Therefore, taking into account that $\hC_{+}$ has four
 eigenvalues
 $(-1,-\frac{\alpha}{2t},-\frac{\beta}{2t},-\frac{\gamma}{2t})$
 and $\hC_{-}$ has two eigenvalues $(0,-\frac{1}{2})$, the generic
 characteristic identity for the split Casimir operator reads
   \be
 \lb{chvog}
 \hC_{\ad} (\hC_{\ad} + \frac{1}{2})(\hC_{\ad} + 1)
 (\hC_{\ad} + \frac{\alpha}{2t})(\hC_{\ad} + \frac{\beta}{2t})
 (\hC_{\ad} + \frac{\gamma}{2t}) = 0 \; .
 \ee
  In the case of the $s\ell(N)$ algebras, the  eigenvalue $(-1/2)$ of the operator
 $\hC_{\ad}$  is doubly degenerated, since
 $ \frac{\gamma}{2t} = 1/2 $; therefore, in identity (\ref{chvog})
one should keep only one factor $ (\hC_{\ad} + \frac{1}{2})$ of two
(compare the identities (\ref{chad1}) and (\ref{chvog})).

We now turn to the discussion of the case of the exceptional Lie algebras.
Note that all exceptional Lie algebras are distinguished  in Table 5 by
the value of the parameter
$-\gamma/(2t)$ equals to $-1/3 $ (all other parameters
of the exceptional Lie algebras
in Table 5 are in agreement with the parameters listed in Table 4
of Section {\bf \ref{uniex}}).
Thus, all exceptional Lie algebras in the three-dimensional space of the Vogel parameters
$ (\alpha, \beta, \gamma) $ lie  in the plane $\alpha = -2$
 on the line:
\be
 \lb{exline}
 3 \gamma = 2 t \;\;\;\;\; \Rightarrow \;\;\;\;\;
 \gamma = 2 \beta - 4 \; .
 \ee
  We chose the coordinates $(\beta, \gamma)$ on this plane.
When the  condition (\ref{exline}) is fulfilled,
the dimension (\ref{unidim02c}),(\ref{dimy2g})  of
the space of the representation  $Y_2(\gamma)$ is zero
in view of the factor $(3 \gamma - 2 t)$
in the numerator of (\ref{unidim02c}).
So the corresponding projector $\proj_{(-\frac{\gamma}{2t})}$ on this space
is also equal to zero and the parameter
$ -\gamma/(2t) $  cannot be an eigenvalue of $\hC_{\ad}$
(in the case of the non-exceptional Lie algebras, this parameter is the eigenvalue of the operator $ \hC_{\ad} $ on the
representation $ Y_2(\gamma) $; see Subsection {\bf \ref{uncha}}). In this case,
in the general characteristic identity (\ref{chvog}) for the operator
$ \hC_{\ad} = (\ad \otimes \ad) (\hC)$,
 the last factor $ (\hC_{\ad} + \frac{\gamma}{2t})$ will be absent
 and the universal characteristic identity
 coincides with (\ref{chexe}):
 \be
 \lb{chvog01}
 \hC_{\ad} (\hC_{\ad} + \frac{1}{2})(\hC_{\ad} + 1)
 (\hC_{\ad} + \frac{\alpha}{2t})(\hC_{\ad} + \frac{\beta}{2t}) = 0 \; .
 \ee

As we showed in Subsection {\bf \ref{uniex}},
 identity (\ref{chvog01}) for the values of the
parameters $\alpha,\beta$ given in Table 4 and Table 5 exactly
reproduces the characteristic identities (\ref{ch-g2}), (\ref{ch-f4}), (\ref{ch-e6}), (\ref{ch-e7}) and (\ref{ch-e8}) for the   split Casimir operator $\hC_{\ad}$
  in the case of the exceptional Lie algebras. Note that
   both algebras $so(8)$ and $s\ell(3)$
(for the latter one has to replace the parameters $\beta\leftrightarrow\gamma$) lie on the line (\ref{exline})
and the characteristic identities (\ref{idso82}) and (\ref{chasl3})
are also given by the generic formula (\ref{chvog01}).
Indeed, for the algebra $s\ell(3)$ we
have $\frac{\gamma}{2t} = \frac{1}{2}$; therefore,
the eigenvalue $(-1/2)$ of the operator $ \hC_{\ad} $ is doubly degenerated and
one of the factors $ (\hC_{\ad} +1/2) $ in (\ref{chvog}) must be omitted. Wherein,
for the algebra $so(8)$ both parameters $\frac{\beta}{2t}$  and  $\frac{\gamma}{2t}$
are equal to the critical value  $\frac{1}{3}$, which gives zero in
denominators of the expressions (\ref{unidim02b}), (\ref{dimy2b})
and (\ref{unidim02c}), (\ref{dimy2g})
for the dimensions $\dim V_{(-\frac{\beta}{2t})}$
and $\dim V_{(-\frac{\gamma}{2t})}$ of the
representations $Y_2(\beta)$ and $Y_2(\gamma)$. However, these
zeros are canceled with zeros coming from the terms
$(3\beta - 2\, t)$ and $(3\gamma - 2\, t)$ in the numerators
of the  expressions for $\dim V_{(-\frac{\beta}{2t})}$, $\dim V_{(-\frac{\gamma}{2t})}$ and these dimensions
 turn out to be $35$, which is consistent with (\ref{so8Rep}).
 Since the eigenvalue $-\frac{\beta}{2t}=-\frac{\gamma}{2t}=-\frac{1}{3}$
 of the operator $\hC_{\ad}$ is doubly degenerated, we must omit
one of the factors $(\hC_{\ad} +1/3)$ in (\ref{chvog})
and this identity is transformed into identity (\ref{chvog01}).

The antisymmetric part $\mathbb{A}(\ad \otimes \ad)$ decomposes
for all simple Lie algebras into a direct sum of two terms
${\sf X}_1$ and ${\sf X}_2$ (see Section {\bf \ref{hCad}}),
one of which ${\sf X}_1$ is the adjoint
representation $\ad$ with the value
of the quadratic Casimir $c_{(2)}^{(\ad)} = 1 $,  and the other representation ${\sf X}_2$ has
the value of the quadratic Casimir $ c_{(2)}^{({\sf X}_2)} = 2$.
The representation ${\sf X}_2$ is reducible for the case of algebras
 $s\ell(N)$ (see Subsection {\bf \ref{slad}})
 and irreducible for all other simple Lie algebras.
The dimension of the representations
${\sf X}_1,{\sf X}_2$ and the corresponding eigenvalues
$ \hat{c}_{(2)}^{(\ad)} $ and
$\hat{c}_{(2)}^{({\sf X}_2)} $   are equal to
(cf. (\ref{XX123}))
 $$
 \begin{array}{c}
\dim {\sf X}_1= \dim \mathfrak{g} , \;\;\;\;\;\;
 \hat{c}_{(2)}^{(\ad)} = -1/2 \; ,  \\ [0.3cm]
\dim {\sf X}_2=\frac{1}{2}\dim \mathfrak{g}\; (\dim \mathfrak{g}-3), \;\;\;\;\;\;
\hat{c}_{(2)}^{({\sf X}_2)} = 0 \; .
\end{array}
$$
The values $\hat{c}_{(2)}^{(\ad)} $ and
$\hat{c}_{(2)}^{({\sf X}_2)} $ agree with the characteristic identity (\ref{idCC}) for the antisymmetrized part of $\hC_{-}$, which
is valid for all simple Lie algebras.

\section*{Acknowledgements}
The authors are thankful to O.V.~Ogievetsky who
draw our attention to the
 relation of the Vogel parametrization and
 characteristic identities of the
 split Casimir operator in the
 adjoint representation
   and to P.~Cvitanovi\'{c}, R.L.Mkrtchyan,
  M.A.Vasiliev for useful comments.
 We are thankful to D.~Lezin for the help
 with calculations  of the projectors (\ref{slProj1})
 at the initial stage. The authors are also grateful to
D.O.~Orlov and N.A.~Tyurin for the explanation of the methods for solving the nonlinear Diophantine equations. A.P.I. acknowledges the support of the Russian Science Foundation, grant No. 19-11-00131.

\bibliographystyle{abbrv}

\end{document}